\documentclass{aa}
\usepackage[dvips]{rotating}
\usepackage{txfonts,graphicx,natbib,epsfig,amssymb}
\usepackage{times,textcomp,longtable,lscape,color}
\usepackage{latexsym,times,amsfonts}
\usepackage{longtable,lscape}
\usepackage{subfigure,array, tabularx}
\newcounter{subtablec}

\newcommand\Ltable{\onecolumn\clearpage\setlength{\LTcapwidth}{\textwidth}}

\usepackage[english]{babel}
\usepackage[T1]{fontenc}

\bibpunct{(}{)}{;}{a}{}{,}

\newcount\longrefs
\def\aap{\ifnum\longrefs=1 {Astron.\ Astrophys.}\else 
                           {A\hbox{\rm \&}A}\fi}
\def\aapr{\ifnum\longrefs=1 {Astron.\ Astrophys.\ Rev.}\else 
                            {A\hbox{\rm \&}AR}\fi}
\def\aaps{\ifnum\longrefs=1 {Astron.\ Astrophys.\ Suppl.}\else 
                            {A\hbox{\rm \&}AS}\fi}
\def\aj{\ifnum\longrefs=1 {Astron.\ J.}\else 
                          {AJ}\fi} 
\def\ao{\ifnum\longrefs=1 {Applied Optics}\else 
                           {Appl.\ Opt.}\fi} 
\def\aspcs{\ifnum\longrefs=1 {Astron.\ Soc.\ Pacific Conf. Series}\else 
                           {ASP Conf.\ Ser.}\fi} 
\def\apj{\ifnum\longrefs=1 {Astrophys.\ J.}\else 
                           {ApJ}\fi} 
\def\apjl{\ifnum\longrefs=1 {Astrophys.\ J. Lett.}\else 
                            {ApJ}\fi} 
\def\aplett{\ifnum\longrefs=1 {Astrophys.\ J. Lett.}\else 
                            {ApJ}\fi} 
\def\apjs{\ifnum\longrefs=1 {Astrophys.\ J. Suppl.}\else 
                            {ApJS}\fi}
\def\apss{\ifnum\longrefs=1 {Astrophys.\ and Space Science}\else 
                            {Ap\hbox{\rm \&}SS}\fi}
\def\araa{\ifnum\longrefs=1 {Ann.\ Rev.\ Astron.\ Astrophys.}\else 
                            {ARA\hbox{\rm \&}A}\fi}
\def\azh{\ifnum\longrefs=1 {Astronomicheskii Zhurnal}\else 
                            {Astron.\ Zhur.}\fi}
\def\baas{\ifnum\longrefs=1 {Bull.\ Am.\ Astron.\ Soc.}\else 
                            {BAAS}\fi}
\def\bain{\ifnum\longrefs=1 {Bull.\ Astronom.\ Institutes Netherlands}\else
                            {Bull.\ Astr.\ Inst.\ Neth.}\fi}
\def\gca{\ifnum\longrefs=1 {Geochim.\ Cosmochim.\ Acta}\else 
                           {Geochim.\ Cosmochim.\ Acta}\fi}
\def\grl{\ifnum\longrefs=1 {Geophys.\ Res.\ Lett.}\else 
                           {Geoph.\ Res.\ Lett.}\fi}
\def\iaucirc{\ifnum\longrefs=1 {IAU Circulars}\else 
                          {IAU Circ.}\fi}
\def\ip{\ifnum\longrefs=1 {in press}\else 
                          {in press}\fi}
\def\jgr{\ifnum\longrefs=1 {J.\ Geophys.\ Res.}\else 
                           {J.\ Geophys.\ Res.}\fi}  
\def\jrasc{\ifnum\longrefs=1 {J.\ Royal Astron.\ Soc.\ Canada}\else 
                           {JRAS Can.}\fi}  
\def\mnras{\ifnum\longrefs=1 {Mon.\ Not.\ Roy.\ Astron.\ Soc.}\else 
                             {MNRAS}\fi} 
\def\nat{\ifnum\longrefs=1 {Nature}\else 
                           {Nat}\fi}
\def\pasj{\ifnum\longrefs=1 {Pub.\ Astron.\ Soc.\ Japan}\else 
                            {PASJ}\fi} 
\def\pasp{\ifnum\longrefs=1 {Pub.\ Astron.\ Soc.\ Pacific}\else 
                            {PASP}\fi} 
\def\physscr{\ifnum\longrefs=1 {Physica Scripta}\else 
                            {Phys.\ Scrip.}\fi} 
\def\planss{\ifnum\longrefs=1 {Planetary \& Space Science}\else 
                            {Plan. \& Space Sci.}\fi} 
\def\procspie{\ifnum\longrefs=1 {Proc.\ SPIE}\else 
                            {Proc.\ SPIE}\fi} 
\def\qjras{\ifnum\longrefs=1 {Quarterly J.\ Royal Astron.\ Soc.}\else 
                            {QJRAS}\fi} 
\def\sa{\ifnum\longrefs=1 {Soviet Astron..}\else 
                               {Sov.\ Astron.}\fi}
\def\skytel{\ifnum\longrefs=1 {Sky \& Telescope}\else 
                            {Sky \& Tel.}\fi} 
\def\solphys{\ifnum\longrefs=1 {Solar Phys.}\else 
                               {Solar Phys.}\fi}
\def\ssr{\ifnum\longrefs=1 {Space Science Rev.}\else 
                               {Space\ Sci.\ Rev.}\fi}

\def\nl{,\ } %%\def\nl{\newline}  %% redefine as \newline for mail addresses

\def\BIRA{BIRA-IASB\nl Ringlaan 3\nl B-1180 Brussel\nl Belgium}

\def\Leuven{Instituut voor Sterrenkunde\nl K.U. Leuven\nl Celestijnenlaan 200D\nl B-3001 Leuven\nl
Belgium}	
\def\Liege{Institut d'Astrophysique et de G\'eophysique\nl Universit\'e de Li\`ege\nl All\'ee
du 6 Ao\^ut 17, B\^at B5c\nl B-4000 Li\`ege\nl Belgium}

\def\Munich{Universit\"atssternwarte M\"unchen\nl Scheinerstr. 1\nl
D-81679 M\"unchen\nl Germany}

\def\Nijmegen{Departement Astrofysica\nl IMAPP\nl Radboud Universiteit
Nijmegen\nl PO Box 9010\nl 6500 GL Nijmegen\nl the Netherlands}

\def\dutch{\def\refname{Referenties}\def\abstractname{Samenvatting}%
  \def\bibname{Bibliografie}\def\chaptername{Hoofdstuk}%
  \def\appendixname{Bijlage}\def\contentsname{Inhoudsopgave}%
  \def\listfigurename{Lijst van figuren}\def\listtablename{Lijst van tabellen}%
  \def\indexname{Index}\def\figurename{Figuur}\def\tablename{Tabel}%
  \def\partname{Deel}\def\enclname{Bijlage(n)}\def\ccname{Ter attentie van}%
  \def\headtoname{Aan}\def\headpagename{Pagina}%
  \def\today{\number\day\space\ifcase\month\or januari\or februari\or maart\or%
     april\or mei\or juni\or juli\or augustus\or september\or oktober\or%
     november\or december\fi \space\number\year}%
  \typeout{
              >>>>> use hlatex209 for Dutch hyphenation <<<<< 
         }}
\newcounter{onefig} \newcounter{fignumber}
\newcount\nocaptions \newcount\nofigures \newcount\figwidth
\newcount\viewgraphs
  \def\paper{}  \def\figlabel{} 
\long\def\nextfig#1{\setcounter{figure}{\value{fignumber}}
  \addtocounter{fignumber}{1}
  \ifnum \viewgraphs=1 \newpage \pagestyle{empty} \fi 
  \ifnum\value{onefig}=0 #1 \fi                 
  \ifnum\value{onefig}=\value{fignumber} #1 \fi}
\def\figwidths#1#2{\ifnum \nocaptions=1 #2mm \else #1mm \fi}  
\def\paper#1{}  %% redefine for separate-figure identification line
\long\def\plotfig#1#2{\ifnum \nofigures=1 \else #2 \fi}
\long\def\captiontext#1{\ifnum \nofigures=1 \raggedright \fi 
   \ifnum \nocaptions=1 \paper
     \ifnum \viewgraphs=0 
       \newline  \mbox{}\hrulefill\mbox{} \newline 
       \newline label:~\{\figlabel\} 
     \fi 
     \else \ifnum \nofigures=0 \fi 
   #1 \fi}

\newcount\panelwidth \newcount\panelheight 
\newcount\bxmin \newcount\bymin \newcount\bxmax \newcount\bymax
\newcount\tbxmin \newcount\tbymin
\newcount\tpanelwidth \newcount\tpanelheight \newcount\tpdif
\def\panelsize #1,#2;{\panelwidth=#1 \panelheight=#2}  
\def\setbb #1,#2;#3,#4;#5,#6;{% UNITS: bp (from ghostview)
  \tbxmin=#1 \tbymin=#2    %% full box (axis titles) lower left corner
  \bxmin=#3 \bymin=#4      %% bare box (ticks only) lower left corner
  \bxmax=#5 \bymax=#6}     %% upper right corner
\def\barepanel #1{%
  \ifnum\panelheight=0 
    \tpdif=\bymax \advance\tpdif by -\bymin
    \multiply \tpdif by \panelwidth
    \tpanelheight=\tpdif
    \tpdif=\bxmax \advance\tpdif by -\bxmin
    \divide \tpanelheight by \tpdif
  \else \tpanelheight=\panelheight \fi
  \epsfig{file=#1,%
     bbllx=\bxmin bp,bblly=\bymin bp,bburx=\bxmax bp,bbury=\bymax bp,clip=,%
     width=\panelwidth mm,height=\tpanelheight mm}}
\def\labelypanel #1{% TeX permits only integer arithmetic, so bp and mm
  \ifnum\panelheight=0 
    \tpdif=\bymax \advance\tpdif by -\bymin
    \multiply \tpdif by \panelwidth
    \tpanelheight=\tpdif
    \tpdif=\bxmax \advance\tpdif by -\bxmin
    \divide \tpanelheight by \tpdif
  \else \tpanelheight=\panelheight \fi
  \tpdif=\bxmax \advance\tpdif by -\tbxmin
  \tpanelwidth=\panelwidth \multiply \tpanelwidth by \tpdif
  \tpdif=\bxmax \advance\tpdif by -\bxmin
  \divide \tpanelwidth by \tpdif
  \epsfig{file=#1,%
    bbllx=\tbxmin bp,bblly=\bymin bp,bburx=\bxmax bp,bbury=\bymax bp,%
    clip=,width=\tpanelwidth mm,height=\tpanelheight mm}}
\def\labelxpanel #1{%
  \ifnum\panelheight=0 
    \tpdif=\bymax \advance\tpdif by -\bymin
    \multiply \tpdif by \panelwidth
    \tpanelheight=\tpdif
    \tpdif=\bxmax \advance\tpdif by -\bxmin
    \divide \tpanelheight by \tpdif
  \else \tpanelheight=\panelheight \fi
  \tpdif=\bymax \advance\tpdif by -\tbymin
  \multiply \tpanelheight by \tpdif
  \tpdif=\bymax \advance\tpdif by -\bymin
  \divide \tpanelheight by \tpdif
  \epsfig{file=#1,%
    bbllx=\bxmin bp,bblly=\tbymin bp,bburx=\bxmax bp,bbury=\bymax bp,%
    clip=,width=\panelwidth mm,height=\tpanelheight mm}}
\def\labelxypanel #1{%
  \ifnum\panelheight=0 
    \tpdif=\bymax \advance\tpdif by -\bymin
    \multiply \tpdif by \panelwidth
    \tpanelheight=\tpdif
    \tpdif=\bxmax \advance\tpdif by -\bxmin
    \divide \tpanelheight by \tpdif
  \else \tpanelheight=\panelheight \fi
  \tpdif=\bxmax \advance\tpdif by -\tbxmin
  \tpanelwidth=\panelwidth \multiply \tpanelwidth by \tpdif
  \tpdif=\bxmax \advance\tpdif by -\bxmin
  \divide \tpanelwidth by \tpdif 
  \tpdif=\bymax \advance\tpdif by -\tbymin 
  \multiply \tpanelheight by \tpdif
  \tpdif=\bymax \advance\tpdif by -\bymin
  \divide \tpanelheight by \tpdif
  \epsfig{file=#1,%
    bbllx=\tbxmin bp,bblly=\tbymin bp,bburx=\bxmax bp,bbury=\bymax bp,%
    clip=,width=\tpanelwidth mm,height=\tpanelheight mm}}

\def\CC{\par \vspace*{-2ex} \footnotesize \baselineskip=8pt \begin{verbatim}}

\long\def\startignore #1\stopignore{}   %% use \startignore....\stopignore

\def\setlistparams{         
  \topsep=0.7ex                 %% ADAPT: parskip=0: 0.7;  parskip=1: -1.2ex
  \itemsep=0.7ex                %% space between items
  \leftmargini=3ex}             %% dashes at beginning of line 
\setlistparams                  %% recall after type changes 

\newcounter{alistindex}       %% problems: a)  b) etc

\newcounter{romenumnr}

\newlength{\minipagewidth}

\newsavebox{\boxcontent}
\newcommand{\ovalhead}[1]{
  \unitlength=1cm
  \sbox{\boxcontent}{\mbox{~~{#1}~~}}
  \begin{center}
    \ifdim\wd\boxcontent>6ex 
    \ifdim\wd\boxcontent<8cm 
    \begin{picture}(8,3) \thicklines     
      \put(4.0,0.8){\oval(8,1.6)} 
      \put(0.0,0.7){\parbox{8cm}{
         \begin{center} \usebox{\boxcontent} \end{center}}}
    \end{picture}
    \else \ifdim\wd\boxcontent<12cm 
    \begin{picture}(12,3) \thicklines     
        \put(6.0,0.8){\oval(12,1.6)} 
        \put(0.0,0.7){\parbox{12cm}{
           \begin{center} \usebox{\boxcontent} \end{center}}}
    \end{picture}
    \else
    \begin{picture}(16,3) \thicklines     
        \put(8.0,0.8){\oval(16,1.6)} 
        \put(0.0,0.7){\parbox{16cm}{
           \begin{center} \usebox{\boxcontent} \end{center}}}
    \end{picture}
    \fi \fi \fi
  \end{center}} 

\newcounter{headnr}            
\newcounter{subheadnr}[headnr]
\newcounter{subsubheadnr}[subheadnr]
\def\head #1\par{
  \stepcounter{headnr}                          %% sets subheadnr = 0 too 
  \vspace{2ex} \noindent                        %% 2ex = space above, no *
  {\bf \theheadnr~~~~#1}\\[1ex] \noindent}      %% 1ex = space below
\def\subhead #1\par{  
  \stepcounter{subheadnr}
  \vspace{1.3ex} \noindent
  {\bf \theheadnr.\arabic{subheadnr}~~~#1}\\[0.3ex] \noindent}
\def\subsubhead #1\par{
  \stepcounter{subsubheadnr}
  \vspace{1.0ex} \noindent
  {\bf \theheadnr.\arabic{subheadnr}.\arabic{subsubheadnr}~~~#1}\\ \noindent}

\font\dropfont= cmr12 scaled \magstep5
\def\dropcap#1#2{{\noindent
    \setbox0\hbox{\dropfont #1}\setbox1\hbox{#2}\setbox2\hbox{(}%
    \count0=\ht0\advance\count0 by\dp0\count1\baselineskip
    \advance\count0 by-\ht1\advance\count0by\ht2
    \dimen1=.5ex\advance\count0by\dimen1\divide\count0 by\count1
    \advance\count0 by1\dimen0\wd0
    \advance\dimen0 by.25em\dimen1=\ht0\advance\dimen1 by-\ht1
    \global\hangindent\dimen0\global\hangafter-\count0
    \hskip-\dimen0\setbox0\hbox to\dimen0{\raise-\dimen1\box0\hss}%
    \dp0=0in\ht0=0in\box0}#2}

\def\Ha{\mbox{H$\alpha$}}

\def\Hb{\mbox{H$\beta$}}

\def\Hg{\mbox{H$\gamma$}}

\def\Hd{\mbox{H$\delta$}}

\def\level #1 #2#3#4{$#1 \: ^{#2} \mbox{#3} ^{#4}$}   

\def\kms{\hbox{km$\;$s$^{-1}$}}

\def\Teff{\hbox{$T_\mathrm{eff}$}}            %% T_eff

\def\logg{\hbox{$\log g$}}                  %% logg
\def\loggc{\hbox{$\log g_c$}}                  %% logg
\def\vinf{\hbox{$\nu_{\infty}$}}             %% vinf
\def\vsini{\hbox{$\nu \sin i$ }}             %% vsini
\def\vmacro{\hbox{$\nu_{\rm macro}$}}        %% vmacro
\def\Mdot{\hbox{$\dot{M}$}}                %% Mdot
\def\Rstar{\hbox{$R_*$}}              %% solar radius
\def\Stromgren{\hbox{Str\"omgren}}              %% Stromgren
\def\logsi{\hbox{$\log$~n(Si)/n(H)}} 
\def\logQ{\hbox{$\log$~Q}} 

\def\mathstacksym#1#2#3#4#5{\def#1{\mathrel{\hbox to 0pt{\lower 
    #5\hbox{#3}\hss} \raise #4\hbox{#2}}}}

\mathstacksym\lta{$<$}{$\sim$}{1.5pt}{3.5pt} % less than approximately
\mathstacksym\gta{$>$}{$\sim$}{1.5pt}{3.5pt} % greater than approximately
\mathstacksym\lrarrow{$\leftarrow$}{$\rightarrow$}{2pt}{1pt} % equilibrium
\mathstacksym\lessgreat{$>$}{$<$}{3pt}{3pt} %% less or greater

\begin{document}

\title{Spectroscopic determination of the fundamental parameters of
66 B-type stars in the field-of-view of the CoRoT satellite}

\author{K.\ Lefever\inst{1,2} \and J.\ Puls\inst{3} \and T.\
  Morel\inst{1,4} \and C.\ Aerts\inst{1,5} \and L.\
Decin\inst{1,}$^{\star\star}$ \and
M.\ Briquet\inst{1,}\thanks{Postdoctoral researcher of the Fund for Scientific
Research of Flanders (FWO)} \institute{\Leuven \and
\BIRA \and \Munich \and \Liege \and \Nijmegen}}

\authorrunning{Lefever et al.}
\titlerunning{Fundamental parameters of B stars in the FoV of CoRoT}

\offprints{Karolien.Lefever@aeronomie.be}

\date{Received / Accepted }

\abstract{} {We aim to determine the fundamental parameters of a sample
of B stars with apparent visual magnitudes below 8 in the field-of-view of the
CoRoT space mission, from high-resolution spectroscopy. }
{We developed an
automatic procedure for the spectroscopic analysis of B-type stars with winds,
based on an extensive grid of FASTWIND model atmospheres.  We use the equivalent
widths and/or the line profile shapes of continuum normalized hydrogen, helium
and silicon line profiles to determine the fundamental properties of these stars
in an automated way. } {After thorough tests, both on synthetic datasets and
on very high-quality, high-resolution spectra of B stars for which we already
had accurate values of their physical properties from alternative analyses, we
applied our method to 66 B-type stars contained in the ground-based archive of
the CoRoT space mission. We discuss the statistical properties of the sample and
compare them with those predicted by evolutionary models of B stars.}
{Our spectroscopic results provide a valuable
starting point for any future seismic modelling of the stars, should they be
observed by CoRoT.}

\keywords{Stars: atmospheres -- Stars: early-type -- Stars: fundamental
parameters -- Methods: data analysis -- Techniques: spectroscopic -- Line:
profiles} 

\maketitle

\section{Introduction}

The detailed spectroscopic analysis of B-type stars has for a long time been
restricted to a limited number of targets. Reasons for this are the a priori
need for a realistic atmosphere model, the lack of large samples with
high-quality spectra, and the long-winded process of line profile fitting, as
the multitude of photospheric and wind parameters requires a large parameter
space to be explored.

The advent of high-resolution, high signal-to-noise spectroscopy in the nineties
led to a renewed interest of the scientific community in
spectroscopic research, and in particular in the relatively poorly understood
massive stars. The establishment of continuously better instrumentation and the
improvement in quality of the obtained spectroscopic data triggered a series of
studies, which led to a rapid increase in our knowledge of massive stars. In
this respect, it is not surprising to note
that this is exactly the period where several groups started to
upgrade their atmosphere prediction code for such stars, see, e.g., CMFGEN -
\citet{Hillier1998}, PHOENIX - \citet{Hauschildt1999}, WM-Basic -
\citet{Pauldrach2001}, POWR - \citet{Grafener2002} and FASTWIND -
\citet{Santolaya1997}, \citet{Puls2005}. Initially, major attention was devoted
to the establishment of a realistic atmosphere model (improvement of atomic
data, inclusion of line blanketing and clumping), rather than analyzing large
samples of stars.

Simultaneously with improvements in the atmosphere predictions, also the number
of available high-quality data increased rapidly, mainly thanks to the advent of
multi-object spectroscopy. At the time of writing, the largest
survey is the VLT-FLAMES Survey of Massive Stars \citep{Evans2005}, containing
over 600 Galactic, SMC and LMC B-type spectra\footnote{additionally, roughly
90 O-stars have been observed.} (in 7 different clusters),
gathered over more than 100 hours of VLT time. The survey not only allowed to
derive the stellar parameters and rotational velocities for hundreds of stars
\citep{Dufton2006, Hunter2008}, but also to study the evolution of surface N
abundances and the effective temperature scales in the Galaxy and Magellanic
Clouds \citep{Trundle2007, Hunter2007, Hunter2008a}.

In preparation of the CoRoT space mission, and almost contemporary with the
FLAMES setup, another large database was constructed: GAUDI (Ground-based
Asteroseismology Uniform Database Interface, \citealt{Solano2005}). It gathers
ground-based observations of more than 1500 objects, including high-resolution
spectra of about 250 massive B-type stars, with the goal to determine their
fundamental parameters as input for seismic modeling (see Section\,\ref{GAUDI}).

The availability of such large samples of B-type stars brings within reach
different types of studies, e.g., they may lead to a significant improvement in
the fundamental parameter calibration for this temperature range and to a
confrontation with and evaluation of stellar evolution models (e.g.,
\citealt{Hunter2008}). The drawback of this huge flood of data, however, is, as
mentioned before, the large parameter space to be explored, which can be quite
time-consuming, if no adequate method is available. It requires a method which
is able to derive the complete set of parameters of stars with a
wide variety of physical properties in an objective way. 

To deal with the large GAUDI dataset, we
investigated the possibility of \textit{automated} spectral
line fitting and we opted for a grid-based fitting method: AnalyseBstar.
In Section\,\ref{method}, we justify our choice for a grid-based method, present
its design and discuss several tests which were applied to check the
performance of the routine. A more detailed description of our methodology can
be found in the (online) appendix.
Section\,\ref{GAUDI} illustrates the
first application of AnalyseBstar to the sample of CoRoT candidate targets in
the GAUDI database and Section\,\ref{interpretation} deals with the physical
interpretation and some statistical properties of the resulting parameters.
Section\,\ref{summary} summarizes the main results obtained in this paper.

\section{Automated fitting using a grid-based method \label{method}}

Spectral line fitting is a clear example of an optimization problem. To
find the optimal fit to a given observed stellar spectrum among a set of
theoretically predicted spectra emerging from stellar atmosphere models,
requires scanning the parameter space spanned by the free parameters of the
stellar atmosphere model. Due to the extent of this parameter space, and with
the goal of analyzing large samples in mind, it is clear that performing the
parameter scan through fit-by-eye is not the best 
option. The `subjective' eye should be
replaced by an `intelligent' algorithm, in such way that the
procedure of finding the optimum fit becomes automatic, objective, fast and
reproducible, even though human intervention can never be excluded completely.

\subsection{General description of our grid-based method}

In contrast to the case of the O stars \citep[e.g.,][]{Mokiem2005}, treating the
entire spectral range B with a single approach
requires additional diagnostic lines besides H and He.
In this region, Si (in its different ionization stages), rather than He, becomes
the most appropriate temperature indicator. 
With this in mind, we have chosen
to develop an automatic procedure which is based on an extensive and refined
grid of FASTWIND models. 
This offers a good compromise between effort, time and precision if an
{\it appropriate grid} has been set up.  The grid should be comprehensive, as
dense as possible and representative for the kind of objects one wants to
analyze. Similar to a fit-by-eye method, a grid-based algorithm will follow an
iterative scheme, but in a reproducible way using a goodness-of-fit
parameter. Starting from a first guess for the fundamental parameters, based on
spectral type and/or published information, improved solutions are derived by
comparing line profiles resulting from well-chosen existing (i.e.\
pre-calculated) grid models to the observed line profiles. The algorithm
terminates once the fit quality cannot be improved anymore by modifying the
model parameters.

We are well aware of the limitations inherent to this method. As soon as a new
version/update of the atmosphere or line synthesis code is released (e.g.,
due to improved atomic data), the grid needs to be updated as well.
The advantage, on the other hand, is
that the line-profile fitting method itself will remain and the job is done with
the computation of a new grid. This is of course a huge work (e.g., seven
months were needed to compute\footnote{180 CPU months were
needed to compute the full grid of almost 265\,000 models. To reduce the
effective computation time, calculations were done on a dedicated Linux cluster
of 5 dual-core, dual-processor computers (3800 MHz processors, sharing 4 Gb RAM
memory and 8 Gb swap memory), amounting to 20 dedicated CPU processors, in
addition to 40 more regularly used institute CPUs (8 of 3800MHz and 32 of
3400MHz). The grid filled
60\% of a terabyte disk, connected to a Solaris 10 host pc.} and check the grid
in \citealt{Lefever2007a}),
but thanks to the fast performance of the FASTWIND code, this should
not really be an insurmountable problem.

A grid-based method fully relies on a static grid and no additional models are
computed to derive the most likely parameters.  Consequently, the quality of the
final best fit and the precision of the final physical parameters are fully
determined by the density of the grid. This underlines, again, the necessity of
a grid which is as dense as possible, allowing for interpolation.  On the other
hand, the grid-method has a plus-point: it is fast. Whereas the analysis with,
e.g., a genetic algorithm approach \citep[e.g.]{Mokiem2005} needs parallel
processors and several days of CPU to treat one target star, our grid-method
will, on average, require less than half an hour on {\it one} computer, because
no additional model computations are required.

\subsection{The code AnalyseBstar}

Our grid-based code, called AnalyseBstar, was developed for the
spectroscopic analysis of B stars with winds. It is written in the Interactive
Data Language (IDL\label{acro_IDL}), which allows for interactive manipulation
and visualization of data. It fully relies on the extensive grid of NLTE model
atmospheres and the emerging line profiles presented in
\citet[ see also \texttt{Appendix\,\ref{section_grid}} for a brief overview of
the considered parameter set]{Lefever2007a} and
is developed to treat large samples of stars in a homogeneous way. We use
continuum normalized H, He and Si lines to derive the photospheric properties of
the star and the characteristics of its wind. Such an automated method is not
only homogeneous, but also objective and robust. A full description of the
design of AnalyseBstar, including the preparation of the input, the iteration
cycle for the determination of each physical parameter and the inherent assumptions,
is given in \texttt{Appendix B}\,\footnote{Manual
available upon request from the authors.}.

The intrinsic nature of our procedure allows us to derive estimates for the
`real' parameters of the star, which in most cases are located in between
two model grid points. In what follows, we refer to them as `interpolated
values'. The corresponding grid values are the parameters corresponding to
the grid model lying closest to these interpolated values.

Furthermore, we will denote the surface gravity as derived from fitting the
Balmer line wings as $\logg$, while the gravity corrected for centrifugal
terms (required, e.g., to estimate consistent masses) will be denoted by
$\loggc$. The (approximate) correction itself is obtained by adding the term
$(\vsini)^2/\Rstar$ to the uncorrected gravity \citep[ and references
therein]{Repolust2005}.

\subsection{Testing the method}

\subsubsection{Convergence tests for synthetic FASTWIND spectra
\label{convergencetests}}

Before applying AnalyseBstar to observed stellar spectra, we tested
whether the method was able to recover the parameters of synthetic FASTWIND
input spectra. To this end, we created several synthetic datasets in
various regions of parameter space, with properties representative for a
`typical' GAUDI spectrum. For more details on the setup of this test dataset
and the results of the convergence tests, we refer to
\texttt{Appendix\,\ref{details_convergencetests}}.
In all cases, the input parameters were well recovered and no convergence
problems were encountered. Minor deviations from the input parameters were as
expected and within the error bars. Also the derived $\vsini$-values, which are
difficult to disentangle from potential macroturbulent velocities, agreed very well with
the inserted values, irrespective of \vmacro. 
This shows that the Fourier Transform method of \citet{Gray1973, Gray1975},
whose implementation by \citet{SimonDiaz2007} we used, indeed allows to separate
both effects (but see below). All together, this gave us enough confidence to
believe that
our procedure will also be able to recover the physical parameters from real
spectral data.

\subsubsection{FASTWIND vs.\ Tlusty: profile comparison and analysis of
Tlusty spectra}

As an additional test, we compared line profiles based on FASTWIND models with
negligible mass loss with their corresponding counterpart in the grid computed
with the NLTE model atmosphere code Tlusty \citep{Hubeny2000},
which allows for a fully consistent NLTE metal line blanketing as well.
This resulted in an overall good agreement, except for low gravity stars, where
differences arise. It concerns differences in the forbidden He\,{\sc
i} components
(which are quire strong in FASTWIND, but almost absent in Tlusty), the Balmer line wings
(which are a bit stronger in FASTWIND than in Tlusty, yielding slightly lower
$\logg$ values) and the Si\,{\sc
ii}/{\sc iii} EW ratio for $\Teff > 18$ kK (which is larger
for FASTWIND than for Tlusty - due to both stronger Si\,{\sc
ii} and weaker Si\,{\sc
iii}
lines -, yielding higher $\Teff$ values).
The reason that FASTWIND predicts stronger Balmer line wings is due
to differences in the broadening functions and an underestimated
photospheric
line pressure, which can affect the photospheric structure of low-gravity
objects. Differences are about or below 0.1~dex, and become negligible
for dwarf stars. Except for the problem with the forbidden components, the
profiles of the cooler models ($\Teff < 18$ kK) compare very well.

To check the uncertainty in the synthetic profiles, we applied AnalyseBstar
to profiles of six prototypical Tlusty models at 3 different temperature points
(15\,000, 20\,000, and 25\,000~K) and at a high and a low gravity.
We added artificial noise to the data and artificially broadened the profiles
with a fixed projected rotational velocity. The results of the comparison can be
found in
Table\,\ref{tlustycomparison}. The input and output fundamental parameters
agree very well within the resulting errors, with only one exception, being the
model with the high $\Teff$ and higher $\logg$, where the temperature is off by
2\,000~K.

Around 20kK, we experience similar difficulties to retrieve a unique
temperature as in the cool B star domain, i.e. depending on the gravity, we
may have only one stage of Si available, in this case Si\,{\sc
iii}. Indeed, whereas
the Si\,{\sc
ii} lines are clearly visible in the higher gravity test case ($\logg$ =
3.00), they are no longer detectable in the lower gravity case ($\logg$ = 2.25).
At this $\Teff$, there are (unfortunately) no obvious Si\,{\sc
iv} lines yet, which
complicates the analysis. Even though, AnalyseBstar is still able to retrieve
the correct effective temperature, albeit with somewhat larger
errors, using the same alternative method as for late B type stars when there
is only Si\,{\sc ii} available (see `method 2' in Section\,\ref{method_1_2}).

\begin{table}[h]
\tabcolsep=4pt
\caption{Result of applying AnalyseBstar to synthetic Tlusty profiles, which are
artificially adapted to obtain a typical SNR of 150 and a projected rotational
broadening
of $\vsini = 50$ km\,s$^{-1}$. \Teff$_{\rm , in}$ and \logg$_{\rm , in}$ are,
respectively, the input effective temperature and gravity of the Tlusty
spectra. \Teff$_{\rm , out}$ and \logg$_{\rm , out}$ represent the best matching
($\Teff$, $\logg$)-combination in the FASTWIND grid.}
\label{tlustycomparison}
\centering
\begin{tabular}{cc|cc}
\hline \hline
\Teff$_{\rm , in}$ & \logg$_{\rm , in}$ & \Teff$_{\rm , out} \pm \Delta$ &
 \logg$_{\rm , out} \pm \Delta$ \\
(kK) & (cgs) & (kK) &  (cgs)\\
\hline
 15 & 1.75 & 15.0 $\pm$ 0.5 & 1.7 $\pm$ 0.05\\
 15 & 3.00 & 15.5 $\pm$ 0.5 & 3.0 $\pm$ 0.10\\
 20 & 2.25 & 20.0 $\pm$ 1.0 & 2.2 $\pm$ 0.10\\
 20 & 3.00 & 21.0 $\pm$ 1.0 & 3.2 $\pm$ 0.20\\
 25 & 2.75 & 25.0 $\pm$ 1.0 & 2.7 $\pm$ 0.05\\
 25 & 3.00 & 27.0 $\pm$ 1.0 & 3.1 $\pm$ 0.10\\
\hline
\end{tabular}
\end{table}

\subsubsection{Comparison with fit-by-eye results for well-studied
pulsators \label{test_pulsators}}

\begin{figure*}
\centering
\begin{minipage}{7in}
\resizebox{3.5in}{!}{\includegraphics{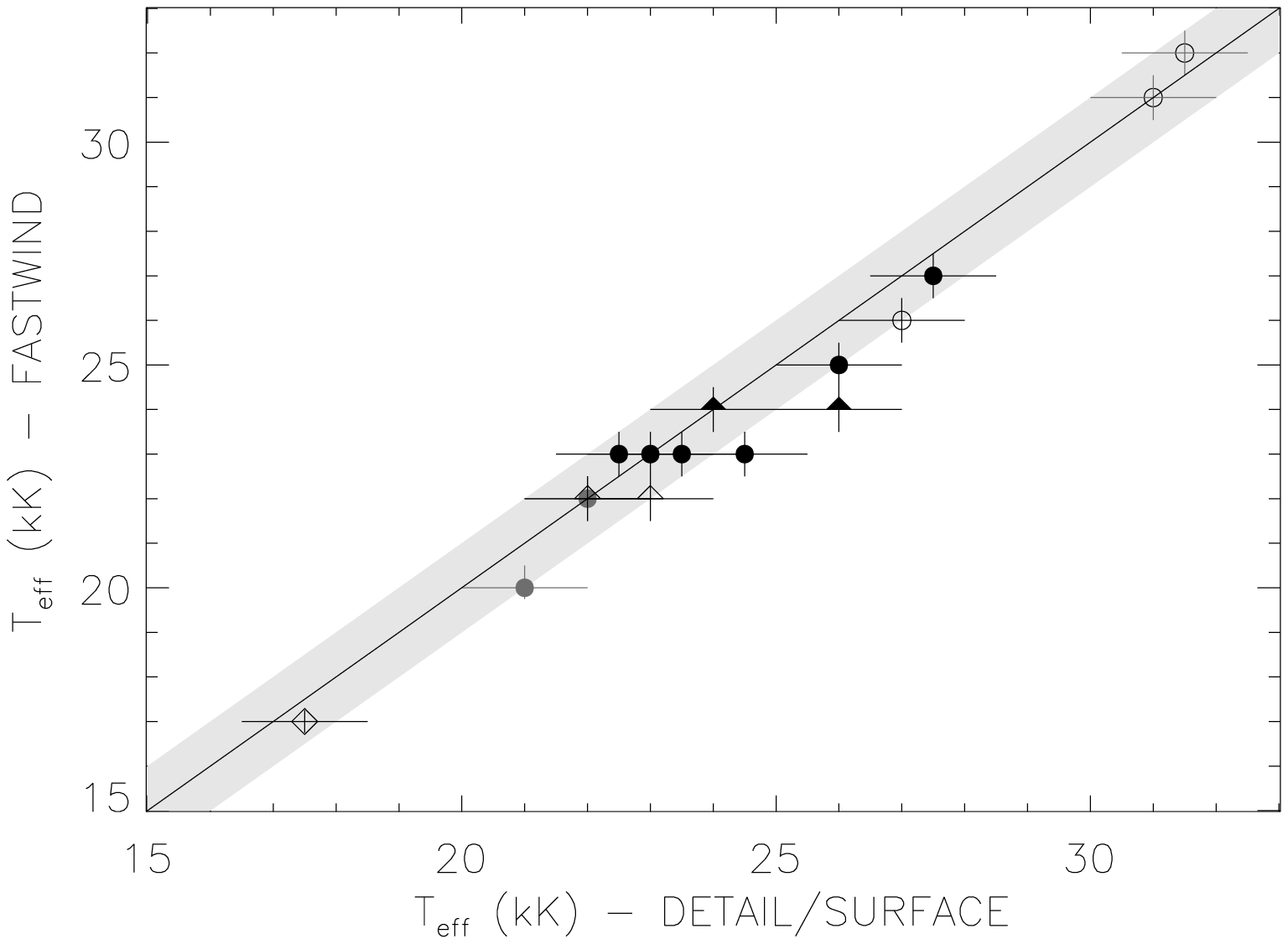}}
\resizebox{3.5in}{!}{\includegraphics{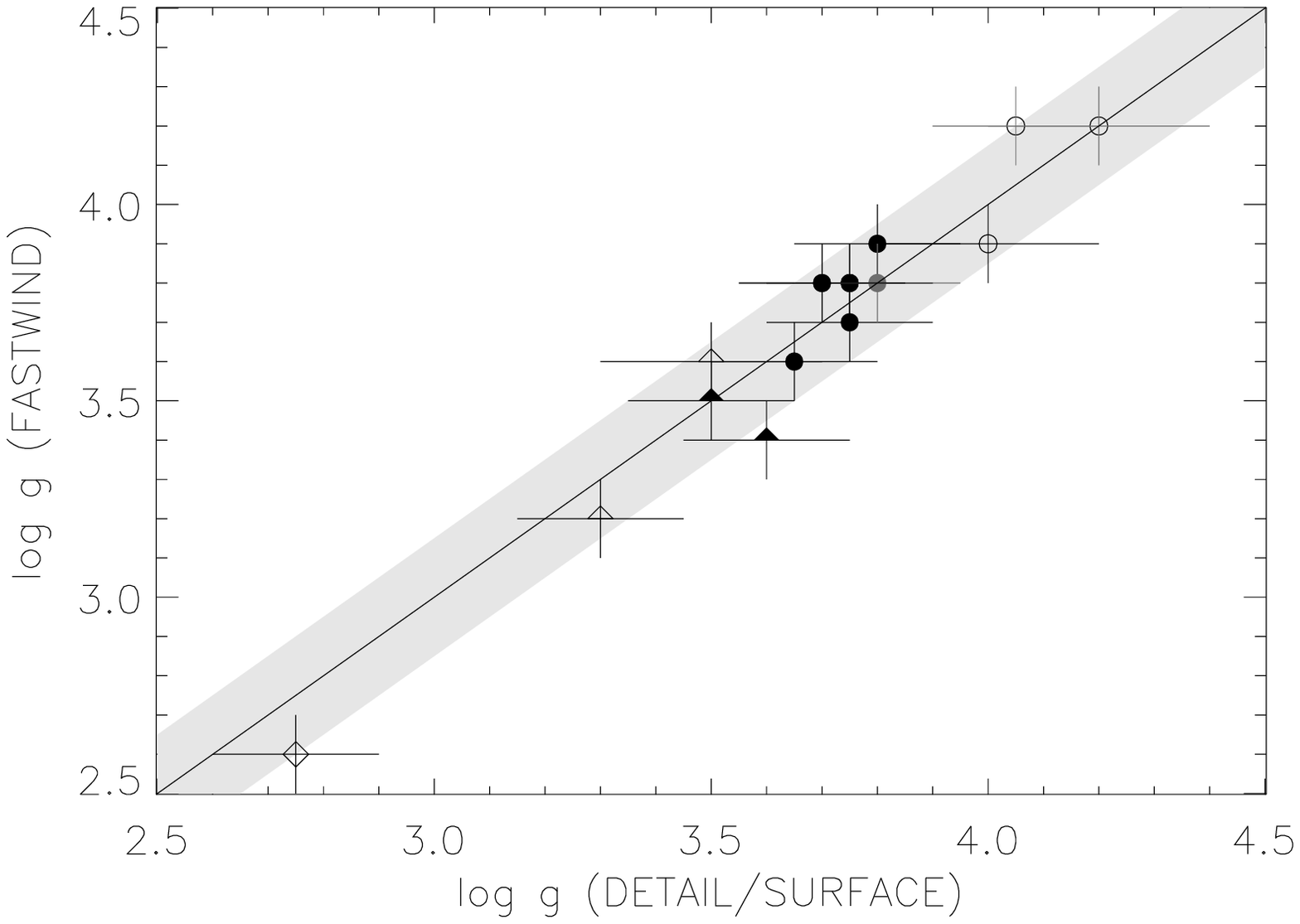}}
\end{minipage}
\caption{\label{comparison_tefflogg} Comparison of the effective temperatures
(left) and surface gravities (right) derived with FASTWIND (AnalyseBstar) with
those found by \citet{Morel2006,Morel2008} and \citet{Briquet2007a} using
DETAIL/SURFACE, for the $\beta$ Cephei stars (black), SPBs (grey) and
some well-studied hot B stars (open symbols).
Supergiants are indicated as diamonds, giants as triangles and dwarfs as
circles. The shaded area around the one-to-one relation (straight line) denotes
the uncertainty on
the derived fundamental parameters of
\citet{Morel2006,Morel2008} and
\citet{Briquet2007a} (i.e., 1\,000~K in \Teff\,and 0.15 in \logg).}
\end{figure*}

Having thoroughly tested that AnalyseBstar indeed converges towards the
optimum solution and that it is able to recover the input parameters of
synthetic spectra, we performed an additional test on real spectra.
We tested our method on a selected sample of high-quality, high-resolution
spectra of pulsating B stars ($\beta$ Cephei and Slowly Pulsating B
stars (SPB)). The (mean) spectra of these stars have a very high SNR, attained
through the addition of a large number of individual exposures (see,
\citealt{Morel2007}). Given that we excluded binaries from our GAUDI sample 
(see Section\,\ref{CoRoT}),
we also ignored $\theta$ Oph and $\beta$ Cru for the test sample. Moreover,
V2052\,Oph is chemically peculiar and was thus excluded, as well as V836\,Cen
for which only one spectrum is available. 
Thus we limited our test sample to those single stars in Morel et al. (2007)
for which numerous high-quality spectra were available.
Because of the high quality of these spectral time series,
they are ideally suited for testing AnalyseBstar. 

\renewcommand{\thetable}{\arabic{table}\alph{subtablec}}
\addtocounter{table}{0}
\setcounter{subtablec}{1}
\begin{table*}[ht!]
\tabcolsep=3pt \small
\caption{\label{parameters_bcep} Fundamental parameters of
well-studied $\beta$ Cephei stars (spectral types from SIMBAD), as
derived from AnalyseBstar: the effective temperature ($\Teff$), the
surface gravity ($\logg$, not corrected for centrifugal acceleration due to
stellar rotation), the helium abundance (n(He)/n(H), with solar value 0.10), the
Si abundance ($\logsi$, with solar value -4.49), the microturbulence ($\xi$),
the projected rotational velocity ($\vsini$) and the macroturbulent
velocity ($\vmacro$). We list the values of the
closest grid model, except for the Si abundance, where the interpolated value
is given. In italics, we display the results from the DETAIL/SURFACE analysis by
\citet{Morel2006,Morel2008}. For 12~Lac, we additionally list the parameters
which we retrieve when forcing $\vmacro$ to be zero while keeping the value for $\vsini$
(indicated between brackets, see text for details). All stars have thin or
negligible winds.\newline
Typical errors for $\Teff$ and $\logg$ are, respectively, 1\,000~K and 0.15
for the comparison data sets of \citet{Morel2006,Morel2008} and
\citet{Briquet2007a}, which is somewhat more conservative than the
errors adopted in this study. 
We adopt typical 1-$\sigma$ errors of 0.10 dex for \logg, 1\,000~K for 
$\Teff > 20\,000~K$ and $\Teff < 15\,000~K$, and 500~K for $15\,000~K \le
\Teff \le 20\,000~K$. The error for the derived Si-abundance, $\logsi$,
has been estimated as 0.15 and 0.20 dex for objects above and below
15\,000~K, respectively (see also Section\,\ref{method_1_2}.)}
\vspace{0.3cm}
\centering
\begin{tabular}{rlllllcrclrc}
\hline \hline 
HD  & alternative & Spectral & \Teff\,& \logg\,& n(He)/n(H) & $\logsi$
& $\xi$ & \vsini & \vmacro \\ 
number & name & Type & (K) & (cgs) & & &(km\,s$^{-1}$) & (km\,s$^{-1}$) &
(km\,s$^{-1}$) \\
\hline
46328 & $\xi^1$ CMa & B0.5-B1IV & 27000 & 3.80 & 0.10 & -4.69
& 6 & 9 $\pm$ 2
&11\\ 
& & & \textit{27500 }& \textit{3.75} & & \textit{-4.87 $\pm$ 0.21} &\textit{6
$\pm$ 2} & \textit{10 $\pm$ 2} & -\\ 
50707 & 15 CMa & B1III & 24000 & 3.40 & 0.10 & -4.79 & 12 & 34 $\pm$ 4
& 38\\ 
& & & \textit{26000 }& \textit{3.60} & & \textit{-4.69 $\pm$ 0.30} &\textit{7
$\pm$ 3} & \textit{45 $\pm$ 3} & - \\ 
205\,021 & $\beta$ Cep & B1IV & 25000 & 3.80 & 0.10 & -4.70 & 6 & 26
$\pm$ 3 & 24\\ 
& & & \textit{26000} & \textit{3.70} & &\textit{-4.89 $\pm$ 0.23}  &\textit{6
$\pm$ 3 }& \textit{29 $\pm$ 2} & -\\ 
44743 & $\beta$ CMa & B1.5III & 24000 & 3.50 & 0.10 & -4.76 & 15 & 19
$\pm$ 4&
20\\ 
& & & \textit{24000} & \textit{3.50 }& & \textit{-4.83 $\pm$ 0.23} &\textit{14
$\pm$3} & \textit{23 $\pm$ 2}& -\\ 
214\,993 & 12 Lac & B1.5IV & 23000 & 3.60 & 0.10 & -4.41 & 6 & 44 $\pm$
6 & 37\\
 & & & & & & [-4.83] & [12] & & [0]\\
& & & \textit{24500} & \textit{3.65} & & \textit{-4.89 $\pm$ 0.27} &\textit{10
$\pm$ 4 }& \textit{42 $\pm$ 4} & -\\ 
16582 & $\delta$ Ceti & B1.5-B2IV & 23000 & 3.90 & 0.10 & -4.80 & 6 &
14 $\pm$
2&0\\ 
& & & \textit{23000} & \textit{3.80} & & \textit{-4.72 $\pm$ 0.29}
&\textit{1$^{+3}_{-1}$} &\textit{14 $\pm$ 1} & - \\ 
886 & $\gamma$ Peg & B1.5-B2IV & 23000 & 3.80 & 0.10 & -4.84 & $<$ 3 &
10 $\pm$
1&0\\ 
& &  & \textit{22500} & \textit{3.75} & & \textit{-4.81 $\pm$ 0.29} &
\textit{1$^{+2}_{-1}$ } & \textit{10 $\pm$ 1} & - \\ 
29248 & $\nu$ Eri & B1.5-B2IV & 23000 & 3.70 & 0.10 & -4.73 & 10 & 21
$\pm$ 3
&39\\ 
& & & \textit{23500 }& \textit{3.75} & & \textit{-4.79 $\pm$ 0.26} &
\textit{10 $\pm$ 4 }& \textit{36 $\pm$ 3} & -  \\ 
\hline
\end{tabular}
\end{table*}
\begin{figure}[h]
\centering
\resizebox{3.5in}{!}{\includegraphics{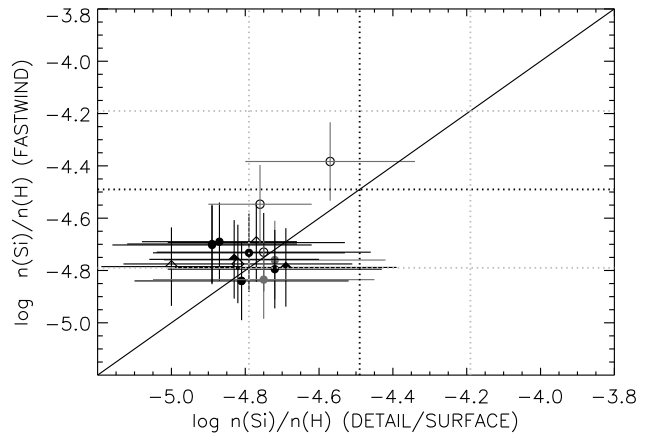}}
\caption{\label{hrd_pulsators} Comparison between the Si abundances derived from
AnalyseBstar and those derived from DETAIL/SURFACE for the set of photometric
targets, used to evaluate the performance of AnalyseBstar. The black dotted
lines represent the solar Si abundance ($\logsi$~=~-4.49), while the grey dotted
lines represent a depletion and enhancement of Si by 0.3 dex (i.e., -4.79 and
-4.19, respectively). Symbols have the same meaning as in
Fig.\,\ref{comparison_tefflogg}. Individual error bars are indicated.}
\end{figure}
\addtocounter{table}{-1}
\addtocounter{subtablec}{1}
\begin{table*}[ht!]
\tabcolsep=3pt \small
\caption{\label{parameters_SPB} Same as for 
Table\,\ref{parameters_bcep}, 
but now for some well-studied SPBs. In italics, we show
the results of the DETAIL/SURFACE analysis by \citet{Briquet2007a}.}
\vspace{0.3cm}
\centering
\begin{tabular}{rllccccrccrc}
\hline \hline
HD & alternative & Spectral & \Teff\,& \logg\,& n(He)/n(H) & $\logsi$
& $\xi$ & \vsini & \vmacro\\ 
number & name & Type & (K) & (cgs) &  & & (km\,s$^{-1}$) &  (km\,s$^{-1}$) &
(km\,s$^{-1}$) \\
\hline
3360 & $\zeta$ Cas & B2IV & 22000 & 3.80 & 0.10 & -4.76 & $<$ 3 & 18
$\pm$ 2 & 13\\
     &  &     & \textit{22000} & \textit{3.70} &  & \textit{-4.72 $\pm$ 0.30} &
\textit{1 $\pm$ 1} & \textit{19 $\pm$ 1} & -  \\ 
85953 & V335 Vel &  B2IV & 20000 & 3.80 & 0.10  & -4.84 & 6 & 30
$\pm$ 1 & 20 \\ 
  & & & \textit{21000} & \textit{3.80}  & & \textit{-4.75 $\pm$ 0.30}&
\textit{1 $\pm$ 1} & \textit{29 $\pm$ 2} & -   \\ 
3379 & 53 Psc & B2.5IV & 20000 & 4.30 & 0.10 & -4.73 & $<$ 3 & 48 $\pm$
8&43\\
74195 & o Vel & B3IV & 16500 & 3.70 & 0.10 & -4.79 & $<$ 3 &18 $\pm$ 2
&18\\
160762 & $\iota$ Her & B3IV & 19500 & 4.10 & 0.10 & -4.86 & 3 & 8$\pm$
2 &0\\
25558 & 40 Tau & B3V & 17500 & 4.00 & 0.10  & -4.72 & 6 & 28 $\pm$ 2 &
31 \\
181558 & HR 7339 & B5III & 15000 & 4.00 & 0.10  & -4.79 & 6 & 17 $\pm$
2 & 0 \\
26326 & HR 1288 & B5IV & 15500 & 3.60 & 0.10 & -4.49 & $<$ 3 & 17 $\pm$
1 &17\\
206540 & HR 8292 & B5IV & 13500 & 3.80 & 0.10 & -4.79 & 3 & 15 $\pm$ 2
& 0 \\
24587 & HR 1213 & B5V & 14500 & 4.00 & 0.10 & -4.79 & $<$ 3 & 25 $\pm$
4 &21\\
28114 & HR 1397 & B6IV & 14000 & 3.50 & 0.10 & -4.79 & $<$ 3 & 21 $\pm$
4 &17\\
138764 & HR 5780 & B6IV & 14500 & 3.90 & 0.10 & -4.51 & 3 & 21 $\pm$ 2
&0\\
215573 & HR 8663 & B6IV & 13500 & 3.80 & 0.10 & -4.49 & $<$ 3 & 8 $\pm$
1 &0 \\
39844 & HR 2064 & B6V & 14500 & 3.70 & 0.10  & -4.39 & $<$ 3 & 16 $\pm$
1 & 0\\
191295 & V1473 Aql & B7III & 13000 & 3.70 & 0.10 & -4.79 & 3 & 16 $\pm$
1 &15\\
21071 & V576 Per &  B7V & 13500 & 3.70 & 0.10  & -4.79 & 3 & 22 $\pm$ 1
& 0\\
37151 & V1179 Ori & B8V & 12500 & 3.80 & 0.10  & -4.79 & $<$ 3 & 20
$\pm$2&0\\
\hline
\end{tabular}
\end{table*}
\addtocounter{table}{-1}
\addtocounter{subtablec}{1}
\begin{table*}[ht!]
\tabcolsep=3pt \small
\caption{\label{parameters_nonconfirmed} Same as for 
Table\,\ref{parameters_bcep}, but now for some well-studied hot B stars.
In italics, we display the independent results from the DETAIL/SURFACE analysis
by \citet{Morel2006,Morel2008}, except for $\theta$~Car and $\tau$~Sco which can
be found in \citet{Hubrig2008}.
For $\tau$~Sco, we additionally compare with the values found 
by \citet{Mokiem2005}
using a genetic algorithm approach (superscript $M$). The
results for two objects overlapping with the study of \citet{Przybilla2008} are
also indicated (superscript $P$).}
\centering
\begin{tabular}{rlllllcrclrc}
\hline \hline
HD & alternative & Spectral & \Teff\,& \logg\,& n(He)/n(H) & $\logsi$
& $\xi$ & \vsini & \vmacro \\ 
number & name & Type & (K) & (cgs) & & & (km\,s$^{-1}$) &(km\,s$^{-1}$) &
(km\,s$^{-1}$) \\
\hline
 93030 & $\theta$ Car & B0Vp & 31000 & 4.20 & 0.15 & -4.38  & 10 & 
108 $\pm$ 3 & 0 \\ 
 & & & \textit{31000} & \textit{4.20} & - & \textit{-4.57 $\pm$ 0.23} &
\textit{12 $\pm$ 4} & \textit{ 113 $\pm$ 8} & - \\
149\,438 & $\tau$ Sco & B0.2V & 32000 & 4.20 & 0.10 & -4.55 & 6 & 10
$\pm$ 2  & 0 \\ 
 & & & \textit{31500} & \textit{4.05} & - & \textit{-4.76 $\pm$ 0.14} & 
\textit{2 $\pm$ 2} & \textit{8 $\pm$ 2} & - \\
 & & & \textit{31900$^{M}$} & \textit{4.15$^{M}$} & \textit{0.12$^{M}$} & -
& \textit{10.8$^{M}$} & \textit{5$^{M}$} & - \\
 & & & \textit{32000$^{P}$} & \textit{4.30$^{P}$} & \textit{0.10$^{P}$} &
\textit{-4.50$^{P}$} & \textit{5$^{P}$} & \textit{4$^{P}$} &\textit{4$^{P}$} \\
36591 & HR 1861 & B1V & 26000 & 3.90 & 0.10 & -4.73  & 6 & 13
$\pm$ 1 & 0 \\ 
 & & &\textit{27000} & \textit{4.00} & - & \textit{-4.75 $\pm$ 0.29} & 
\textit{3 $\pm$ 2} & \textit{16 $\pm$ 2} & - \\
 & & & \textit{27000$^{P}$} & \textit{4.12$^{P}$} & \textit{0.10$^{P}$} &
\textit{-4.52$^{P}$} & \textit{3$^{P}$} & \textit{12$^{P}$} &-\\
52089 & $\epsilon$ CMa & B1.5-B2II/III & 22000 & 3.20 & 0.10 & -4.69 &
15 & 32
$\pm$ 2 & 20 \\ 
 & & & \textit{23000} & \textit{3.30} & - & \textit{-4.77 $\pm$ 0.24} &
\textit{16 $\pm$ 4} & \textit{28 $\pm$ 2} & - \\
35468 & $\gamma$ Ori & B2II-III & 22000 & 3.60 & 0.10 & -4.79 & 10 & 46
$\pm$ 8
& 37 \\ 
 & & & \textit{22000} & \textit{3.50} & - & \textit{-5.00 $\pm$ 0.19} &
\textit{13 $\pm$ 5} & \textit{51 $\pm$ 4} & - \\
51309 & $\iota$ CMa & B2.5Ib-II & 17000 & 2.60 & 0.10 & -4.78 & 15 & 27
$\pm$ 4
& 39 \\ 
 & & & \textit{17500} & \textit{2.75} & - & \textit{-4.82 $\pm$ 0.31} &
\textit{15 $\pm$ 5} & \textit{32 $\pm$ 3} & - \\
\hline
\end{tabular}
\end{table*}
\addtocounter{table}{0}
\addtocounter{subtablec}{-3}

The $\beta$\,Cephei stars and
two of the SPB stars were analyzed in detail by
\citet{Morel2006,Morel2008} and \citet{Briquet2007a}, respectively. 
These authors used the latest version of the NLTE line formation codes
DETAIL and SURFACE \citep{Giddings1981, Butler1984a}, in combination with
plane-parallel, fully line-blanketed LTE Kurucz atmospheric models
(ATLAS9, \citealt{Kurucz1993a}), to determine (by eye) the atmospheric
parameters and element abundances of low-luminosity class objects with
negligible winds.
The outcome of the comparison with their results are summarized in
Tables\,\ref{parameters_bcep} to \ref{parameters_nonconfirmed}, and in
Figs.\,\ref{comparison_tefflogg} and \ref{hrd_pulsators}.
Although the effective temperatures derived from FASTWIND tend to
be, on the average, slightly below the ones derived from DETAIL/SURFACE, 
it is clear that, within the error bars, FASTWIND and DETAIL/SURFACE give
consistent results for $\Teff$ and $\logg$ (Fig.\,\ref{comparison_tefflogg}) as
well as for $\logsi$ (Fig.\,\ref{hrd_pulsators}),
On the other hand, the \vsini\
values derived by \citet{Morel2008} are typically slightly above those derived
by us. This is readily understood as their values implicitly include the
macroturbulent velocity, which has not been taken into account as a separate
broadening component. In those cases where we derive a vanishing \vmacro,
the \vsini\ values are in perfect agreement.
\citet{Aerts2009} recently suggested that macroturbulence might be explained in
terms of collective pulsational velocity broadening due to the superposition of
a multitude of gravity modes with low amplitudes (see also \texttt{Appendix
\ref{vmacro_cycle}}). In this respect, we expect stars for which $\vmacro$
differs from zero to be pulsators. 

For few overlapping objects, we could also compare our results with
those from independent studies by \citet{Mokiem2005}, based on FASTWIND
predictions for H and He lines and using a genetic algorithm approach, and
by \citet{Przybilla2008}, using ATLAS9/DETAIL/SURFACE, allowing for a
comparison of H, He, and Si. The consistency with both studies is quite
good, except for the Si abundance of HD~36591 as determined by Przybilla et
al. For this object, \citet{Hubrig2008} have obtained a value close to
ours, using the same ATLAS9/DETAIL/SURFACE code. Ad hoc, we cannot judge the
origin of this discrepancy. 

For two targets, HD~85953 and $\tau$~Sco,
somewhat larger differences in the derived microturbulence are
found. For HD~85953, this difference is compensated by an opposite difference in
the derived Si abundance. For $\tau$ Sco, on the other hand, 
the derived microturbulence (and also
the Si abundance) is in perfect agreement with the values provided by
\citet{Hubrig2008} and \citet{Przybilla2008}
within the defined error bars, whereas
the value for the microturbulence derived by \citet{Mokiem2005} differs
substantially. As they did not derive the Si abundance, we cannot judge on the
difference in Si abundance this would have caused.

An even more detailed test can be done for stars with seismically determined
values of the fundamental parameters. 12 Lac is such a star. It is known as a
non-radial pulsator with at least 11 independent oscillation frequencies
\citep{Handler2006}. The fact that 12 Lac is a rich pulsator clearly shows up in
the line profiles, which are strongly asymmetric.  Especially the Si\,{\sc
iii}
triplets (Si\,{\sc
iii}~4552-4567-4574 and Si\,{\sc
iii}~4813-4819-4829) are skew.
Fig.\,\ref{12Lac_vmacro} shows the observed spectrum (black), which is a
combination of 31 exposures.
Using the Fourier Transform method, we find the
projected rotational velocity to be 44 km\,s$^{-1}$, which is slightly above the
36 km\,s$^{-1}$ derived from modelling the line profile variations
\citep{Desmet2009}.
Including macroturbulence in the spectral line fitting,
results in the following parameters: \Teff\ = 23\,000~K $\pm$ 1\,000~K, \logg\ =
3.6 $\pm$ 0.1, $\xi$ = 6 $\pm$ 3 km\,s$^{-1}$, solar He and Si abundances
(solid grey lines in Fig.\,\ref{12Lac_vmacro}). 
Actually, the derived surface gravity is
in perfect agreement with the seismically determined value of $\logg$
(3.64-3.70) determined by \citet{Desmet2009}.

Except for the Si abundance, our results are also
in agreement with \citet[ see
Table\,\ref{parameters_bcep}]{Morel2006}, who derived \Teff\ =
24\,500~K $\pm$ 1\,000~K, \logg\ = 3.65 $\pm$ 0.15, $\xi$~=~10 $\pm$ 4
km\,s$^{-1}$ and  
depleted Si abundances as seem to be typical for B dwarfs in the solar
neighborhood. To account for the pulsational
broadening, we needed to include a macroturbulence of 37 km\,s$^{-1}$. 
Forcing the
macroturbulence to be zero while keeping the derived value of $\vsini$
leaves the main physical parameters ($\Teff$, $\logg$
and He abundance) unaltered, while the microturbulent velocity and the Si
abundance change to compensate for the change in profile shape:
the microturbulence becomes larger (12 km\,s$^{-1}$,
broader lines) and the Si abundance becomes lower (depleted, $\logsi \approx
-4.83$). As expected, this depleted Si abundance \textit{is}
in agreement with \citet{Morel2006}, who did not include $\vmacro$ in their
analysis. When evaluating the fit quality, we
find, besides the expected mismatch in the Si line wings, also a discrepancy in
the line cores of He (dashed grey lines in Fig.\,\ref{12Lac_vmacro}).  Following
\citet{Aerts2009}, this might be improved by accounting for the pulsational
broadening introduced by the collective set of detected oscillations in 12 Lac,
as this would result in a shape intermediate between a rotation and a Gaussian
profile. However, the only way one can account for this, is by using an
appropriate time series of spectra. The above example illustrates that we need
to account for pulsational broadening to correctly assess the Si abundance and
microturbulence. In this sense, it is advisable to work with `$\vmacro$' as a
substitute for the time-dependent broadening in the case of `snapshot'
spectra.

\begin{figure*}[ht!]
\centering
{\resizebox{\hsize}{!}{\includegraphics{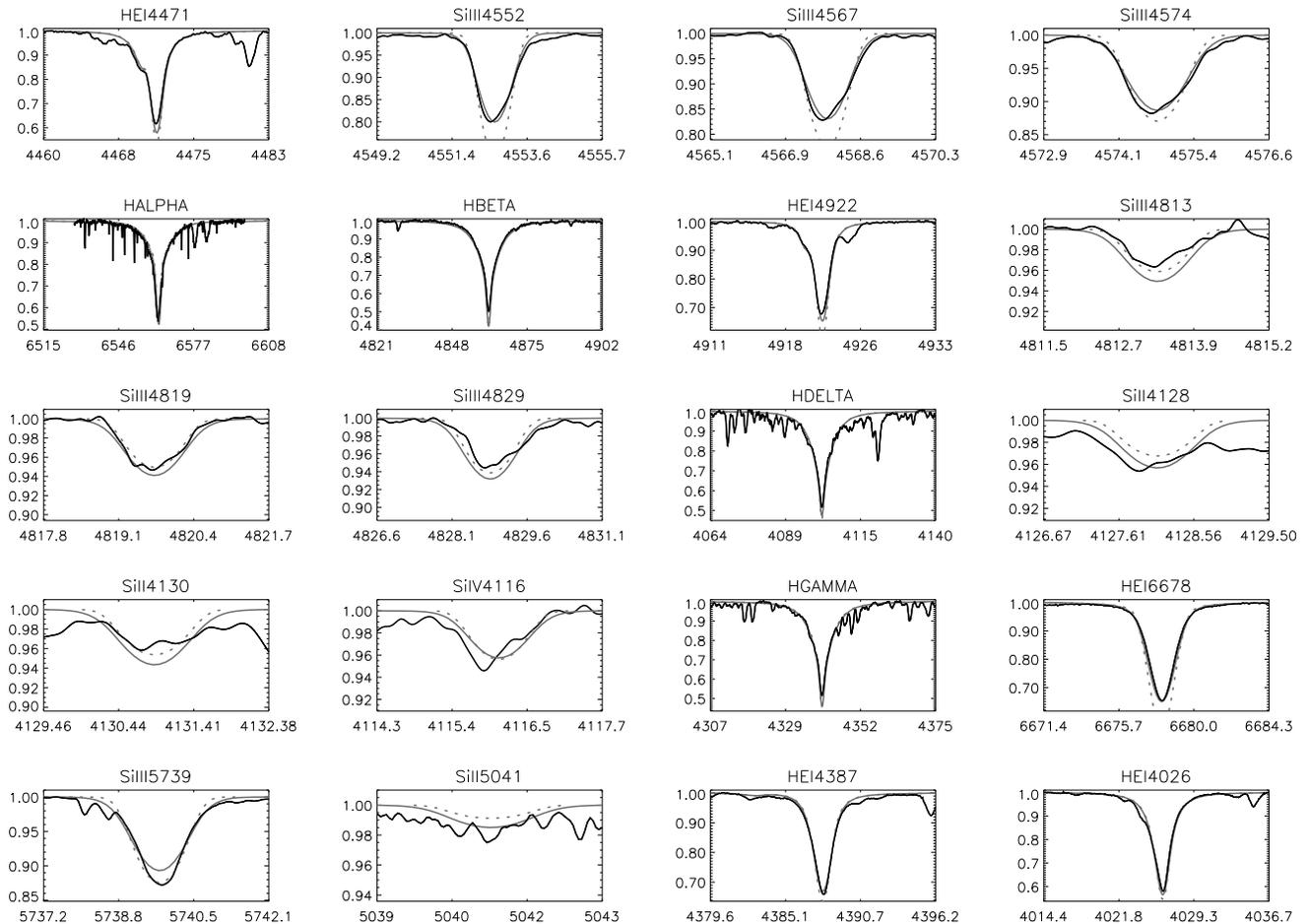}}}
\caption{\label{12Lac_vmacro} Illustration of the effects of time-independent
broadening on the derived parameters for 12 Lac (B2 III). The
observed line profiles (solid black lines) are fitted using AnalyseBstar.
Inclusion of time-independent broadening (by means of
macroturbulence) yields the solid grey fit, with optimum values of
\vsini\ = 44 km\,s$^{-1}$ and \vmacro\ = 37 km\,s$^{-1}$. Forcing the
macroturbulence to be zero leaves all parameters but the Si abundance and the
microturbulence unaltered. The resulting fit is represented by the grey dashed
lines.}
\end{figure*}

\section{Application to the GAUDI B star sample\label{GAUDI}}

\subsection{CoRoT \& GAUDI\label{CoRoT}}

The French-led European space mission CoRoT (Convection, Rotation, and planetary
Transits\label{acro_COROT}; see ``the CoRoT Book'', \citealt{Fridlund2006}) was
launched successfully on December 27th, 2006. The observational setup of 
the CoRoT
seismology programme (observations of a small number of bright stars over a very
long period) made target selection a crucial issue.  Due to the acute shortage
of available information on the potential CoRoT targets, the need for additional
data was high. Therefore, an ambitious ground-based observing program for more
than 1500 objects was set up, under the leadership of C.\ Catala at Meudon, to
obtain \Stromgren\ photometry ($uvby\beta$) as well as high-resolution
spectroscopy (FEROS, ELODIE, SARG, CORALIE, GIRAFFE, coud\'e spectrograph at the
2m telescope in Tautenburg, CATANIA).  These data were collected in an extensive
catalogue, maintained at LAEFF (\label{acro_LAEFF}Laboratorio de 
Astrof\'{\i}sica
Espacial y F\'{\i}sica Fundamental) and baptized \textit{GAUDI}: Ground-based
Asteroseismology Uniform Database Interface\label{acro_GAUDI}
\citep{Solano2005}.

The Be stars within the GAUDI database
were studied by \citet{Fremat2006} and \citet{Neiner2005} and are omitted here,
since the FASTWIND code is not developed to treat stars with a circumstellar
disc. Also double-lined spectroscopic binaries (SB2\label{acro_SB2}) were
omitted from our sample, as their combined spectra make an accurate fundamental
parameter estimation impossible as long as their flux ratios are not known.
Among the GAUDI sample, we discovered a
few candidate spectroscopic binaries, which were not known to be SB2s. It
concerns HD~173003 (B5), HD~181474 (B5), HD~42959 (B8), HD~50982 (B8, variable
star), HD~51150 (B8, clearly asymmetric profiles), HD~181761 (B8), HD~45953 (B9,
variable star) and HD~46165 (B9). 

For the few available SARG data, only the default normalized spectra were
inserted into the database and we failed to get hands on the raw data, which
would be needed for a careful rerectification. Indeed, the poor continuum
rectification is especially clear from H$\beta$, but also shows up in other
line profiles. Moreover, the spectral coverage of the SARG
spectra is too small to deduce any useful information. Therefore, we excluded
all SARG spectra from our sample. The same accounts for the CATANIA spectra,
for which the spectral quality was too poor for a detailed spectral analysis,
so also these spectra were omitted. 
We thus restrict this paper to FEROS and ELODIE spectra, having a
resolution of 48\,000 and 50\,000, respectively.

The standard FITS data of the 
FEROS and ELODIE spectra in the GAUDI
database 
contain information about both the normalized spectrum, resulting from
the pipeline reduction, and the unnormalized spectrum. Since
the quality of normalization turned out to be
insufficient for our detailed analyses, one of us (TM) 
redid the normalization 
for all spectral ranges around the diagnostic lines
in a uniform way using
IRAF\footnote{\label{acro_IRAF}IRAF (Image Reduction and
Analysis Facility) is distributed by the National Optical Astronomy
Observatories, operated by the Association of Universities for Research in
Astronomy, Inc., under cooperative agreement with the National Science
Foundation, USA.} to ensure a homogeneous treatment of the sample.

Among the GAUDI sample, we find, as expected, a lot of fast rotators. Their
spectra usually contain insufficient line information due to their high
projected
rotational velocities. Often, only the Balmer lines and the strongest He\,{\sc
i}
lines (i.e.\ He I 4026, 4471 and 4922) can be detected, while all Si lines are
lost. The stars that were discarded for this reason are HD~182519 (B5, $>$300
km\,s$^{-1}$), HD~50751 (B8, $>$200 km\,s$^{-1}$), HD~56006 (B8, $>$150
km\,s$^{-1}$), HD~45515 (B8 V, $>$190 km\,s$^{-1}$), HD~182786 (B8,
$>$160 km\,s$^{-1}$), HD~50252 (B9 V, $>$140 km\,s$^{-1}$), HD~169225 (B9,
$>$160 km\,s$^{-1}$), HD~171931 (B9, $>$250 km\,s$^{-1}$), HD~174836 (B9,
$>$140 km\,s$^{-1}$), 176258 (B9 V, $>$160 km\,s$^{-1}$), HD~179124 (B9 V,
$>$270 km\,s$^{-1}$), and HD~45760 (B9.5 V, $>$190 km\,s$^{-1}$).
For some of the GAUDI stars, we were able to fit the spectrum despite the high
projected rotational velocity, but their resulting parameters cannot be very
reliable.
This concerns HD~51507 (B3 V, 148 km\,s$^{-1}$), HD~45418 (B5, 237 km\,s$^{-1}$),
HD~46487 (B5 Vn, 265 km\,s$^{-1}$), HD~178744 (B5 Vn, 224 km\,s$^{-1}$),
HD~43461 (B6 V, 210 km\,s$^{-1}$), HD~44720 (B8,
160 km\,s$^{-1}$), HD~49643 (B8 IIIn, 296 km\,s$^{-1}$), HD~173370 (B9V, 282
km\,s$^{-1}$,
double-peaked $\Ha$ profile), HD~179124 (B9 V, 278 km\,s$^{-1}$), and HD~181690 (B9
V, 189 km\,s$^{-1}$). Their parameters can be found in
Table\,\ref{table_parameters}.

For other targets, we have only spectra of insufficient quality available
which cannot be used for spectral line fitting purposes. It concerns HD~48691
(B0.5 IV), HD~168797 (B3 Ve), HD~178129 (B3 Ia), HD~57608 (B8 III, possible
instrumental problem), HD~44654 (B9), HD~45257 (B9), and HD~53204 (B9).

For a few stars, we have both an ELODIE and a FEROS spectrum available, e.g.,
for HD~174069. From inspection of the spectra, it was immediately clear that
there are differences in the line profiles, in particular the wings
of the Balmer lines are much less pronounced in the ELODIE spectra (see
Fig.\,\ref{FEROS_ELODIE}). Also other stars, for which we have both a FEROS and
an ELODIE spectrum available, show the same discrepancy, so there seems to
be a systematic effect. We fitted both spectra and came up with a
different set of parameters. The difference in $\logg$ is large ($>$ 0.5 dex)
and certainly worrisome.
To investigate this problem, we looked up the spectra
in the FEROS and ELODIE archives and 
realized that the merging of the ELODIE spectra by the pipeline is far 
less accurate than for the FEROS spectra. This is due to an inaccurate
correction for the blaze function in the case of the ELODIE pipeline. Due to
this, we also redid the merging of the orders for the ELODIE spectra,
before the normalization.
This led to a much better
agreement with the available FEROS spectra.
We thus advise future users
against working with the merged and normalized
spectra from GAUDI, but rather to go back
to the original spectra in the FEROS and ELODIE archives, and not only redo the
normalization but even the
merging in the case of the ELODIE spectra.
Unfortunately, not all stars had a spectrum available in the ELODIE archives
though, e.g., for HD~47887 (B2 III), HD~52559 (B2 IV-V), HD~48977 (B2.5 V),
HD~50228 (B5), HD~57291 (B5), HD~51892 (B7 III), HD~52206 (B8) and HD~53202
(B9), we could not find the original spectrum, so we left out these stars
from our analysis. For HD~43317 (B3 IV), the archival spectrum was not usable.

\begin{figure}[t!]
\centering
\begin{minipage}{3.4in}
\resizebox{\hsize}{!}{\includegraphics{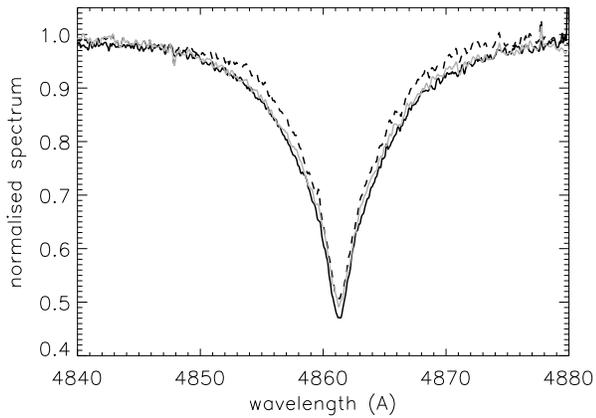}}
\end{minipage}
\caption{\label{FEROS_ELODIE} HD~174069: Significant (and apparently
systematic) discrepancies are observed between the line profiles of the FEROS
(solid black line) and the ELODIE (dashed black line) spectrum from the GAUDI
database. This leads to considerable discrepancies in the derived stellar
parameters. In grey, we show the normalized ELODIE spectrum obtained from the
ELODIE archives, which shows a much better agreement with the available FEROS
spectra.}
\end{figure}

Finally, for 17 of the remaining stars, we could not find a
satisfactory fit, and we decided to leave them out of the sample.  One
possible explanation might be that they are single-lined binaries (see also
\citealt{Massey2009} who encountered similar problems when analyzing a large
sample of LMC/SMC O-stars).

All together, this, unfortunately, reduces our sample a lot, and we are only
left with a bit more than a third of the targets initially in the database: 66
out of 187 objects.
Moreover, most of the objects are late B-type stars. Fig.\,\ref{histogram}
shows a histogram with the breakdown of the sample following spectral subtype.
The ratio marked above each bin indicates the number of analyzed targets
compared to the total amount that was available before the selection
procedure.

\subsection{Analysis results}

\subsubsection{Example fits}

To illustrate the obtained quality of the final fits, we show in
Figs.\,\ref{coolgiant} to \ref{hotdwarf} some examples of the resulting spectral
fits for three very different stars included in GAUDI: 
a `cool' giant, a hot supergiant and a middle
type dwarf. 

\scriptsize
\Ltable
\tabcolsep=3pt
\begin{longtable}{rllcccrcrccl}
\caption{\label{table_parameters} Stellar parameters of the analyzed
GAUDI B type stars. The first two columns give the HD number of the star and
the spectral type (SpT) taken from SIMBAD. Then we list the effective
temperature ($\Teff$) of the model which was found by AnalyseBstar to match the
observations best, the surface gravity corrected for centrifugal acceleration
($\loggc$, with corresponding uncertainty $\Delta \logg$), the He abundance
(n(He)/n(H), where 0.10 is solar), the interpolated Si abundance when 
available, otherwise the closest grid Si abundance ($\logsi$, where -4.49 is
solar, -4.19 is enhanced and -4.79 is depleted) and the microturbulence
($\xi$); $\vsini$ is the projected rotational velocity as obtained by
applying the implementation of the Fourier Transform method of
  \citet{Gray1973, Gray1975} by \citet{SimonDiaz2007}. In some
cases, we need an additional broadening to explain the line profiles. Therefore,
we list the required macroturbulent velocity (\vmacro) and indicate the
stars for which we suspect they might be pulsators from the fact that they have
a significant \vmacro. We attribute a flag to each star, to represent the
obtained fit quality or problems we encountered during the fitting procedure:
1~-~well-fitted, 2~-~probably a wrong spectral type (about 1\,000~K or more away
from the calibrations, proposed spectral type indicated in the
`remarks'-column), 3~-~no or hardly any Si~lines available, mostly due to 
fast rotation, 4~-~not very reliable fit (for other reasons than fast rotation,
see `remarks'), 5~-~He weak/Si strong, 6~-~fitting of $\Ha$, $\Hb$, and $\Hg$
line profiles complicated due to very broad wings which abruptly
change into very narrow line cores ($\logg$ may be less accurate). In the last
column, a few additional comments are given.
The formal uncertainty on the effective temperature is 1,000~K for 
$\Teff > 20,000$~K and $\Teff < 15\,000~K$, and 500~K for 
$\Teff < 15,000~K \le \Teff \le 20\,000~K$. The formal uncertainty on the
microturbulence and on the He and Si abundance is half of the grid step in most
cases: 2~km\,s$^{-1}$ and 0.025 dex, respectively. For objects below $15\,000~K$ we
adopt somewhat largers errors (3-$\sigma$ deviation corresponding to 2 grid
steps): 3 km\,s$^{-1}$, 0.033 and 0.2 dex, respectively.}\\
\hline
% &  &  &  &  &  &  &  &  &  &  & \\
HD number & SpT & \Teff &  \loggc & n(He)/n(H) & $\logsi$ &  $\xi$ &
\vsini &
$\vmacro$ & suspected  & flag & remarks \\ 
 &  & (K) &  (cgs) &  &  & (km\,s$^{-1}$) & (km\,s$^{-1}$) & (km\,s$^{-1}$) & pulsator & & \\ 
% &  &  &  &  &  &  &  &  &  & & &  \\ 
\hline
\endfirsthead
\caption{Continued.}\\
\hline
% &  &  &  &  &  &  &  &  &  &  & & \\ 
HD number & SpT & \Teff & \loggc & n(He)/n(H) &  $\logsi$ & $\xi$ & \vsini
& $\vmacro$ &
suspected  & flag & remarks \\ 
 &  & (K)  & (cgs) &  &  & (km\,s$^{-1}$) & (km\,s$^{-1}$) & (km\,s$^{-1}$) & pulsator & & \\ 
% &  &  &  &  &  &  &  &  &  &  & & \\ 
\hline
% &  &  &  &  &  &   &  &  & & &  \\ 
\endhead
\hline
 &  &  &  &  &   &  &  &  & &  \\ 
48434 & B0III & 28000 %& 27984 
& 3.11 $\pm$ 0.10 & 0.10 & -4.44 & 15 &  62 $\pm$ 3 & 31 & yes & 4 &
ELODIE, poor quality spectrum, He~lines too strong\\ 
 &  &  &  &  &   &  &  &  & & & Balmer lines not well fitted due to bumps in
blue wings\\ 

52382 & B1Ib & 23000 %& 23228 
& 2.71 $\pm$ 0.10 & 0.15 & -4.67 & 20 &  56 $\pm$ 5 & 53 & yes & 4 &  P
Cygni profile $\Ha$ not reproduced,\\
 &  &  &  &  &   &  &  &  & & & $\Hb$ and $\Hg$ not well fitted\\ 

170580 & B2V & 20000 %& 20135 
& 4.10 $\pm$ 0.10 & 0.10 & -4.81 & 6 &  11 $\pm$ 1 & 0 & no & 1 & N(He)/N(H)
lower than solar\\

44700 & B3V & 17500 %& 17450
 & 3.80 $\pm$ 0.10 & 0.10 & -4.93 & 6 &  8 $\pm$ 1 & 0 & no & 1 & N(He)/N(H)
lower than solar\\ 

181074 & B3 & 20000 %& 19843 
& 3.64 $\pm$ 0.10 & 0.10 & -4.35 & 12 & 133 $\pm$ 7 & 0 & no & 2 & probably
B2, fast rotator\\

45418 & B5 & 16000  & 4.0 $\pm$ 0.20 & 0.10 & -4.49 & 3 &  237 $\pm$ 28
& 0 &
no & 3 & \\

46487 & B5Vn & 14500  & 3.69 $\pm$ 0.10 & 0.15 & -4.49 & 3 & 
265 $\pm$        5 & 0 & no & 3 & \\ 

48215 & B5V & 14500 & 3.82 $\pm$ 0.10 & 0.10 & -4.49 & 3 & 
98 $\pm$        3 & 48 & yes & 1 & only Si~II \\

54596 & B5 & 20000 %& 20445 
& 3.31 $\pm$ 0.05 & 0.10 & -4.48 & 6 &  69 $\pm$        6 & 53 & yes & 2 &
only Si~III, probably B2\\

58973 & B5 & 15000 &  3.43 $\pm$ 0.20 & 0.10 & -4.49 & 3 & 
94 $\pm$        1 & 50 & yes & 1 &  \\

177880 & B5V & 14500 &  3.81 $\pm$ 0.10 & 0.10 & -4.49 & 3 & 
49 $\pm$        1 & 23 & yes & 1 & \\

178744 & B5Vn & 14500 &  3.71 $\pm$ 0.20 & 0.10 & -4.49 & 12 & 
224 $\pm$        5 & 0 & no & 3 & \\

43461 & B6V & 13500 & 3.35 $\pm$ 0.10 & 0.10 & -4.79 & 3 & 
210 $\pm$        5 & 0 & no & 3 & \\

48807 & B7Iab & 12500 %& 12300
 & 2.00 $\pm$ 0.10 & 0.10 & -4.28 & 6 &  24 $\pm$ 1 & 27 & yes & 1 & \\

51360 & B7III & 13500 %& 13619 
& 3.12 $\pm$ 0.10 & 0.10 & -3.93 & 3 &  73 $\pm$ 1 & 55 & yes & 1 &
very high Si abundance\\

42677 & B8 & 11000 & 3.55 $\pm$ 0.20 & 0.10 & -4.79 & 6 & 
121 $\pm$  14 & 104 & yes & 2 & probably B9\\

44720 & B8 & 14000 &  3.96 $\pm$ 0.10 & 0.10 & -4.19 & 3 & 
160 $\pm$ 5 & 0 & no & 2 & probably B5 or B6\\

45153 & B8 & 11000 & 3.66 $\pm$ 0.10 & 0.10 & -4.49 & 3 & 
142 $\pm$ 4 & 0 & no & 2 & probably B9\\

45284 & B8 & 15000 &  4.10 $\pm$ 0.10 & 0.10 & -4.19 & 3 & 
52 $\pm$   5 & 42 & yes & 2 & SPB with magnetic field, skew Si~II lines,
probably B5\\

45397 & B8 & 11500 &  3.59 $\pm$ 0.10 & 0.20 & -4.79 & 6 & 
156 $\pm$ 12 & 0 & no & 3 & \\

45515 & B8V & 11000 & 4.18 $\pm$ 0.10 & 0.15 & -4.49 & 3 & 
190 $\pm$ 5 & 0 & no & 2, 3 &  probably B9\\

46616 & B8 & 17500 %& 17480 
& 4.20 $\pm$ 0.50 & 0.10 & -4.24 & 3 &  8 $\pm$  1 & 7 & yes & 5 & He-weak
star\\

47964 & B8III & 11500 & 3.01 $\pm$ 0.20 & 0.10 & -4.79 & 3 & 
49 $\pm$        1 & 44 & yes & 1 & \\

48497 & B8 & 14000 %& 13937
 & 3.70 $\pm$ 0.10 & 0.10 & -4.58 & 3 & 14 $\pm$ 1 & 13 & yes & 4, 6 & 
N(He)/N(H) lower than solar\\

49481 & B8 & 11000 &  2.70 $\pm$ 0.10 & 0.20 & -4.79 & 3 & 
9 $\pm$        1 & 13 & yes & 2, 6 & probably B9\\

49643 & B8IIIn & 14500 & 3.88 $\pm$ 0.10 & 0.10 & -4.19 & 3 & 
296 $\pm$        2 & 13 & no & 3 &\\

49886 & B8 & 10000 &  3.30 $\pm$ 0.20 & 0.10 & -4.79 & 6 & 
8 $\pm$ 1 & 16 & yes & 4, 6 & \\

49935 & B8 & 12500 &  3.33 $\pm$ 0.20 & 0.10 & -4.79 & 6 & 
92 $\pm$        9 & 72 & yes & 1 & $\logg$ may be too low?\\

50251 & B8V & 11500 &  3.00 $\pm$ 0.20 & 0.10 & -4.79 & 3 & 
9 $\pm$        1 & 17 & yes & 1 & emission line around Si~III~4813 and
Si~IV~4212\\

50513 & B8 & 11500 &  4.04 $\pm$ 0.10 & 0.10 & -4.79 & 6 & 
112 $\pm$        5 & 80 & yes & 1 & \\

55793 & B8 & 11000 & 3.05 $\pm$ 0.05 & 0.20 & -4.79 & 3 & 
101 $\pm$        1 & 56 & yes & 2 & probably B9, $\logg$ may be too low?\\

56446 & B8III & 11500 &  3.21 $\pm$ 0.20 & 0.20 & -4.19 & 3 & 
229 $\pm$       32 & 13 & no & 3 & \\

170795 & B8 & 15500 & 4.01 $\pm$ 0.10 & 0.10 & -4.49 & 3 & 
79 $\pm$        7 & 37 & yes & 2 & probably B5\\

171247 & B8IIIsp & 10000 &  2.83 $\pm$ 0.10 & 0.10 & -4.19 & 6 & 
66 $\pm$        2 & 20 & yes & 4 & peculiar line behaviour\\

173673 & B8 & 10000 &  2.90 $\pm$ 0.10 & 0.20 & -4.49 & 3 & 
25 $\pm$        2 & 26 & yes & 4, 6 & too low $\logg$?\\

179761 & B8II-III & 12500 & 3.30 $\pm$ 0.05 & 0.10 & -4.49 & 3 & 
17 $\pm$        1 & 12 & yes& 1 &  \\

180760 & B8 & 17000 &  4.05 $\pm$ 0.10 & 0.10 & -4.49 & 6 & 
167 $\pm$        3 & 0 & no & 3 & wrong spectral type?\\

44321 & B9 & 11000 &  3.62 $\pm$ 0.05 & 0.10 & -4.79 & 6 & 
90 $\pm$        3 & 30 & yes & 1 & \\

44354 & B9 & 13000 &  3.94 $\pm$ 0.05 & 0.15 & -4.79 & 10 & 
121 $\pm$        5 & 72 & yes & 2 & probably B7 or B8\\

45050 & B9V & 11500 &  3.86 $\pm$ 0.10 & 0.10 & -4.79 & 10 &
140 $\pm$       47 & 0 & no & 3 & \\

45516 & B9 & 13000 &  3.90 $\pm$ 0.10 & 0.10 & -4.49 & 6 & 
281 $\pm$       55 & 12 & no & 3 & \\

45657 & B9 & 11000 &  4.24 $\pm$ 0.10 & 0.10 & -4.79 & 3 & 
153 $\pm$       36 & 0 & no& 3 &  \\

45709 & B9 & 11000 &  3.95 $\pm$ 0.30 & 0.20 & -4.79 & 10 & 
231 $\pm$       24 & 10 & no & 3 & \\

45975 & B9 & 11000 &  3.51 $\pm$ 0.10 & 0.15 & -4.79 & 3 & 
54 $\pm$        1 & 56 & yes & 1 & $\vmacro$ too high\\

46138 & B9 & 12000 &  3.93 $\pm$ 0.05 & 0.10 & -4.19 & 3 & 
100 $\pm$        5 & 0 & no & 4 &\\

46886 & B9 & 11000 &  3.00 $\pm$ 0.10 & 0.10 & -4.79 & 3 & 
16 $\pm$        3 & 23 & yes & 1 & $\logg$ too low?\\

47278 & B9 & 10000 &  3.60 $\pm$ 0.05 & 0.10 & -4.79 & 3 & 
31 $\pm$        4 & 25 & yes & 2 & rather A0\\

48808 & B9 & 12000 &  3.21 $\pm$ 0.10 & 0.10 & -4.79 & 3 & 
47 $\pm$       14 & 63 & yes & 1 & $\vmacro$ too high, rather B8\\

48957 & B9 & 12000 &  3.10 $\pm$ 0.10 & 0.10 & -4.79 & 3 & 
24 $\pm$        2 & 23 & yes & 4 & $\Teff$ too low?, wrong spectral type?\\

49123 & B9 & 10000 &  3.41 $\pm$ 0.05 & 0.10 & -4.79 & 3 & 
45 $\pm$        2 & 0 & no & 2 & rather A0\\

52312 & B9III & 11000 &  3.04 $\pm$ 0.05 & 0.20 & -4.49 & 3 & 
175 $\pm$        5 & 0 & no & 3 & \\

53004 & B9 & 11000 &  3.91 $\pm$ 0.20 & 0.10 & -4.79 & 6 & 
51 $\pm$        7 & 54 & yes & 1 & \\

54761 & B9 & 11500 &  3.11 $\pm$ 0.10 & 0.10 & -4.79 & 6 & 
53 $\pm$        1 & 71 & yes & 1 & Balmer line cores slightly refilled, $\logg$
too high?\\

54929 & B9 & 10000 &  3.11 $\pm$ 0.10 & 0.10 & -4.49 & 3 & 
46 $\pm$        1 & 107 & yes & 4 & fast rotator, difficult to fit, too high
$\vmacro$\\

56613 & B9 & 13000 &  3.92 $\pm$ 0.20 & 0.10 & -4.79 & 6 & 
91 $\pm$        7 & 36 & yes & 2 & rather, B7 or B8\\

172850 & B9 & 11000 &  3.62 $\pm$ 0.10 & 0.10 & -4.49 & 3 & 
77 $\pm$        4 & 0 & no & 1 & \\

173370 & B9V & 11500 &  3.46 $\pm$ 0.10 & 0.10 & -4.79 & 15 & 
282 $\pm$        8 & 0 & no & 4 & very fast rotator, no Si~lines
available,double-peaked $\Ha$ profile,\\
 &  &  &  &  &  &  &  &  & & & $\Hb$ refilled on both sides of the line core\\

173693 & B9 & 10000 &  3.32 $\pm$ 0.10 & 0.10 & -4.79 & 3 & 
60 $\pm$        1 & 22 & yes & 2 & rather A0\\

174701 & B9 & 12500 &  3.68 $\pm$ 0.10 & 0.15 & -4.49 & 3 & 
170 $\pm$        2 & 10 & no & 3 & \\

175640 & B9III & 10000  & 3.20 $\pm$ 0.10 & 0.15 & -4.49 & 3 & 
7 $\pm$        1 & 10 & yes & 2 & rather A0\\

176076 & B9 & 10000 &  3.41 $\pm$ 0.10 & 0.10 & -4.79 & 3 & 
42 $\pm$        2 & 0 & no & 2 & rather A0\\

176158 & B9 & 13500 &  3.64 $\pm$ 0.10 & 0.15 & -4.79 & 3 & 
121 $\pm$       11 & 0 & no & 2 & rather B8\\

179124 & B9V & 12000 &  3.45 $\pm$ 0.10 & 0.10 & -4.49 & 15 & 
277 $\pm$        6 & 0 & no & 3 & \\

181440 & B9III & 11500 &  3.61 $\pm$ 0.50 & 0.10 & -4.49 & 3 & 
55 $\pm$        7 & 32 & yes & 1 & \\

181690 & B9V & 13000 &  3.42 $\pm$ 0.10 & 0.10 & -4.49 & 6 & 
189 $\pm$        5 & 0 & no & 3 & \\

182198 & B9V & 11000 &  3.10 $\pm$ 0.10 & 0.10 & -4.79 & 3 & 
23 $\pm$        1 & 12 & yes & 1 & \\ 
\hline
\end{longtable}
%\end{landscape}
\normalsize

\twocolumn

The corresponding physical parameters are listed in
Table\,\ref{table_parameters}, which summarizes the derived stellar
parameters of the analysed GAUDI B type stars. All other spectral line fits,
and their corresponding parameters, can be found at
http://www.ster.kuleuven.be/\verb|~|karolien/AnalyseBstar/Bstars/.

\begin{figure}[t!]
\centering
\begin{minipage}{3.4in}
\resizebox{\hsize}{!}{\includegraphics{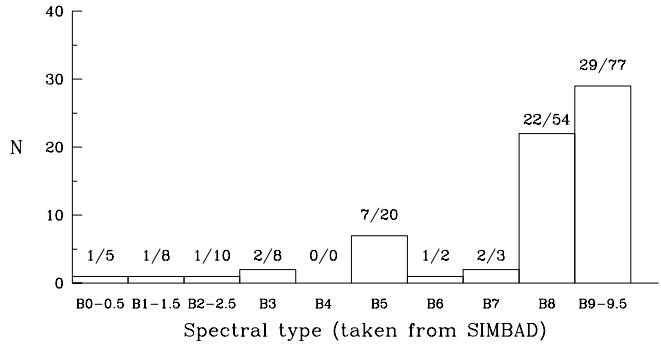}}
\end{minipage}
\caption{\label{histogram} Histogram of the analyzed stars within the GAUDI
B star sample. The ratio marked above each bin indicates the number of analyzed
targets compared to the total amount that was available before the selection
procedure, e.g., 1/5 means that there were 5 objects with this spectral type in
the database, but that, for certain reasons, we were only able to analyze 1 of
them. In total, 66 out of 187 objects could be analyzed.}
\end{figure}

\subsubsection{Comments on individual stars}

In what follows, we discuss a few of the sample stars individually, because
of certain particularities,

\paragraph{HD~45284.} The derived effective temperature is in complete
agreement with what was found by \citet{Hubrig2006}. This star is an SPB
exhibiting a magnetic field. The combination of the effective temperature and
gravity resulting from Geneva photometry \citep{Hubrig2006} positions the star
just outside the main sequence towards the high-gravity side. Now, with our
newly derived $\logg$, it falls within the expected instability domain of the
SPB stars.

\paragraph{HD~46616 (B8)} is a clear example of a He-weak star: all He\,{\sc
i}-lines
have extremely low equivalent widths compared to `normal' B stars. Even though
the hydrogen and Si lines fit nicely, the He lines are predicted too strong in
our grid models\footnote{the lowermost considered He-abundance in our grid is
n(He)/n(H)=0.1, see Section\,\ref{section_grid}.}.

\paragraph{HD~48106 (B8)} is oxygen poor based on the non-detection of the
O~I~7771-7775 triplet and is either He weak or Si strong. We were unable to fit
both the He and Si lines at the same time.

\paragraph{For HD~46340 (B8),}only He\,{\sc
i}~4471 and Si\,{\sc
ii}~5041-5056 are visible.
This was only sufficient to make a very rough estimate of the fundamental
parameters: \Teff\ $\approx$ 10\,000~K, \logg\ $\approx$ 4.1, negligible wind,
solar He abundance.

On the hot side of the B-type domain, the analysis of the
early type stars (e.g., HD~172488 (B0.5~V), HD~52918 (B1~V) and HD~173198
(B1~V)) show similar difficulties. The iterative procedure tends to yield too
low temperatures due to the lower weight of the Si\,{\sc
iv} and He\,{\sc
ii} lines as we have
only one weak line. As only Si\,{\sc
iii} is reliable and He\,{\sc
i} is not much affected by
the effective temperature in this range, we are not able to derive trustworthy
results.

\begin{figure*}[t!]
\centering
\begin{minipage}{7.5in}
\resizebox{7in}{!}{\includegraphics{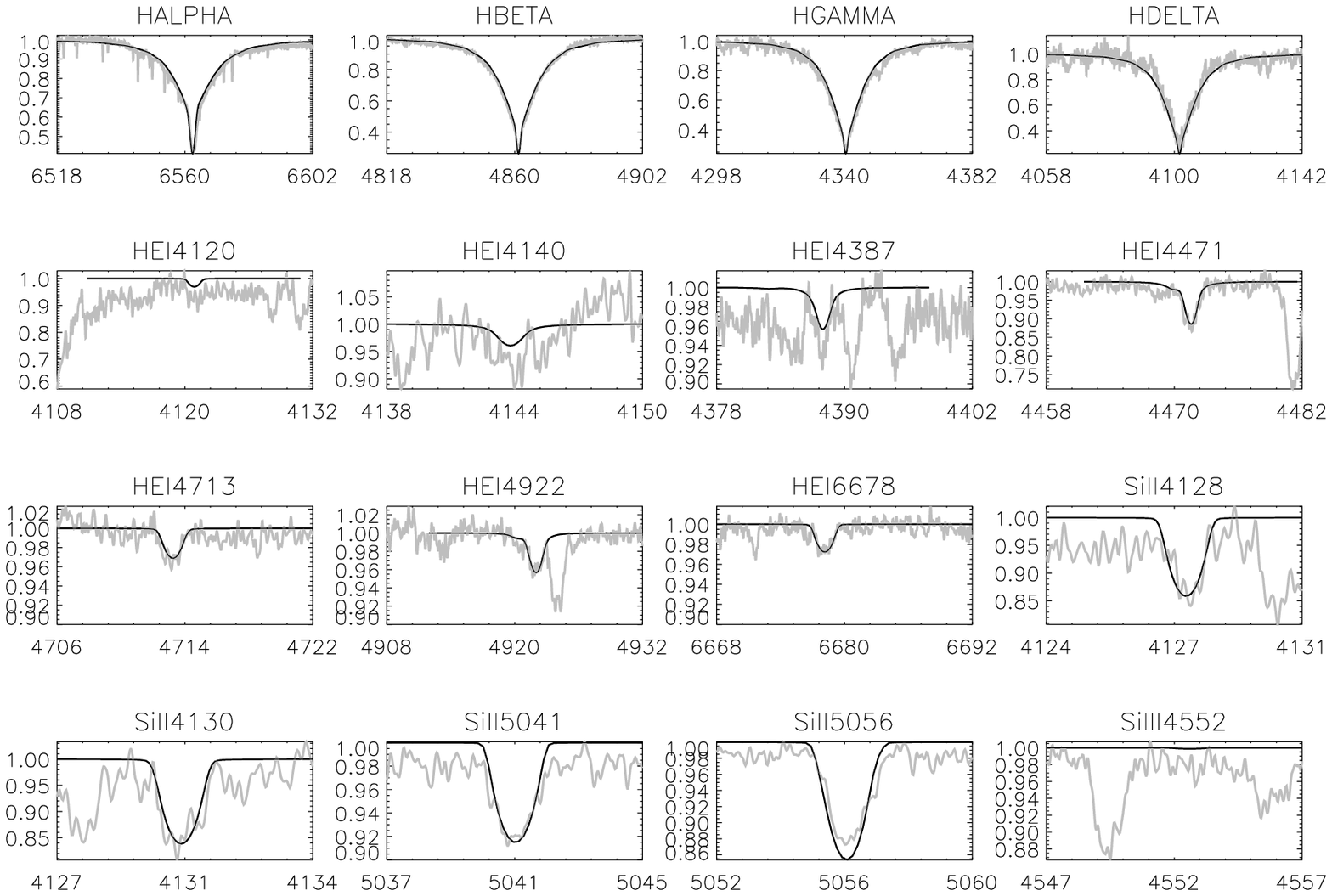}}\\
\resizebox{7in}{!}{\includegraphics{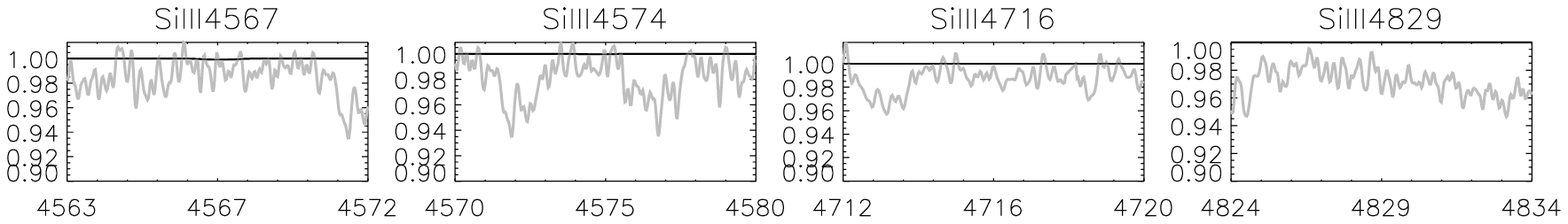}}
\end{minipage}
\caption{\label{coolgiant} Example of the line profile fits for \textit{a `cool' 
giant} which has, meanwhile, been observed by CoRoT (Miglio et al., in
  preparation): HD~181440 (B9~III).
Despite the low quality of the data, and the fact that we have no more Si\,{\sc
iii} left in this temperature region, we are still able to obtain a
satisfying solution.}
\end{figure*}

\begin{figure*}[t!]
\centering
\begin{minipage}{7.5in}
\resizebox{7in}{!}{\includegraphics{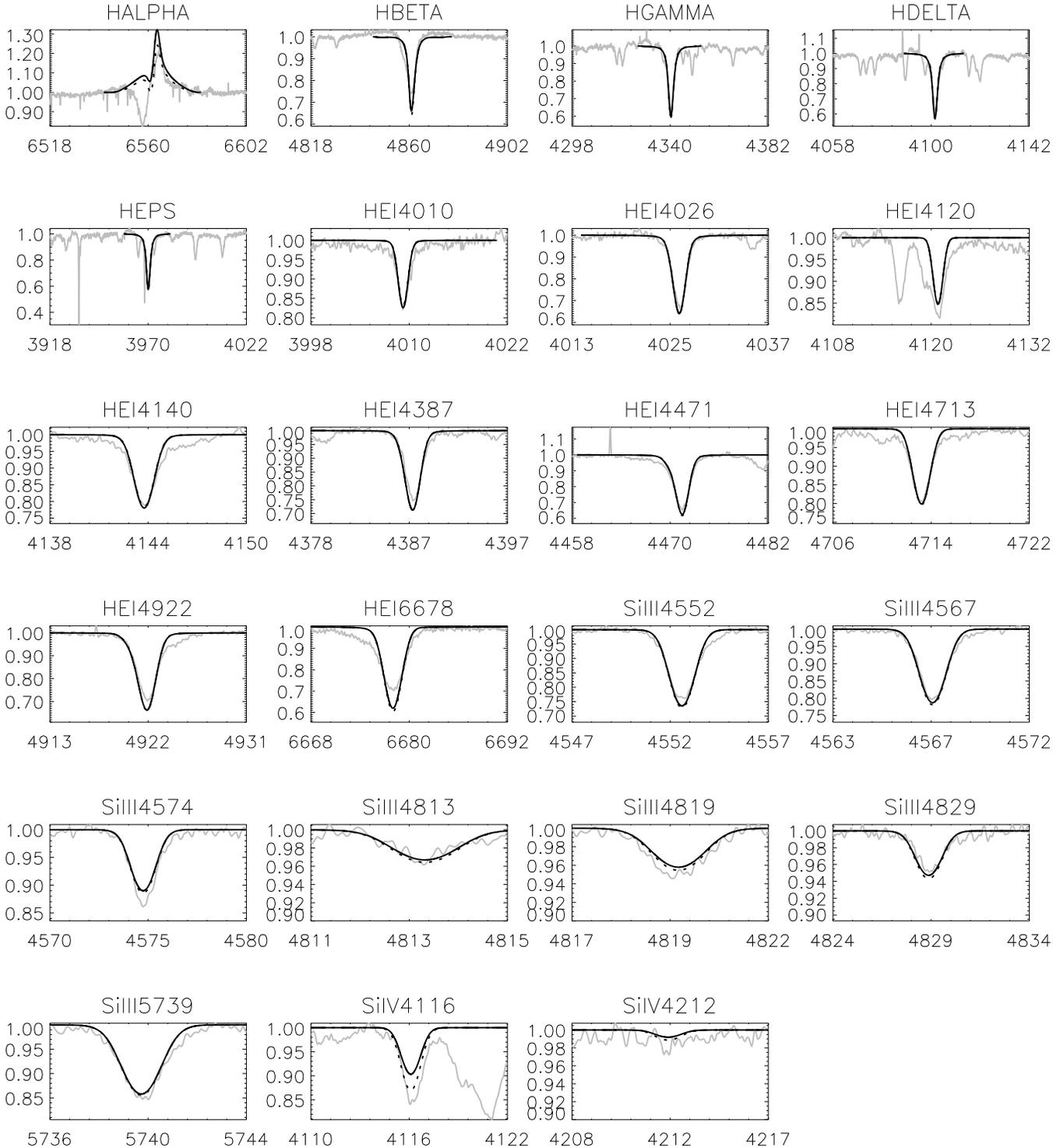}}
\end{minipage}
\caption{\label{hotsupergiant} Example of the line profile fits for the
only \textit{hot supergiant} in our sample: HD~52382 (B1Ib).
While the peak of $\Ha$ is reasonably well reproduced, the blue-ward absorption
trough cannot be fitted at the available grid combinations of wind strength
parameter and wind velocity field exponent. Note that no interpolation 
is performed for these quantities, and that the wind is considered as
unclumped, which might explain the obvious mismatch (e.g.,
\citealt{Puls2006}, in particular their Fig.~7).
The Balmer lines show bumps in the wings. It is not clear whether
this is real and due to the strong wind, or if this is a spectral artefact. It
also arises in the only cool supergiant in our sample. The interpolated Si
abundance ($\logsi = -4.67$) is higher than the closest grid abundance ($\logsi
= -4.79$, plotted here), which explains the observed discrepancies in the Si
line cores. {In dotted lines, we show the line profiles as they
would appear with the interpolated parameters.}}
\end{figure*}

\begin{figure*}[t!]
\centering
\begin{minipage}{7.5in}
\resizebox{7in}{!}{\includegraphics{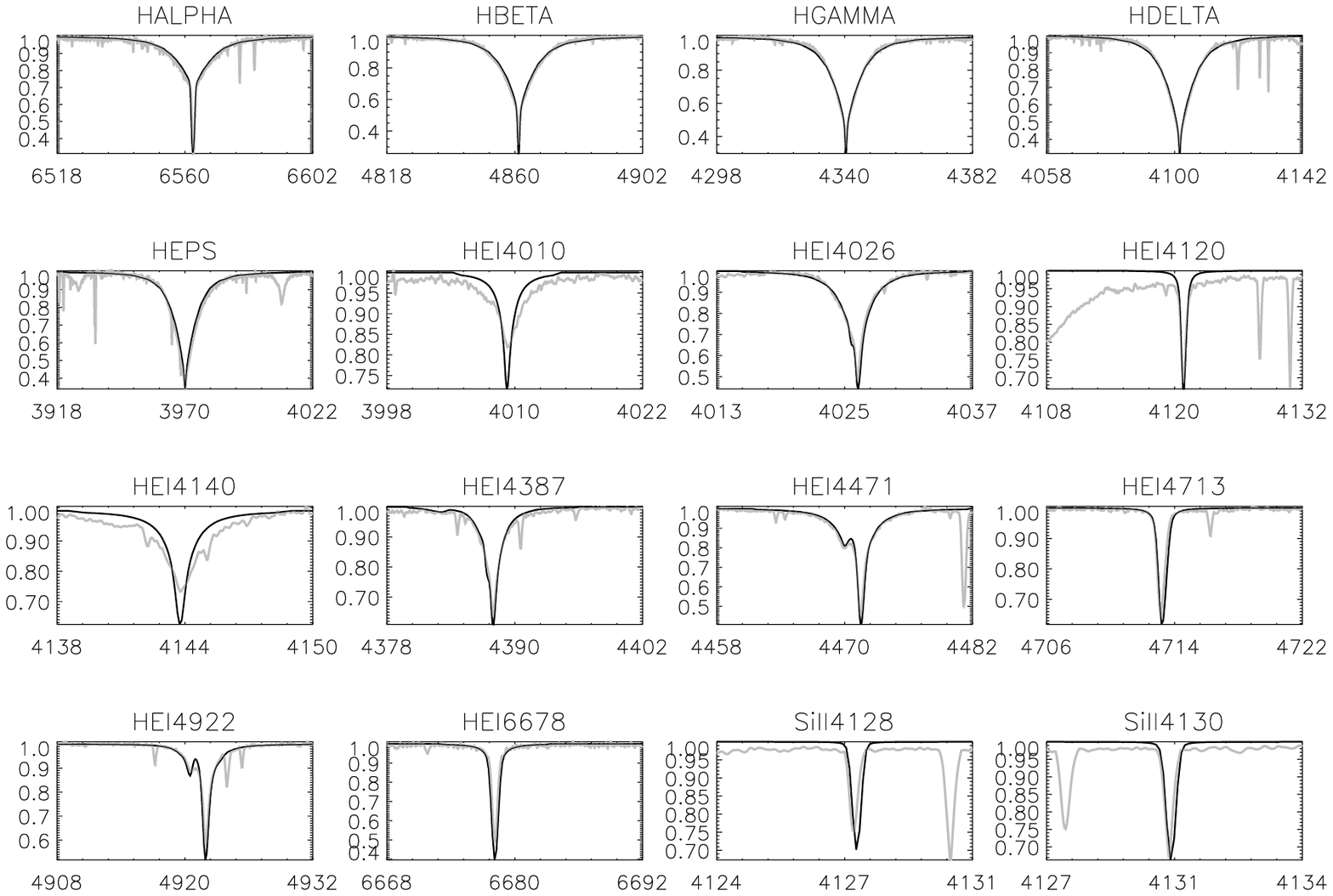}}
\resizebox{7in}{!}{\includegraphics{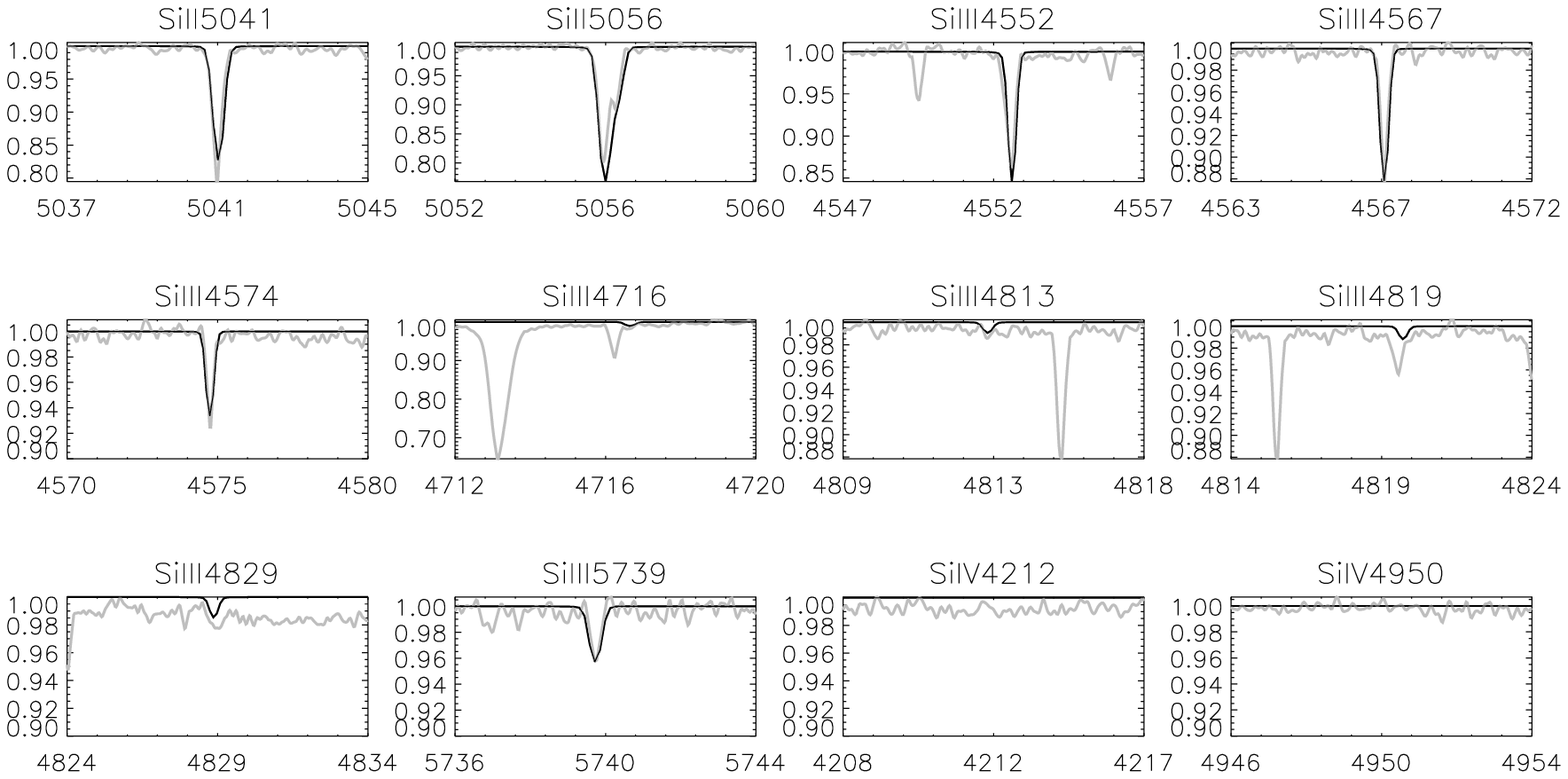}}
\end{minipage}
\caption{\label{hotdwarf} Example of the line profile fits for an
\textit{intermediate hot dwarf}: 
HD~44700 (B3~V). The strong lines on the blue side of
Si\,{\sc iii}~4716 and Si\,{\sc iii}~4819 are easily mistaken for the Si lines
themselves. This is the reason why we always indicate the exact position of the
transition during the full preparation process. The misfit of He\,{\sc i}~4010
and 4140 is due to incomplete broadening functions of these lines. These lines
were not used during the fitting procedure, but only as a double check
afterwards (see \texttt{Appendix\,\ref{lineselection}}). We observe a similar
behavior in several other stars. All other lines, even the weakest, fit very
well. The shape of He\,{\sc i}~4471 is even perfect.}
\end{figure*}

\paragraph{HD~173370 (B9~V)} shows a double peaked $\Ha$ profile. This
morphology is typical for a very fast rotator (in this case, $\vsini \approx$
280 km\,s$^{-1}$), seen almost equator-on. The disk-like feature also strongly affects
the $\Hb$ and $\Hg$ profiles.\\

We also found a few chemically peculiar stars among the B stars in the GAUDI
database. They are HD~44948 (B8 Vp), HD~45583 (B8), HD~46616 (B8) and HD~44907
(B9). They have a very rich spectrum, with a forest of sharp spectral
lines, complicating the continuum determination and the fitting process, as
almost all lines are blended. The He lines of HD~45583 are extremely weak
and those of HD~46616 even completely vanished. Following Landstreet
(2007), HD~45583 is a periodically variable Ap star. The line spectrum is
completely distorted, hence too complicated to fit. The Si\,{\sc
ii} lines are very
strong, which indeed points towards a B8 or even B9 star.

\section{Statistical properties of the sample and physical interpretation
\label{interpretation}}

\subsection{Effective temperature scale}

To derive a reliable calibration for the effective temperature as a function of
the spectral subclass for B dwarfs, we need enough stars for which we have
accurate information for both parameters. Unfortunately, this prerequisite is
not fulfilled for our sample, as can be seen from the very large scatter around
the existing spectral-type-$\Teff$-calibrations, presented in
Fig.\,\ref{TEFF_SPT}. The scatter is due to uncertain effective temperatures
for some targets and wrong spectral type designation for others. Indeed, for
several stars, we encountered problems in the analysis, primarily because of 
the absence of Si lines for many fast-rotating late B type stars. These stars
are too cool to have Si\,{\sc iii} present in their line spectrum, and the
Si\,{\sc ii} lines are completely smoothed, so that they become hardly
detectable. Even when they are still visible, the two lines of the Si\,{\sc
ii}~4128-4130 doublet are so heavily blended that they cannot be used any
longer. In these cases, the He\,{\sc i} lines are the only lines that can be
used to estimate the effective temperature, which are, therefore, not reliable
enough to use in the derivation of an accurate temperature calibration. On
the other hand, several stars for which we were able to derive accurate values
for the effective temperature turned out to have completely wrong spectral
types. We caution the use of spectral types as they are quite often derived from
low-resolution and low-quality spectra.
Flags and remarks on the fit quality of individual targets were added in
Table\,\ref{table_parameters}.

\begin{figure}[t!]
\begin{minipage}{3.7in}
\hspace{-0.8cm}\resizebox{\hsize}{!}{\includegraphics{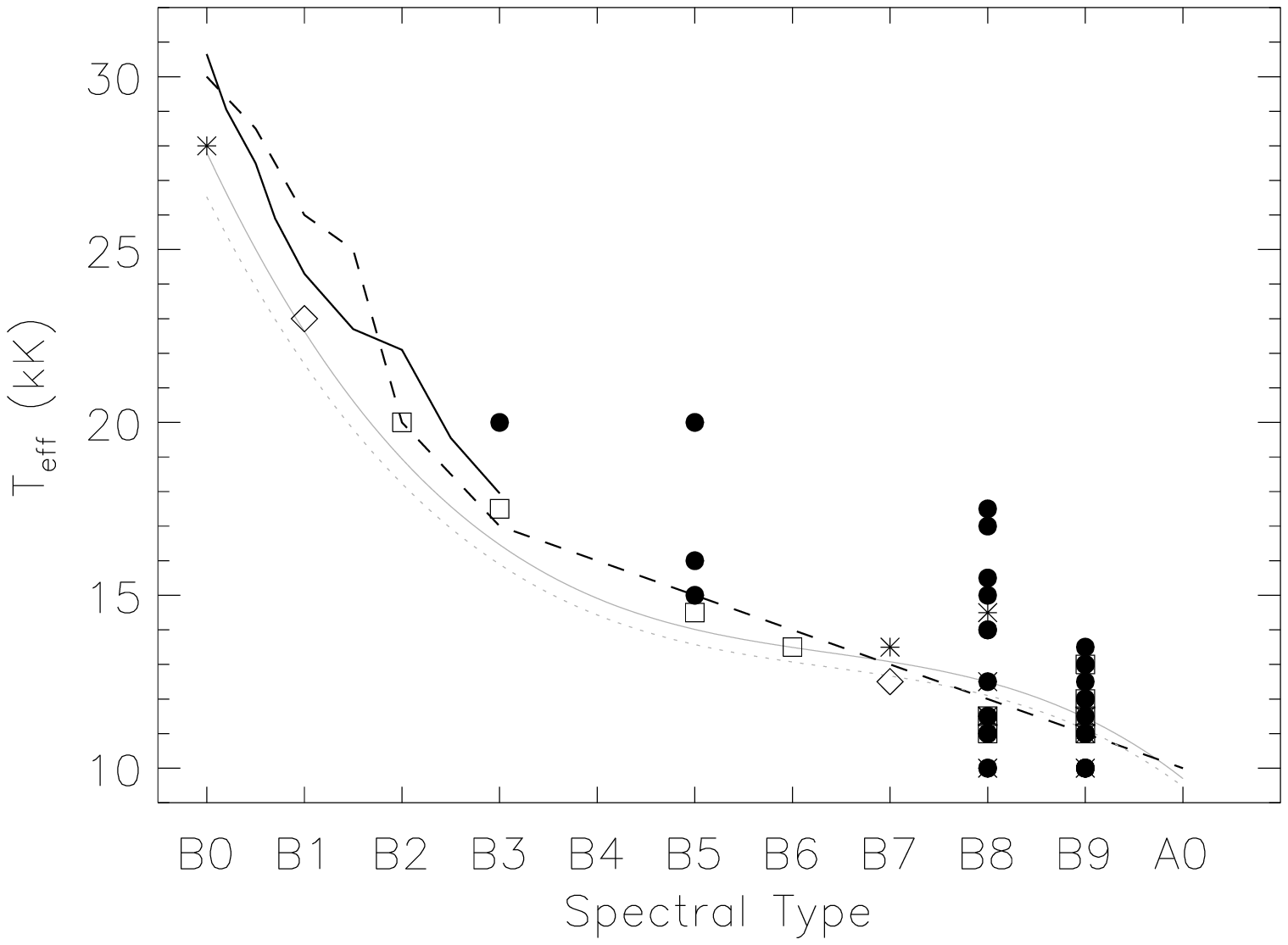}}
\end{minipage}
\caption{\label{TEFF_SPT} Effective temperature as a function of spectral
type. We compare the position of our sample stars to existing temperature
scales: for Galactic dwarfs, for spectral
types up to B3 \citep[solid black line]{Trundle2007}, for
dwarfs in the entire B type domain \citep[dashed black line]{Crowther1998}, 
for Galactic B supergiants \citep[dotted grey line]{Lefever2007} and
for Galactic and SMC B supergiants \citep[solid
grey line]{Markova2008}. Due to wind blanketing effects, the
calibrations for the B supergiants lie at lower temperatures than the
calibrations for B dwarfs. The supergiants (l.c. I) are indicated by diamonds,
the (sub)giants (l.c. II and III) by asterisks and the (sub)dwarfs (l.c. IV and
V) by squares. Filled circles indicate the stars for which we do not have any a
priori information on luminosity class.
It can be seen that spectral types taken from the literature can be quite
inaccurate (see text).}
\end{figure}

\subsection{The stars in the ($\Teff$, $\loggc$)-diagram}

Fig.\,\ref{HRD_GAUDI} shows the ($\Teff$, $\loggc$)-diagram of the analyzed
GAUDI stars, along with the position of instability strips. The latter were
computed with the evolutionary code CL\'ES \citep{Scuflaire2008} and the
pulsation code
MAD \citep{Dupret2002}, respectively. These strips were computed for a
core-overshoot parameter of 0.2 times the local pressure scale height
\citep{Miglio2007a,Miglio2007b}. It was indeed found from seismic modeling of B
type stars that core overshooting occurs, on average with such a value
\citep{Aerts2008b}. We thus compare the position of the GAUDI stars with the
most recent models tuned by asteroseismology.

For 40\% of the GAUDI stars (26 out of 66), we have information on the
luminosity class. All but one (i.e. 9 out of 10) with known luminosity
class II or III are indeed situated beyond the 
Terminal Age Main Sequence (TAMS hereafter),
so we confirm them to be giants.  The only exception
is HD~49643, a star of spectral type B8\,IIIn, i.e.\ with spectroscopic lines
which may originate from nebulosity.  The star is a fast rotator with
a $\vsini$ of almost 300 km\,s$^{-1}$. The absence of Si lines due to the high
rotation implies less reliable estimates for $\Teff$ and $\logg$ than for the
bulk of the sample stars. Thus, its derived gravity of 3.70\,dex might be
compatible with its previous classification as a giant.

Three of the 14 stars with luminosity class IV or V turn out to be giants, i.e.
they have -- within the error bars -- a surface gravity $\loggc$ below 3.5. It
concerns HD~43461 (B6V), HD~50251 (B8V), and HD~182198 (B9V), of
which HD~43461 is a fast rotator and has, therefore, less reliable parameters
due to the absence of Si~lines. Regarding the other two stars,
there is a certain possibility that
they are fast rotators as well, observed almost pole-on.
Due to centrifugal effects, the stellar surface would become distorted, and
effective temperature (due to gravity darkening) and surface gravity
increase towards the pole. If observed pole-on, both quantities might be
underestimated with respect to their average values.
On the other hand, also the main sequence gets extended
compared to standard models when rotation is taken into account
(e.g., \citealt{Maeder2000}).
Similar results to ours were obtained by \citet{Hempel2003}, who found 6 out
of their 27 sample stars with spectral type from B6V to B9V to have a
$\logg$ below 3.5.

Depending on the input physics, stellar evolution theory shows that
roughly 10 -- 20\% of the B8-B9 stars are in the giant phase, but only 2 to
5\% of B0-B3 stars (e.g.,\,\citealt{Prialnik2000, Scuflaire2008}).
>From the 40 stars without luminosity class treated by us, 11 turn out to be
beyond
the main sequence. From the 14 dwarfs, 3 turn out to be giants as well, and from
the 10 giants, one turned out to be a dwarf instead. Thus our
sample contains 23 giants, 19 of which have spectral type B8-B9. As our
sample contains in total 51 stars of spectral type B8-B9, this means
that the percentage of late type giants in our sample (37\%) is about twice
the expected one. Only five stars in our sample have spectral types earlier
than B3, one of which is a giant, which is too low number statistics to verify
the expected percentage.
We found 20 of the 40 stars without luminosity class to have $\vsini$ above 90
km\,s$^{-1}$, and for 8 of them even $\vsini > 150\,$ km\,s$^{-1}$. This fast
rotation may again explain why we observe twice the number of expected giants
when comparing with standard models. Rotation was also found to be a necessary
ingredient in evolutionary models in order to explain the number of observed
giants in the well-studied open clusters h and $\chi$Per
(\citealt{Vrancken2000}, in particular their Fig. 3).

\begin{figure}[t!]
\centering
\begin{minipage}{3.4in}
\resizebox{\hsize}{!}{\includegraphics{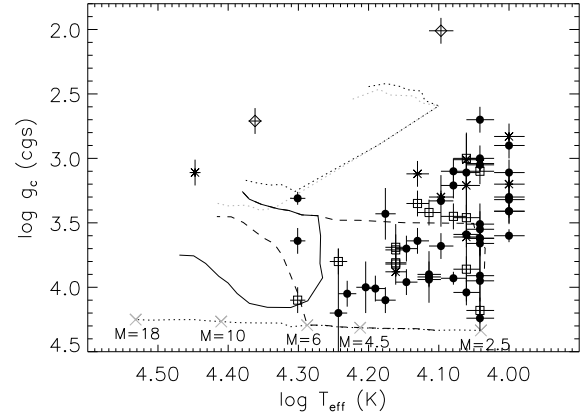}}
\end{minipage}
\caption{\label{HRD_GAUDI} ($\Teff$, $\loggc$) diagram of the analyzed GAUDI B
stars, with the corresponding 2$\sigma$-error bars. Symbols are the same as in
Fig.\,\ref{TEFF_SPT}. The dotted line represents the ZAMS, and 
five initial ZAMS masses - in
$M_{\odot}$ - have been indicated. The most recent theoretical
instability domains for the $\beta$ Cephei (thick solid line) and the SPB stars
(dashed lines) for main-sequence models, with a core overshooting value of 0.2
times the local pressure scale height, are shown
\citep{Miglio2007a,Miglio2007b}, together with the instability domains for
post-TAMS models with $\ell$ = 1 (grey dotted) and $\ell$ = 2 (black dotted)
g-modes computed by \citet{Saio2006}. The TAMS is shown as the
low-gravity edges of the SPB and $\beta$ Cephei instability domains around
$\loggc = 3.5$.}
\end{figure}

\subsection{The macroturbulent and rotational velocities}

When comparing the rotational broadening with macroturbulence, we observe
a systematic trend of increasing $\vmacro$ with increasing $\vsini$ for those stars for
which some pulsational behavior may be expected, i.e., for the stars where
$\vmacro$ cannot be neglected (filled circles in Fig.\,\ref{vsini_vmacro}).
This is consistent with \citet[ and references therein]{Markova2008}, who found
that, in almost all cases, the size of the macroturbulent velocity was similar
to the size of the rotational velocity. On the other hand, they also found
$\vsini$ and $\vmacro$ to decrease towards later subtypes, being about a factor
of two lower at B9 than at B0.5, and that, in none of their sample stars, rotation
alone was able to reproduce the observed line profiles. We cannot confirm 
either of both statements, as we have found, on the contrary, many cases with zero
macroturbulence, and many late B type stars with very high macroturbulent
velocities. The trend connecting $\vsini$ and $\vmacro$ suggests that both
broadening mechanisms \textit{are} difficult to disentangle. All
`$\vmacro$'-values seem to be
at least one third of $\vsini$, as can be derived from the position of the
filled circles above the dashed line in Fig.\,\ref{vsini_vmacro}. From a certain
projected rotational velocity on (i.e., around $\vsini$ = 120 km\,s$^{-1}$, dotted
vertical line), it is impossible to separate both effects, as the pulsational
behavior (represented by $\vmacro$) completely disappears in the projected
rotational broadening. Following the earlier argumentation that stars which have
a non-negligible macroturbulent velocity may have the largest pulsational
amplitudes, we investigated the behavior of $\vmacro$ in the ($\Teff$,
$\logg$)-plane (see Fig.\,\ref{vmacro_teff_logg}). On a global scale, the
$\vmacro$-values show a random spread.  The position of some individual cases is
intriguing, e.g., the
supergiant HD~52382 (log \Teff\ $\approx$ 4.36, \logg\ $\approx$2.71).  The
star's position is compatible with the occurrence of
gravity modes excited by the opacity mechanism in evolved stars and may belong
to a recently found class of pulsating B-type supergiants
\citep{Lefever2007}. As pointed out by \citet{Aerts2009}, these gravity-mode
oscillations may result in pulsational line broadening, hence the high
$\vmacro$-value we find.  HD~48807 (B7\,Iab) and HD~48434 (B0\,III) also show
macroturbulence and may belong to the same class.  Some of the dwarfs in the
sample show macroturbulence, e.g., HD~48215 (B5\,V), HD~177880 (B5\,V) and
HD~50251 (B8\,V) might be g-mode pulsators in the SPB instability domain.  Many
stars with $\logg$ below 3.5 show a reasonable to large $\vmacro$, which may
imply them to be pulsators. It is noteworthy that all instability computations
for mid to late B stars so far have been computed for non-rotating models and
were stopped artificially at the TAMS from the argument that no star is expected
to be found in the Hertzsprung gap. B8 to B9 stars, however, cross this gap at a
much lower pace than stars with spectral type earlier than B7 and our
observational spectroscopic results point out that it would be worthwhile to
extend the instability strip computations past the TAMS.

Finally, the magnitude of the microturbulence seems to be randomly distributed
in the ($\Teff$, $\logg$)-diagram, and in no way related to the effective
temperature nor the surface gravity (see Fig.\,\ref{xi_teff_logg}).

\begin{figure}[t!]
\centering
\begin{minipage}{3.4in}
\resizebox{\hsize}{!}{\includegraphics{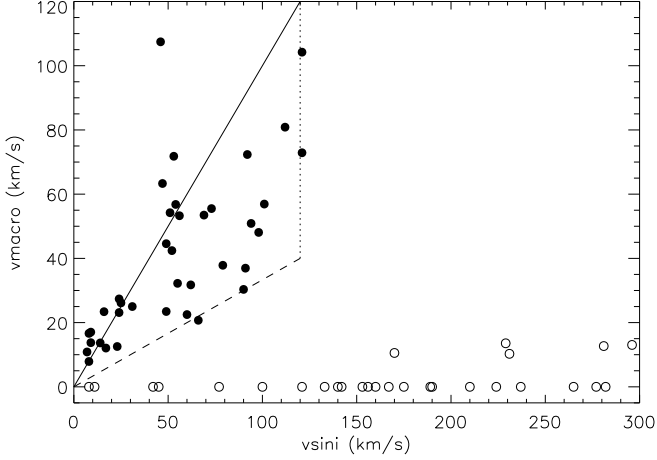}}
\end{minipage}
\caption{\label{vsini_vmacro} Comparison between the macroturbulence and the
projected rotational velocity $\vsini$. The solid line is the 1-1 relation,
while the dashed line represents the relation $\vmacro/ \vsini = 1/3$. Those
stars which we consider as non-pulsating (i.e., which have a negligible \vmacro)
are indicated by open circles. The filled circles represent suspected
pulsators (i.e., they have a significant \vmacro).}
\end{figure}

\begin{figure}[t!]
\centering
\begin{minipage}{3.4in}
\resizebox{\hsize}{!}{\includegraphics{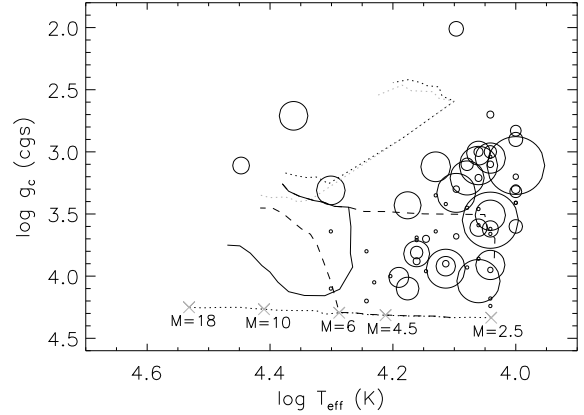}}
\end{minipage}
\caption{\label{vmacro_teff_logg} Same as Fig.\,\protect\ref{HRD_GAUDI}, but
representing the values of $\vmacro$ for the GAUDI B star sample. The size of
the symbols is proportional to the $\vmacro$-value.}
\end{figure}

\begin{figure}[t!]
\centering
\begin{minipage}{3.4in}
\resizebox{\hsize}{!}{\includegraphics{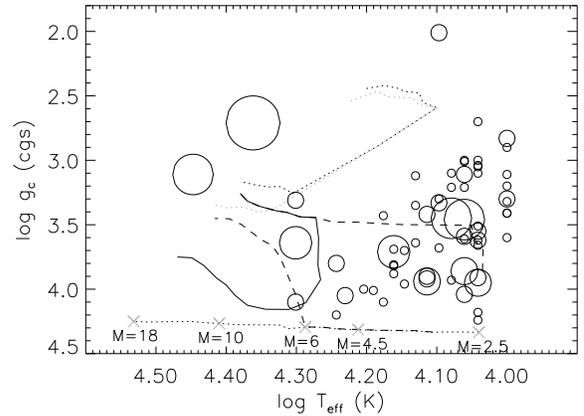}}
\end{minipage}
\caption{\label{xi_teff_logg} Same as Fig.\,\protect\ref{vmacro_teff_logg}, but
the size of the symbols is now proportional to the microturbulent velocity $\xi$
of the GAUDI B star sample.}
\end{figure}

\section{Future use of our results \label{summary}}

The GAUDI database was set up during the preparation of the CoRoT space mission.
We applied our developed automated tool for the determination of the
fundamental parameters to numerous B stars in this database.  
A recent
independent study used the same database to determine the abundances of the
89 B6-B9 stars in GAUDI, using a different approach based on LTE atmosphere
models without wind and LTE occupation numbers. 
We refer to \citet{Niemczura2009} for the underlying
assumptions, but mention here that a mild depletion of both Fe and Si,
with average values of the iron and silicon abundances of 7.13$\pm$0.29 dex and
7.22$\pm$0.31 dex, respectively, was found for their sample. This result was
obtained from spectrum synthesis in the LTE approximation, 
by fixing the effective temperatures from photometric calibrations, the
microturbulence at 2\,km\,s$^{-1}$ and ignoring any macroturbulence in the line
profile fits.
The mild Si depletion is in agreement with our results for the
Si abundance of the late B stars treated by us.

Meanwhile the first data of CoRoT were released to the public.  Two of our
sample stars were already observed by the satellite.  It concerns 
HD~181440 (B9~III) and HD~182198 (B9~V). The spectra of these
stars were available in the GAUDI database and their parameters are
listed in Table\,\ref{table_parameters}. Their seismic analysis is under way
(Miglio et al., in preparation).

We also point out that one of the primary targets of CoRoT is HD~180642
(B1.5~II-III), which is a mildly N-enriched, 
apparently slowly rotating $\beta$ Cephei
star, discovered by \citet{Aerts2000b}.  \citet{Morel2007a} determined its
fundamental parameters from high-resolution spectroscopy and derived $\Teff$ =
24\,500 $\pm$ 1\,000~K and logg = 3.45 $\pm$ 0.15, $\xi$ = 12 $\pm$
3~km\,s$^{-1}$. By means of AnalyseBstar, we found the model with $\Teff$ =
24\,000 $\pm$ 1\,000~K, logg = 3.40 $\pm$ 0.10 and $\xi$ = 13 $\pm$
2~km\,s$^{-1}$ to be the best fitting one, so our results agree very well with
theirs.  We thus classify this star as a
pulsator very close to the TAMS. It will be very
interesting to know if this spectroscopic result will be confirmed by the
seismic modeling of 
oscillation frequencies derived from the CoRoT space photometry, because
this would imply one of the few $\beta\,$Cephei stars very close to the end
of the core-hydrogen burning stage.

At the time of launch, the CoRoT space mission was nominally approved to operate
through 2009, but, given its excellent performance, the operations will probably
be prolonged. It is therefore to be expected that several more B-type stars in
GAUDI will be observed by the mission in the near future.

\begin{acknowledgements}

The research leading to these results has received funding from the European
Research Council under the European Community's Seventh Framework Programme
(FP7/2007--2013)/ERC grant agreement n$^\circ$227224 (PROSPERITY), as well
as from the Research Council of K.U. Leuven grant agreement GOA/2008/04. The
authors wish to thank Alex de Koter cordially for insightful discussions and
helpful comments throughout the project. The authors would like to
thank the anonymous referee for valuable comments and suggestions on how to
improve this paper.
\end{acknowledgements}

\bibliographystyle{aa}
\normalsize
\bibliography{ref}

\Online

\appendix
\twocolumn
\section{Formal convergence tests for synthetic FASTWIND
spectra -- some details\label{details_convergencetests}}

Before applying AnalyseBstar to real, observed spectra, we first tested whether
the method is able to recover the input parameters of synthetic spectra.
For this purpose, we have created several synthetic datasets in various regions
of parameter space. We will not dwell on discussing all of them here, but we
have chosen three specific examples, each representative for a different type of
star:\\
- Dataset A: a B0.5 I star with a rather dense stellar wind and a strongly
enhanced helium abundance,\\
- Dataset B: a B3 III star with a weak stellar wind and a depleted Si
abundance\\
- Dataset C: a B8 V with a very thin stellar wind.\\

Each dataset has been convolved with both a rotational and a
macroturbulent broadening profile. The projected rotational velocity adopted is
50~km\,s$^{-1}$ for each dataset, characteristic for a `typical' slow rotator. 
In the case of low and intermediate resolution spectra, there is also the
possibility to carry out an additional Gaussian, instrumental convolution. This
is, however, not necessary for spectra with such high resolution as FEROS and
ELODIE.
Artificial (normally distributed) noise was added to
mimic a real spectrum with a mean local SNR of 150, which is more or less the
minimal local SNR obtained for the GAUDI sample.

Finally, we have gone through the full process of the preparation of the spectra
as if it were real data, i.e., defining the EW of all available lines,
measuring the SNR to account for the errors on the EW and fixing the
projected rotational velocity. Since we are dealing with synthetic data,
obviously no normalization was required.\newline
Table\,\ref{formaltest_table1} lists the input parameters for the three
synthetic datasets as well as the output parameters, obtained from the
application of AnalyseBstar. 
For each dataset, the second column gives the derived `interpolated'
values, while the third column lists the parameters of the closest (best
fitting) grid model. In the ideal case, both should be exactly the same as the
input parameters.
This depends, however, on how well the equivalent widths were
measured and minor deviations occur as expected.\newline
Besides these three cases, in which we started from a synthetic model which is
one of the grid points, we have additionally created three synthetic datasets
for which the parameters lie in between different grid points. 
In analogy to datasets A to C, their profiles have been convolved with both
a rotational and macroturbulent broadening profile and artificial noise was
added.
The input parameters and the results of the analysis have been added to
Table\,\ref{formaltest_table1} as datasets D, E and F.\newline 
A third set of models (G, H, I) consists of models computed for parameters not
included in the grid, but within the grid limits, and this time without noise,
with the aim to test the predictive power of our method, independent of the
source of errors introduced by the noise. The rotational velocity applied for
these test models is 30 km\,s$^{-1}$, and the macroturbulent velocity is 15 km\,s$^{-1}$. We
find an overall very good agreement of the closest grid model and the
interpolated values with the input parameters. 

Although the models given in Table\,\ref{formaltest_table1} were selected as the best fitting
models, the program additionally came up with some other models, which
agree with the input model within the errors, introduced by the artificial noise
and the errors in the determination of the equivalent widths (datasets A to
F) and/or the various interpolations (datasets D to I). For instance, 
for dataset
I, AnalyseBstar came up with a second possibility with slightly different
parameters, even closer to the input data: $\logg$ = 1.7, $\xi$ = 6.5 km\,s$^{-1}$,
$\vmacro$ = 14.6 km\,s$^{-1}$, $N(He)/N(H)$ = 0.12, but with $\logsi$ = -4.49 (the other
parameters are the same as the ones for the model in the table). As the Si
abundance is overestimated compared to the input value, the Si lines are overall
a bit too strong and the final fit is slightly worse than the one for the model
in the table.

In all nine cases, the input parameters are well recovered.
Typically 10 to 150 models are selected and considered 
during the analysis cycle. The number depends not only on the choice of the
initial parameters, but also on the accuracy of the measured quantities and the
applied procedure (hence, the temperature domain, see below).
We considered numerous other test cases, which are not included in
this text, but for which the input parameters were equally well recovered.
Slight deviations from the input parameters are as expected. Also in the
cool temperature domain (datasets C, F and I), where our alternative
optimization `method 2' is required (see Section\,\ref{method_1_2}), we are
still able to deduce reliable parameter estimates, at least for the synthetic
models. 
The results for the Si abundance in datasets F clearly show the problem
for estimating the abundances when only one ion is present: the estimation of
the Si abundance was, for this particular test case, not equal to the input
value. When we have only Si~\,{\sc ii} lines, the results cannot be as
secure as when we have two ionisation stages of Si. This problem has been
accounted for by adopting somewhat larger errors on the derived parameters, for
objects with $\Teff < 15\,000~K$ (see Table\,\ref{table_parameters}).
\newline
When the observed object is close to one of the borders of the grid, the method
takes `the border' as the closest match. This will, consequently, lead to
larger
errors in some of the other parameters.

The derived $\vsini$-value, which is difficult to disentangle from the
macroturbulent velocity, agrees very well with the inserted values, irrespective
of \vmacro.  This shows that the Fourier Transform method 
indeed allows to separate both effects.
All together, this gives us confidence that our method is working reliably 
and that the procedure will also be able to recover the true physical parameters
from real spectra.

The macroturbulent velocity is the only fit parameter for which significant
deviations arise, if \vmacro\ is low ($<$ 30 km\,s$^{-1}$). Larger macroturbulences are
well recovered, however. Especially for datasets C, E and I, where \vmacro\,is
only 10-15~km\,s$^{-1}$, the deviation is large (of the order of 15 to 30~km\,s$^{-1}$). 
To understand this discrepancy, we first had a look at the line profiles, which
show that the fit is almost perfect for this \vmacro\ (see\,Fig\,\ref{lp1}).
After verifying the fit quality, we also verified that the discrepancies do not
arise from our applied procedure, by performing several tests on simulated
spectra, in which we left \vmacro\,as the only free parameter. In all cases, the
macroturbulent velocity was recovered very well, with deviations within
5~km\,s$^{-1}$. 
The (sometimes significant) deviations in $\vmacro$ can be understood
as a compensation for differences between the interpolated values of
the `observed' line profiles and the closest grid model. Indeed, as shown by 
\citet{Aerts2009}, small deviations in the wings 
of the Si profiles can result in wrong estimates of 
\vmacro\ and \vsini, because these two velocity fields are hard to 
disentangle whenever one of them is small.
Thus, the values derived for $\vmacro$ should be treated with caution.
This does not affect the derivation of the other parameters,
however, since a convolution with the \vmacro-profile preserves the EW of the
lines. It is the EW, and not the line profile shape, which is used for the
derivation of $\Teff$, $\xi$ and $\logsi$. 
The other parameters (gravity, mass-loss rate, and wind velocity law) remain
unaffected because they are deduced from the Balmer lines, which are not
very sensitive to the \vmacro\ values due to the dominance of the  
Stark broadening.
Thus we find a very good agreement between their input and output
values (Table\,\ref{formaltest_table1}).

\begin{table*}[ht!]
{
%\tiny
\caption{\label{formaltest_table1}
Input parameters  for the synthetic models (IN) are compared to the actual
output parameters (OUT) obtained through the application of AnalyseBstar to
these synthetic data: the effective temperature, $\Teff$, the surface gravity,
$\logg$, the wind strength parameters, $\logQ$ (see below), the wind velocity
exponent, $\beta$, the He~abundance, n(He)/n(H), the Si~abundance, $\logsi$, the
microturbulent velocity, $\xi$, the projected rotational velocity, $\vsini$, the
macroturbulent velocity, $\vmacro$, the time elapsed in seconds during the run
of AnalyseBstar, and the number of different models that have been checked
throughout the procedure. The wind strength parameter $\logQ$ is given by
$\rm log\,\dot{M}/ (v_{\infty}\, R_{\ast})^{1.5}$, where $\dot{M}$ is the
mass-loss rate, $v_{\infty}$ the terminal wind velocity, and $R_{\ast}$ the
stellar radius.
In our grid, we allow for 7 different values, represented by a character:
$\logQ$ =
-14.30 (O), -14.00 (a), -13.80 (A), -13.60 (b), -13.40 (B), -13.15 (C), -12.70
(D).\newline
Datasets A, B, and C are models which lie exactly on a grid point, whereas
dataset D, E, and F are models which lie in between the
gridpoints. Artificial noise was added to the synthetic line profiles in
these two datasets, to simulate a real spectrum. The models for datasets G, H,
and I also lie in between the gridpoints, but no artificial noise was added  
and a lower $\vsini$ (30 \kms\ instead of 50 \kms) was used 
in order to test the predictive capabilities of the method.
The `interpolated' values derived for the parameters, as well as
the best fitting (i.e.\ closest) grid model are given. The parameters in
italics are those for which no real interpolation was made, but for which the
most representative grid value is chosen. The longer computation times for the
cooler models are due to the different method applied in this range.}
\centering
\begin{tabular}{c|c|c|c}
\hline 
\begin{tabular}{r}
\\
\\
\\
\\
\hline
fit parameter\\
\hline
\\
\Teff\,(K)\\
\logg\,(cgs)\\
${\rm log}\, Q$ (char)\\
$\beta$\\
n(He)/n(H)\\
$\logsi$\\
$\xi$ (km\,s$^{-1}$)\\
$\vsini$ (km\,s$^{-1}$)\\
$\vmacro$ (km\,s$^{-1}$)\\
time (s)\\
checked models\\
\hline
\\
\hline
\\
\Teff\,(K)\\
\logg\,(cgs)\\
${\rm log}\, Q$ (char)\\
$\beta$\\
n(He)/n(H)\\
$\logsi$\\
$\xi$ (km\,s$^{-1}$)\\
$\vsini$ (km\,s$^{-1}$)\\
$\vmacro$ (km\,s$^{-1}$)\\
time (s)\\
checked models\\
\hline
\\
\hline
\\
\Teff\,(K)\\
\logg\,(cgs)\\
${\rm log}\, Q$ (char)\\
$\beta$\\
n(He)/n(H)\\
$\logsi$\\
$\xi$ (km\,s$^{-1}$)\\
$\vsini$ (km\,s$^{-1}$)\\
$\vmacro$ (km\,s$^{-1}$)\\
time (s)\\
checked models\\
\hline
\end{tabular} &
\begin{tabular}{rrr}
\\
IN & OUT & OUT \\
 & inter- & grid\\
 & polated & \\
\hline
& Dataset A & \\
\hline
& \\
 23,000  & 22,600   & 23,000\\
 2.7   &  \textit{2.7}  & 2.7   \\
 C     &  \textit{C}    & C     \\
 2.0   &  \textit{2.0}  & 2.0     \\
 0.20  &  0.18  & 0.20  \\
 -4.49 &  -4.49 & -4.49 \\
 10    &  10.2  & 10   \\
 50    &  48 $\pm$ 2  & 48 $\pm$ 2  \\
 20    &  29 $\pm$ 4 & 29 $\pm$ 4  \\
 142   & &\\
 10    & &\\
\hline
 & Dataset  D &  \\
\hline 
&&\\
 21,120  & 20,985  & 21,000\\
 3.98   & \textit{4.0}   & 4.0  \\
 b     &  \textit{b}     & b    \\
 1.42   & \textit{2.0}   & 1.2  \\
 0.14  & 0.11   & 0.10  \\
 -4.85 & -4.80  & -4.79 \\
 12  & 15.7   & 15  \\
 50    & 52 $\pm$ 6  &  52 $\pm$ 6\\
 40  & 33 $\pm$ 5   & 33 $\pm$ 5  \\
 174   & &\\
 10    & &\\
\hline
 & Dataset  G &  \\
\hline 
&&\\
 24,800  & 24,650 & 25,000 \\
 3.33  & \textit{3.3} & 3.3  \\
 B     & \textit{B} & B  \\
 1.1  & \textit{1.2} & 1.2  \\
 0.11  & 0.12 & 0.10  \\
 -4.52 & -4.48 & -4.49 \\
 13    & 13.5 & 12  \\
 30    & 30 $\pm$ 1 & 30 $\pm$ 1 \\
 15    & 12 $\pm$ 4 &  12 $\pm$ 4 \\
 2119  & &\\
 66    & &\\
\hline
\end{tabular} &
\begin{tabular}{rrr}
\\
IN & OUT & OUT \\
 & inter- & grid\\
 & polated & \\
\hline
& Dataset B & \\
\hline
& \\
 18,000&  18,100 & 18,000 \\
 3.3   &  \textit{3.3}  & 3.3  \\
 A     &  \textit{b}    & b     \\
 1.5   &  \textit{1.2}  & 1.2   \\
 0.10  &  0.08  & 0.10  \\
 -4.79 & -4.81 & -4.79 \\
 15  &  15  & 15 \\
 50    & 48 $\pm$ 4  &  48 $\pm$ 4  \\
 30  &  33 $\pm$ 2  & 33 $\pm$ 2 \\
 494   & &\\
 21    & &\\
\hline
 & Dataset  E &  \\
\hline 
&&\\
 15,100& 15,030 & 15,000 \\
 1.83  & \textit{1.80}   & 1.8  \\
 A     & \textit{A}      & A   \\
 2.8   & \textit{3.0}    & 3.0  \\
 0.08  & 0.09   & 0.10  \\
 -4.23 & -4.16  & -4.19 \\
 11  & 10.6   & 10 \\
 50    & 50 $\pm$ 7 &  50 $\pm$ 7 \\
 10    & 39 $\pm$ 7   & 39 $\pm$ 7 \\
 392   & &\\
 26    & &\\
\hline
 & Dataset  H &  \\
\hline 
&&\\
 18,600  & 18,392 & 18,500\\
 2.78  & \textit{2.8} & 2.8 \\
 O     & \textit{A} & A\\
 1.0   & \textit{0.9} & 0.9 \\
 0.14  & 0.13 & 0.15 \\
 -4.25 & -4.22 & -4.19\\
 4     & 3.3 & 3\\
 30    & 30 $\pm$ 1 & 30 $\pm$ 1\\
 15    & 20 $\pm$ 7 & 20 $\pm$ 7\\
 2284  & &\\
 67    & &\\
\hline
\end{tabular} &
\begin{tabular}{rrr}
\\
IN & OUT & OUT \\
 & inter- & grid\\
 & polated & \\
\hline
& Dataset C & \\
\hline
& & \\
 13,000  &  \textit{13,000} & 13,000 \\
 4.2   &  \textit{4.2}   & 4.2   \\
 O     &  \textit{a}     & a    \\
 0.9   &  \textit{0.9}   & 0.9   \\
 0.10  &  0.08  & 0.10  \\
 -4.79 &  \textit{-4.79} & -4.79 \\
 6.0   &  7.3   & 6.0 \\
 50    &  51 $\pm$ 1 &  51 $\pm$ 1  \\
 10    &  30 $\pm$ 1  & 30 $\pm$ 1 \\
 2344   & &\\
 128    & &\\
\hline
 & Dataset  F &  \\
\hline 
&&\\
 11,880 & \textit{11,500} & 11,500 \\
 2.43   & \textit{2.3}    & 2.3 \\
 a      & \textit{O}      & O \\
 1.02   & \textit{0.9}    & 0.9 \\
 0.18   & 0.20   & 0.20 \\
 -4.49  & \textit{-4.19}  & -4.19\\
 9      & 6.3   & 6 \\
 50     & 54 $\pm$ 1 &  54 $\pm$ 1 \\
 70     & 67 $\pm$ 7    & 67 $\pm$ 7 \\
 1254   & &\\
 95     & &\\
\hline
 & Dataset  I &  \\
\hline 
&&\\
 10,910 & \textit{11,000} & 11,000\\
 1.72   & \textit{1.8} & 1.8\\
 C      & \textit{C} & C\\
 2.10   & \textit{2.0} & 2.0\\
 0.11   & 0.09 & 0.10\\
 -4.67  & \textit{-4.79} & -4.79\\
 7      & 8.6 & 10\\
 30     & 30 $\pm$ 1 & 30 $\pm$ 1\\
 15     & 0 & 0 \\
 356    & &\\
 15     & &\\
\hline
\end{tabular}
\end{tabular}
}
\end{table*}

\begin{figure*}[t!]
\centering
\begin{minipage}{\hsize}
\resizebox{\hsize}{2.8in}{\includegraphics{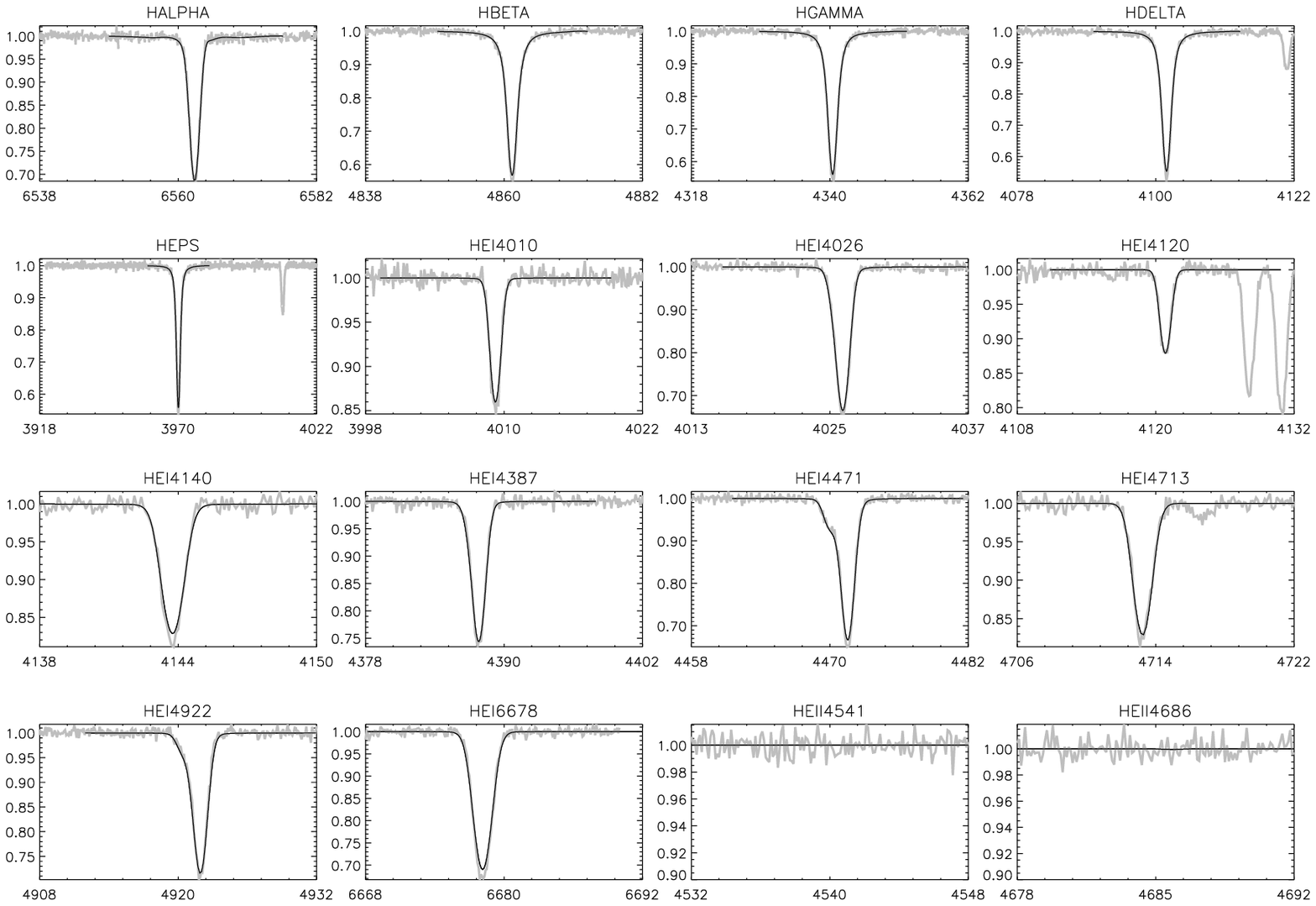}}
\resizebox{\hsize}{2.8in}{\includegraphics{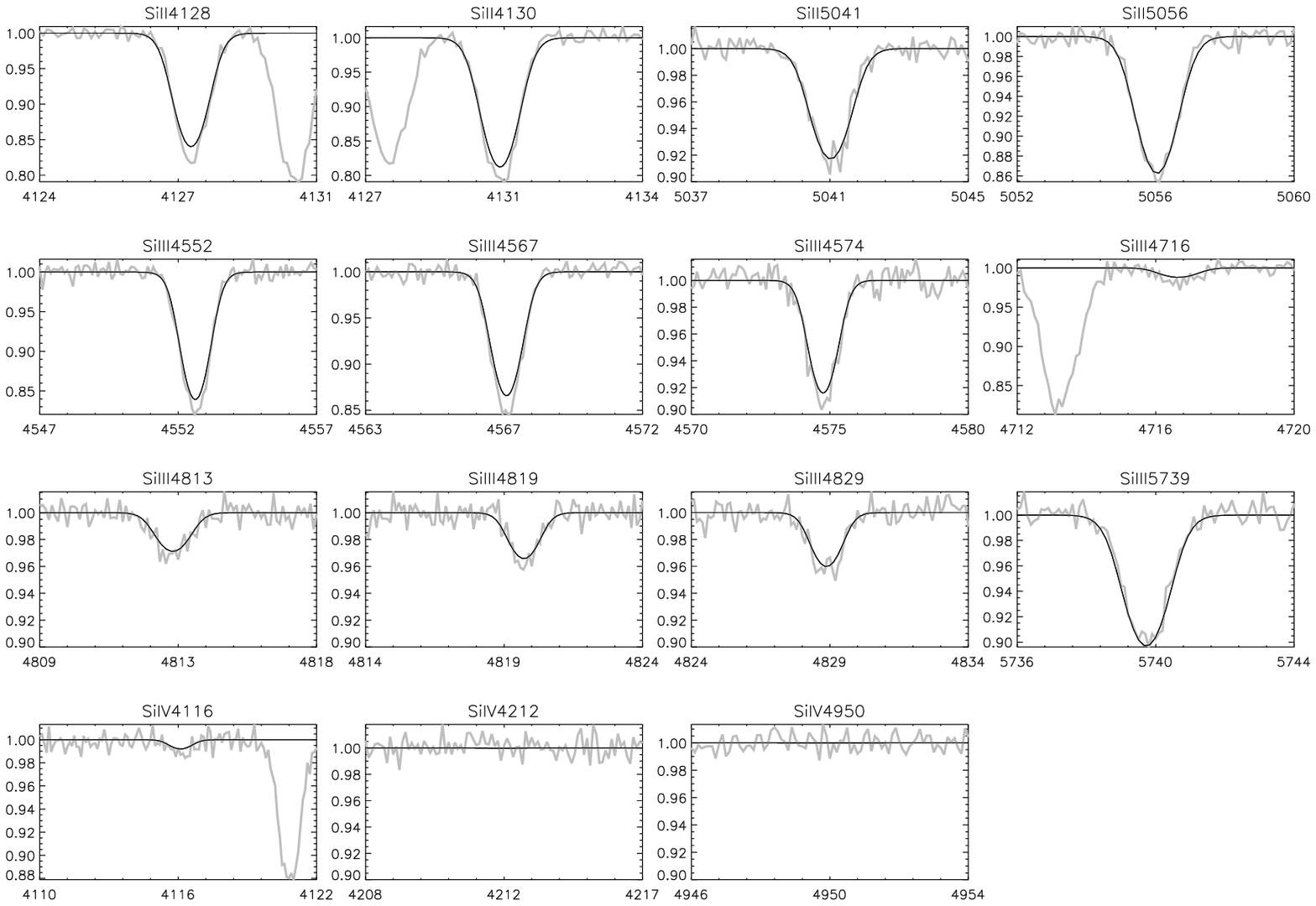}}
\end{minipage}
\caption{\label{lp1} Example of the fit quality for dataset E. The slight
discrepancy in the cores of the Si lines arises from the difference in Si
abundance between the interpolated value ($\logsi$ = -4.16) and the closest grid
value ($\logsi$ = -4.19)}
\end{figure*}

\section{Methodology of AnalyseBstar}

In what follows, we give a detailed description of AnalyseBstar.
In Fig.\,\ref{flowchart}, a flowchart of the full program is drawn
schematically. The reader is advised to follow this context diagram throughout
the further description of AnalyseBstar.

\subsection{Preparation of the input \label{prep}}

No automatic method can fully replace the `by-eye' procedure,
and this is also true for AnalyseBstar. A few steps still require human
intervention. This especially concerns the preparation of the spectra and the
input for the main program, as will be shown in the following.

\paragraph{Observation of peculiarities.}
The first thing to be done when starting the analysis is to inspect
the spectrum regarding any `abnormalities', such
as line behavior resulting from nebulae, disks, binarity or other
peculiarities. Special features can and should be detected through eye
inspection.

\paragraph{Merging and normalization of the spectra.\label{mergingprocess}}
When dealing with \'echelle spectra, like the ELODIE and FEROS spectra in the
GAUDI database, we first need to merge the spectral orders before the spectrum
can be normalized. This is a very delicate and non-trivial process. The edges of
the orders are often noisy, due to the decrease in sensitivity of the CCD near
its edges, and are therefore removed. Spectral lines falling inconveniently
on the edge of one spectral order or in the overlap region of two consecutive
orders, thus resulting in a cut-off within the spectral line, are not reliable
to derive accurate stellar information. They can, in the best case, only be used
for a consistency check afterwards.
After the spectra have been merged, they are normalized through interactive
interpolation of continuum regions. This is done over a large wavelength
coverage, where sufficient continuum regions are present, in order to avoid line
corruption which will propagate in the derivation of the fundamental parameters.
The normalization especially proves to be very important for the determination
of the gravity, which depends sensitively on the Balmer line wings.

\paragraph{Radial velocity correction.}
The spectra need to be corrected for radial velocity shifts. Shifting the
spectrum to rest wavelength automatically has not been included in the current
version yet.

\paragraph{Line selection.\label{lineselection}} 
As a standard procedure, we compute a number of line profiles in the optical
from the underlying model atmosphere.
These standard line profiles can be
used directly. Computing additional line profiles is, however, straightforward.
The selection of lines to be used during the spectral analysis in AnalyseBstar
is gathered from a file which contains a default (but easily editable) list
of currently used optical lines (see below).
We distinguish between optical lines that will be used for the prediction of the
physical parameters during the spectral line fitting procedure and those that
will not be used explicitly for reasons of uncertainties, either in the
theoretical predictions or in the observed spectrum (due to problems
in merging or normalization). The goodness-of-fit of the latter lines is only
checked to investigate systematic differences between theory and observations,
which can be useful input for considerations regarding further improvements of
atomic data and/or atmospheric models. They are H$\epsilon$, He\,{\sc i}~4010,
4120, 4140 and Si\,{\sc iv}~4950.
The lines that \textit{will} be used in the fitting procedure for their
predictive power are the following: the Balmer lines $\Ha$, $\Hb$,
H$\gamma$ and H$\delta$, the neutral and singly ionized He lines He\,{\sc
i}~4026, 4387, 4471, 4713, 4922, 6678, He\,{\sc ii}~4541 and He\,{\sc ii}~4686,
and the
Si lines in their three different ionization stages Si\,{\sc ii}~4128-4130,
Si\,{\sc ii}~5041-5056, Si\,{\sc iii}~4552-4567-4574, 4716, 4813-4819-4829,
5739, Si\,{\sc iv}~4116 and Si\,{\sc iv}~4212. He\,{\sc ii}~4200 was excluded as
it can only safely be used for O-type stars and not for B-type stars. At least
for dwarfs, it only appears at temperatures above about 25\,000~K, where it is
still overruled by N\,{\sc ii}, while at hotter temperatures, He\,{\sc ii}~4200
is for 50\% blended by N\,{\sc iii}\footnote{See
http://www.lsw.uni-heidelberg.de/cgi-bin/websynspec.cgi}.

\begin{figure*}[t!]
\centering
\begin{minipage}{4in}
\resizebox{4in}{!}{\includegraphics{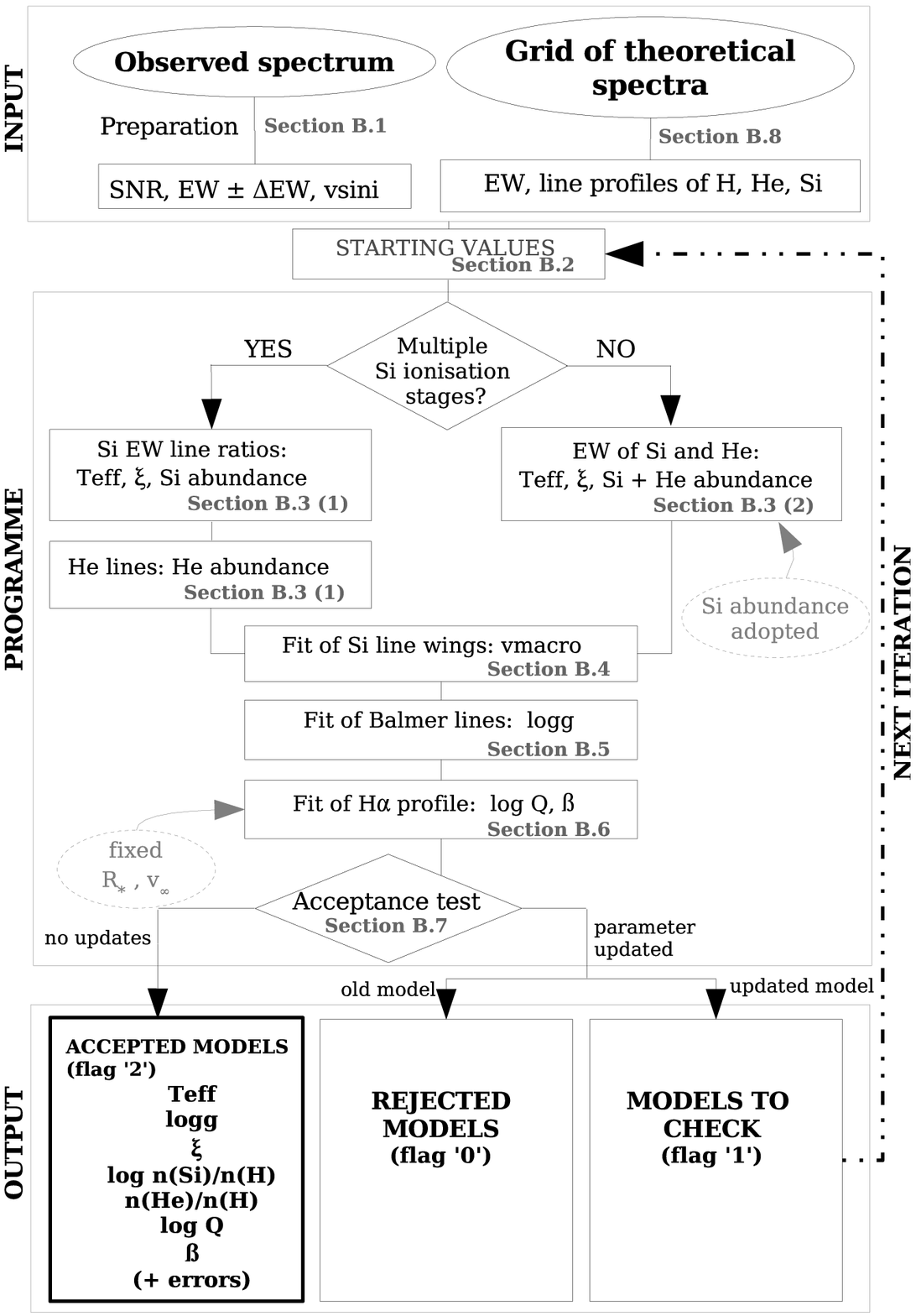}}
\end{minipage}
\caption{\label{flowchart}\label{page_flowchart} Context diagram of
AnalyseBstar. For a detailed explanation, see text.}
\end{figure*}

\paragraph{Signal-to-noise measurement.}
To estimate the error on the equivalent width as accurate as possible, we need
to determine the signal-to-noise ratio, SNR. Since the SNR can vary a lot with
wavelength, we chose to calculate the \textit{local} SNR for each line
separately, by using a continuum region close to the line. This is done manually
through the indication of the left and right edges of the continuum interval.
The SNR is then given by the mean flux $\overline{F_c}$ divided by the standard
deviation $\sigma(F_c)$ of the flux in this interval, i.e., $${\rm SNR} =
\overline{F_c} / \sigma(F_c).$$

\paragraph{Observed equivalent widths and their errors} (cf. Fig.\,\ref{EWdet}).
The equivalent widths of most He and Si lines can be measured by a Gaussian
non-linear least squares fit to the observed line profiles, using the
Levenberg-Marquardt algorithm \citep{Levenberg1944, Marquardt1963}. We
thoroughly tested whether this was a justifiable approach by comparing, for
multiple line profiles of stars with completely different properties, the
equivalent widths computed from a Gaussian fit to the observed line profiles
with those computed from a proper integration of the line pixel values. This
comparison resulted in a minimal difference between both approaches. Some lines
obviously cannot be fitted by a Gaussian, e.g., the He~lines with a strong
forbidden component in their blue wing or Stark-broadened lines such as those
from hydrogen.
The EW of such lines are determined by proper integration. In all other cases,
where the line profile \textit{is} well-represented by a Gaussian fit, the
the observed EW is given by the integral of the Gaussian profile.\newline
The main source of \textit{uncertainty in the EW determination} is
introduced by the noise and its influence on the exact position of the
continuum within the noise. To account for this we 
multiplied the continuum by a factor (1 $\pm$ 1/SNR),
performed a new Gaussian fit
through both shifted profiles and measured the difference in EW. Note that, by
considering a constant factor over the full line profiles (using the signal at
continuum level), we implicitly give an equal weight to each wavelength point,
even though, in reality, 1/SNR may be slightly larger in the cores, since it is
proportional to $\sqrt{1/S}$, with S the obtained signal at these wavelengths.
When the user deems it necessary, (s)he can apply a correction for the continuum
level by a local rerectification, e.g., in the case that the
normalization performed over a large wavelength range is locally a bit offset.
If the offset is different on the blue and red side of the line, then the factor
should be wavelength dependent. Therefore, we allow for a linear rerectification
by a factor $a\lambda + b$. Another source of uncertainty in the EW is
introduced by using a Gaussian fit to estimate the EW. Each of the above factors
have been included in the total error budget of the EW.

\begin{figure}[ht!]
\centering
\begin{minipage}{3.2in}
\resizebox{3.0in}{!}{\includegraphics{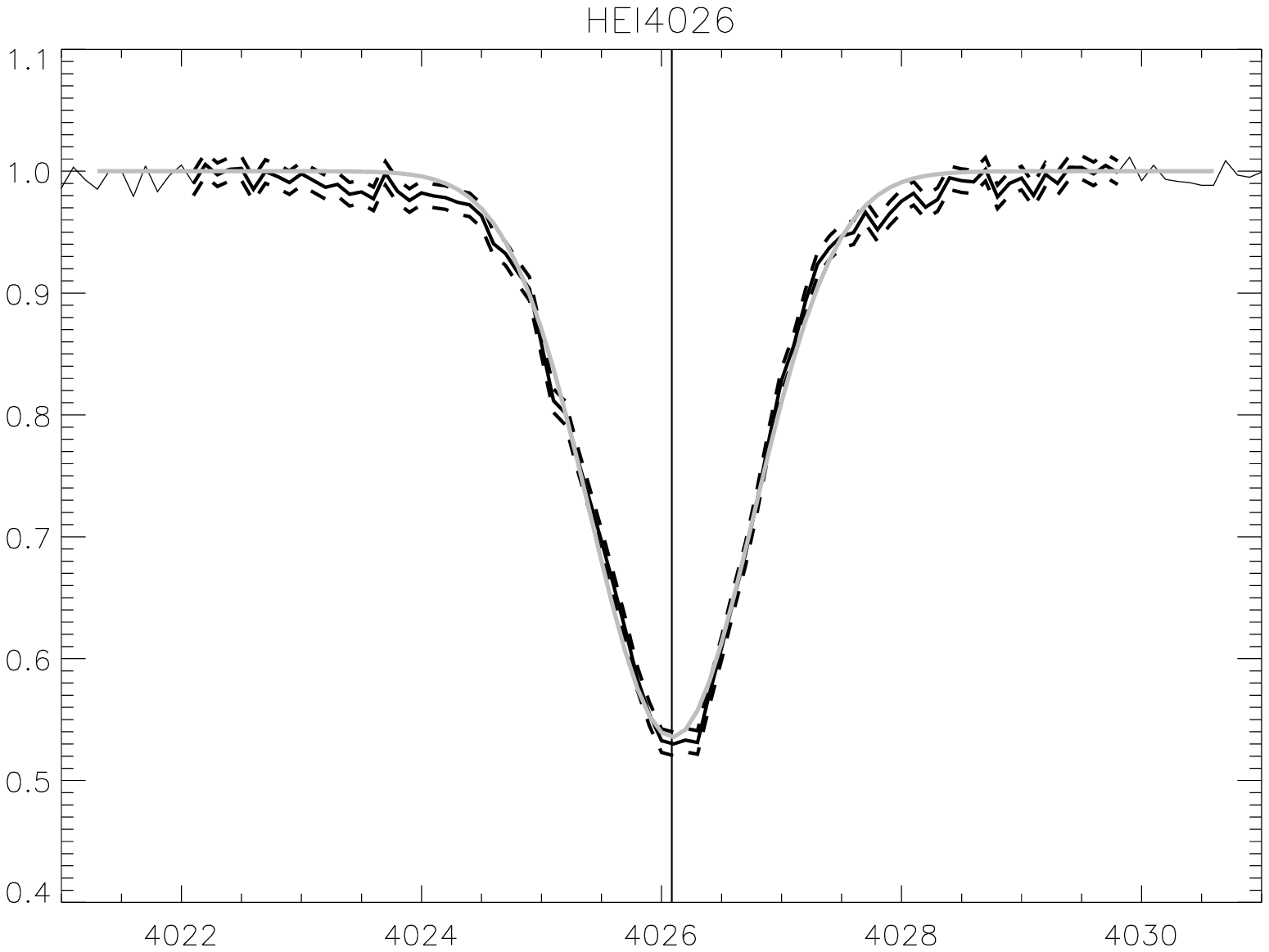}}
\end{minipage}
\caption{\label{EWdet} Illustration of the EW determination for He\,{\sc
i}~4026. After the manual identification of the wavelength interval of the
spectral line (thick black profile), a Gaussian fit to the observed line profile
is made to determine the EW of the line (thick grey profile). Also indicated
are: the center of the line (vertical line) and the `shift' in flux, upwards and
downwards (dashed line profiles), used to account for the noise level in the
determination of the EW. Also to these line profiles a Gaussian fit is made, but
these are omitted here for the sake of clarity.}
\end{figure}

\paragraph{Determination of $\vsini$}(see Fig.\,\ref{illustration_vsini}).
The determination of the projected rotational velocity is the last step 
in the preparation
of the input for AnalyseBstar. We use the semi-automatic tool developed by
\citet{SimonDiaz2007} to derive the projected rotational velocities (see
Fig.\,\ref{illustration_vsini}). We refer
to their paper for a thorough discussion.
We use the following, least blended, metallic lines for the determination
of $\vsini$: Si\,{\sc ii}~5041, Si\,{\sc iii}~4567, 4574, 4813, 4819, 4829,
5739, C\,{\sc ii}~4267, 6578, 6582, 5133, 5145, 5151 and O\,{\sc ii}~4452. 
We carefully checked visually that blended lines were not taken into account.
The projected rotational velocity \vsini\,of the star is calculated as
the mean of the values derived from each individual line, and its uncertainty
as the standard deviation.

\begin{figure}[t!]
\centering
\begin{minipage}{3.4in}
\resizebox{\hsize}{!}{\includegraphics{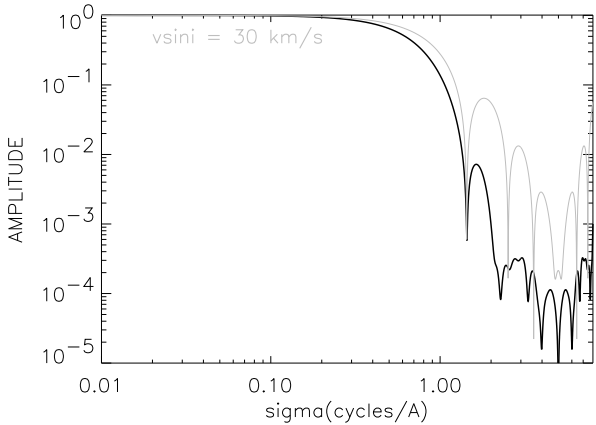}}
\end{minipage}
\caption{\label{illustration_vsini} Illustration of the determination of
\vsini\,from the first minimum in the Fourier transform of the selected line
profile, as implemented and described by \citet{SimonDiaz2007}. The user
can make several attempts and finally decide which \vsini\,gives the best match.
The difference in the slope of the first decay gives an indication of the
macroturbulence. The Fourier transform of the
observed line profile is indicated in black. The user indicates the first
minimum in the  black profile, which gives the projected rotational velocity.
The Fourier transform of the rotational profile at this $vsini$ is indicated in
grey.}
\end{figure}

\subsection{Fit parameters and their starting values}

Once all preparation is finished, the automatic procedure to obtain accurate
values for all physical fit parameters can be started. This procedure follows an
iterative scheme as outlined in Fig.\,\ref{flowchart}. In each iteration, the
fundamental parameters are improved in
the following order. First of all, the effects of the effective temperature, the
Si abundance and the microturbulence on the Si lines are separated. Then, the He
abundance is fixed using all available He lines. Next, the macroturbulent
velocity is determined from well-chosen, user supplied Si lines. Then, the
surface gravity is determined from the wings of $\Hg$, $\Hd$ and, in the case of
weak winds, also $\Hb$, after which the wind strength $\logQ$ ($Q$ = $\dot{M}/
(v_{\infty}\, R_{\ast})^{1.5}$, with $\dot{M}$, the mass loss, $v_{\infty}$, the
terminal wind velocity, and $R_{\ast}$, the stellar radius) and the wind
velocity exponent $\beta$ are determined in parallel using $\Ha$.
Obviously, to start the iterative procedure, we need an initial guess for each
of these parameters. This initial value can either
be user supplied or standard. If no value is set by the user, then the following
initial values are considered. The initial effective temperature will be
determined from the spectral type of the star. Immediately after the rotational
and macroturbulent velocities have been determined, a start value for the
surface gravity is derived by running the subprocedure for the
determination of the gravity (see Section\,\ref{logg_cycle}).
For the He and the Si abundance, we have taken the lowest value as start value.
We initialize the wind parameters at the lowest values, which means
negligible wind, at $\rm log\,Q$ = -14.30 and $\beta$ = 0.9. These are
reasonable assumptions for dwarfs, which comprised the largest part of our
sample. These starting values can easily be adapted when samples of stars with
different stellar properties are aimed at. The macroturbulent velocity is
initially zero, whereas for the microturbulent velocity we took a medium
value of all considered possibilities, i.e., 10 km\,s$^{-1}$.
The convergence speed depends on how far away the initial values are from the
final solution.

\begin{figure}
\centering
\begin{minipage}{3.4in}
\resizebox{3.0in}{!}{\includegraphics{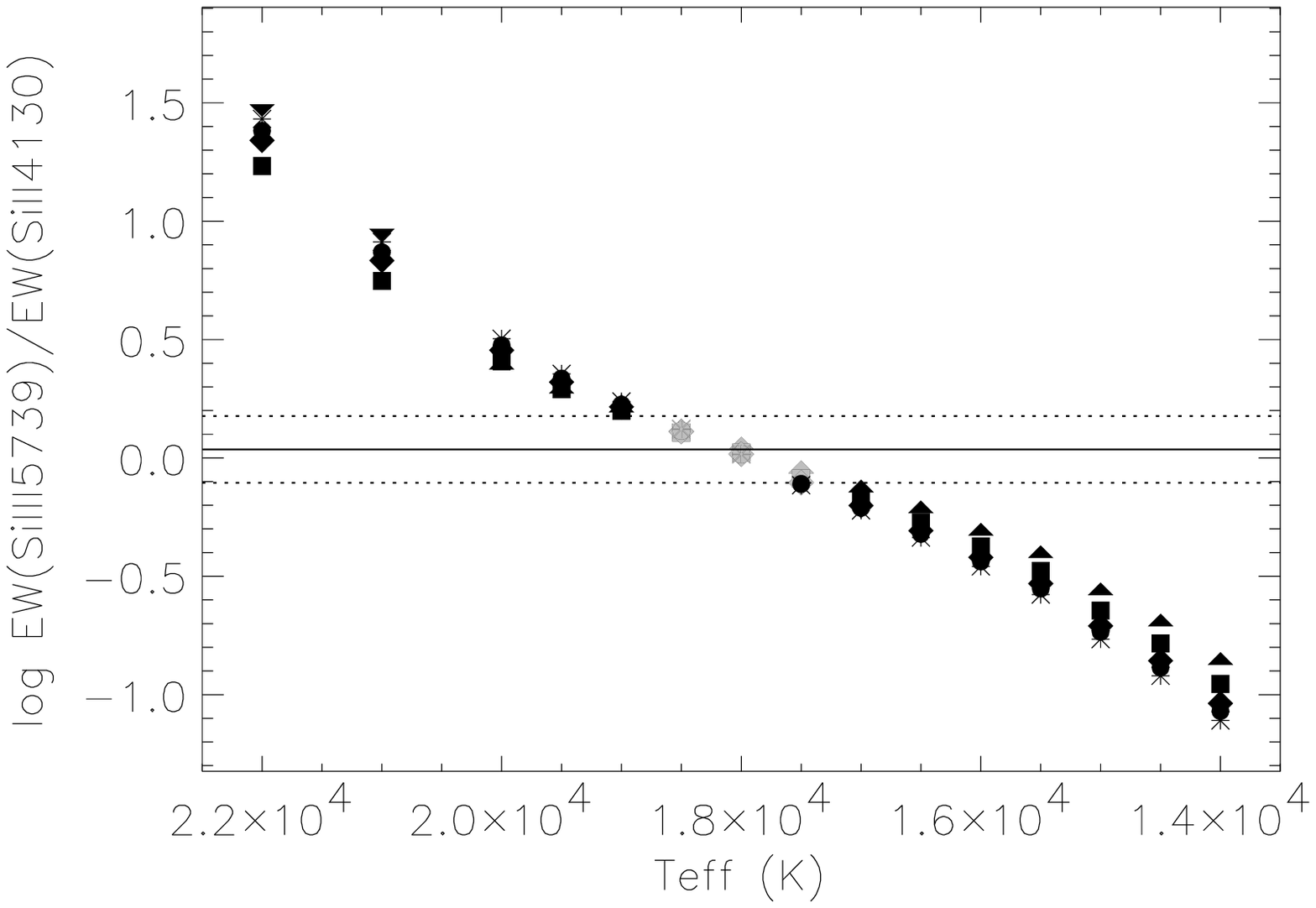}}
\end{minipage}
\begin{minipage}{3.4in}
\vspace{-0.5cm}
\resizebox{3.0in}{!}{\includegraphics{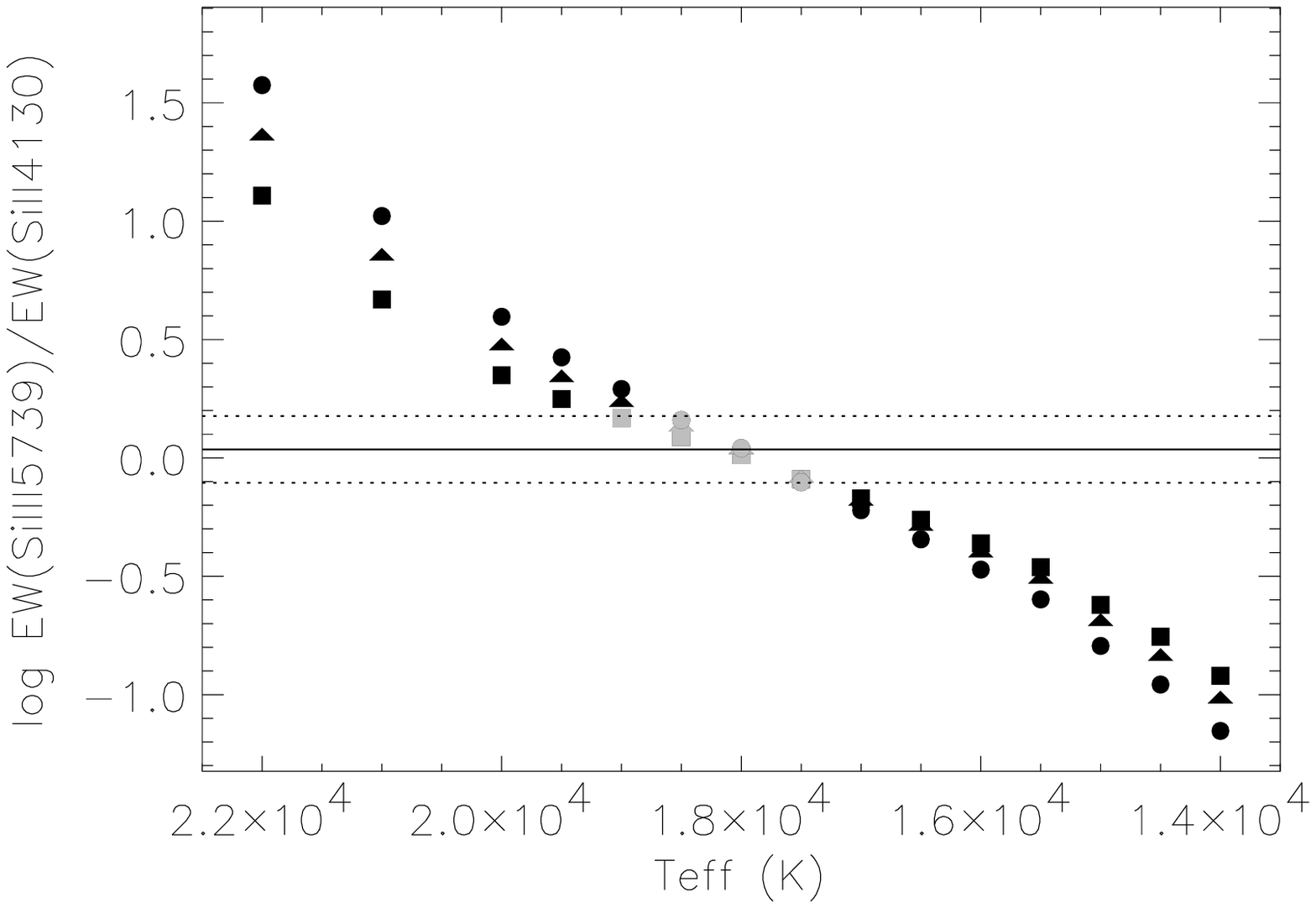}}
\end{minipage}

\caption{Synthetic simulation of the effect of the microturbulence (top) and the
Si abundance (bottom) on the logarithmic EW ratios of Si lines of different
ionization stages. This example gives the (synthetic) EW ratio of Si\,{\sc
iii}~5739 to Si\,{\sc ii}~4130 against a selected range of effective
temperatures, for a fixed surface gravity ($\logg$ = 2.5) and wind parameters
($\rm \Mdot$ = $0.144.10^{-7} \rm M_{\odot}$/yr, $\vinf$ = 500~km\,s$^{-1}$, $\beta$ =
0.9). The ratios are
evaluated for the different possibilities of the microturbulence, for a
Si abundance fixed at -4.79 (upper panel: $\xi$ = 3, 6, 10, 12, 15 and 20
km\,s$^{-1}$; triangle up, square, diamond, circle, asterisk, triangle down,
respectively) and for the different possibilities of the Si abundances, for
a microturbulence fixed at 10 km\,s$^{-1}$
($\logsi$ = -4.79, -4.49 and -4.19: circle, triangle up and
square, respectively). The horizontal lines show
the observed EW ratio and its corresponding uncertainty region. Acceptable EW
ratios, which fall within these boundaries, are indicated in grey.} 
\label{AnalyseBstar_teff}
\end{figure}

\subsection{Determination of the effective temperature, microturbulence and
abundances \label{method_1_2}}

The Si lines serve multiple purposes. By using a well-defined scheme (somewhat
similar to the conventional `curve of growth' method \citep{Gray1976} and
described, e.g., by \citealt{Urbaneja2004}), we are able to separate the
effects of the effective temperature, the Si abundance and the
microturbulence on the line profiles, and derive an accurate value for them.
Our method automatically determines from the observed spectrum which lines
will be used for this purpose. The lines should be well visible
(i.e.\,clearly distinguishable from the noise in the continuum) and the
relative error on their equivalent width is typically below 10 to 15\%,
depending on the combination of the temperature and the specific line
considered. Whenever the relative error on its equivalent width exceeds
75\%, the spectral line will be excluded in the following procedures. Such
large errors are a rare exception and only occur for the weakest lines,
which almost disappear into the noise level.\newline Basically, we can
discern two different cases, namely when multiple and consecutive ionization
stages of Si are available, or when there is only one ionization stage of
Si. Each needs a different approach, as we will explain now.

\paragraph{Method 1: In case there are multiple ionization stages of Si available,}
we can, to first order, put aside the effect of the Si abundance by
considering the EW ratios of two different Si ionization stages, as the Si
abundance affects all Si lines in the same direction.
For each effective temperature point in the grid, the ratios of the observed
equivalent widths of Si\,{\sc iv} to Si\,{\sc iii} and/or Si\,{\sc iii} to
Si\,{\sc ii} are compared to those obtained from the model grid, given a certain
\logg\,and wind parameters (see Fig.\,\ref{AnalyseBstar_teff}).
For each combination of Si\,{\sc iv}/{\sc iii} and/or Si\,{\sc iii}/{\sc ii}
lines, this results in a range of `acceptable' effective temperatures for which
the observed EW ratio is reproduced, within the observed errors (indicated as
grey symbols in Fig.\,\ref{AnalyseBstar_teff}). From the different line ratios, 
slightly different temperatures may arise. Under the assumption that
\logg\,and the wind parameters are perfectly known, the combination of all these
possibilities constitutes the set of acceptable effective temperatures.

Once we have a list of \textit{acceptable} temperatures, we can derive for
each temperature the best microturbulent velocity and the best Si abundance,
by analyzing the variation in equivalent width of different lines as a function
of these quantities.
In what follows, we describe the full process, step by step:

\begin{itemize}
\item \textit{Step 1: Deriving the abundances for each microturbulence, line
by line}

In this first step, the observed EW of each Si line is compared to the set of
theoretically predicted EWs corresponding to the available combinations of Si
abundance and microturbulence in the grid (see Fig.\,\ref{step1}).
For each microturbulent velocity (and for each Si line), we look for the range
of Si abundances that reproduce the observed EW, within the observed errors.
This is done through linear interpolation. Note that this gives us only an
approximative estimate of the Si abundance. Indeed, as can be seen in
Fig.\,\ref{step1}, the change in EW with increasing microturbulence is
marginal for weak lines, but increases for stronger spectral lines,
indicating that we may no longer be in the linear part of the curve of growth.

\item \textit{Step 2: Deriving the mean abundance for each microturbulence and
the slope of the best fit}

In a second step we compare, for each microturbulence, the abundances (with the
derived uncertainties) found in step 1 to the observed EW of each line (see
Fig.\,\ref{step2}). Since a star can only have one Si abundance, we should find
the same Si abundance from all the different Si lines, i.e. the slope of the
best fit to all the lines should be zero. From the different abundances derived
from each line, we can calculate the mean abundance for each microturbulence.

\item \textit{Step 3: Deriving the `interpolated' microturbulent velocity}

In a third step, we investigate the change of the slope when varying the 
microturbulence (see upper panel Fig.\,\ref{step3}). Through
linear interpolation, we find the microturbulence for which the slope would
be zero (i.e. for which the Si abundance derived from each line separately would
be the same). This is the estimated `interpolated' microturbulence.

\item \textit{Step 4: Deriving the `interpolated' abundance}

Using now the relation between the microturbulence and the mean abundance
derived in step 2, we interpolate to find the estimated `interpolated'
Si abundance at the estimated `interpolated' microturbulence found in step 3
(see lower panel Fig.\,\ref{step3}).

\item \textit{Step 5: Deriving the `interpolated' effective temperature}

Now that we obtained the estimated `interpolated' values for the
microturbulence and Si abundance, we can go back to
Fig.\,\ref{AnalyseBstar_teff} and interpolate in two dimensions to derive a
value for the effective temperature, which reproduces the observed EW.  The mean
of the effective temperatures from each line ratio will then be accepted as a
value for the `interpolated' effective temperature of the star.

\item \textit{Step 6: Determination of the `closest' grid values}

For the next steps in the automatic procedure, we will need the closest grid
values to these estimated `interpolated' values rather than the real values
themselves. These closest values will constitute a new entry in the list of
possibilities. If the real value for the Si abundance or microturbulence falls
exactly between two grid points, both are added to the list.
\end{itemize}
We repeat these steps for each `possible' grid temperature,
initially derived from Fig.\,\ref{AnalyseBstar_teff} (grey symbols), and
obtain in this way a set of new possible solutions that optimally reproduce the
Si lines.
Temporarily fixing the values for $\Teff$, $\xi$, $\logsi$ in the way
described above leads to a new estimate for the He abundance, denoted as
n(He)/n(H). As for the Si abundance, we only consider three different
values. We apply a similar method as for the determination of the Si
abundance, in the sense that we derive the best-suited abundance from each line
separately, and take the mean value as the `interpolated' value.
Figs\,\ref{he_abun1} and \ref{he_abun2} illustrate this process. Note that we
intrinsically assume that the microturbulence is the same throughout the
atmosphere, i.e. that there is no radial stratification. In this way, the
microturbulence, necessary to account
for the broadening in the He lines, will be assumed to be the same as the one
derived from the Si lines.
Note also that the uncertainty in the derived He abundance will be larger when
the He abundance is lower than 0.10. Indeed, in this case we are forced
to extrapolate to a region where it is unclear how the dependence of the
equivalent width with abundance will change. The decrease towards abundances
lower than solar may be steeper or follow a logarithmic trend, in which case
the predicted values would be underestimated. In this case we can only
state that the solar abundance is an upper limit of the true abundance.
Values lower than the primordial He abundance of $\sim$0.10 can no
longer be considered physical, except when diffusion effects start to play a
role and He settles. In that case, the He abundance observed at the
stellar surface could be lower. Such chemically peculiar stars are referred to
as He weak stars and are usually high-gravity objects. Helium settling was not
taken into account in our models.

\begin{figure*}
\centering
{\resizebox{5.5in}{!}{\includegraphics{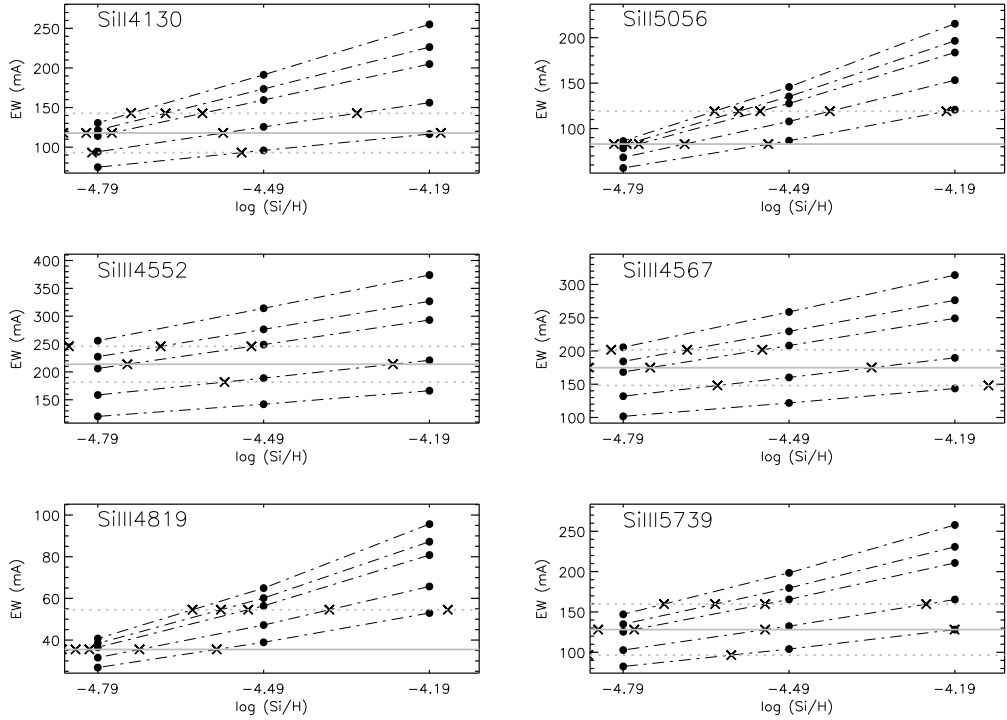}}}
\caption{\label{step1} Step 1 (for each \Teff\,and for each line): For each of
the five considered microturbulent velocities (5 dash-dotted lines, from low EW
to high EW: 3, 6, 10, 12, 15 km\,s$^{-1}$), we derive the Si abundance that
reproduces the observed EW (horizontal grey lines) of the spectral line through
linear interpolation. The Si abundances and the corresponding upper and lower
limit are represented by the crosses.
The displayed example is a synthetic simulation.
Only a selection of lines is shown. Usually many more lines can be used. In
this example, no Si\,{\sc iv} was ``observed'', so only Si\,{\sc ii} and
Si\,{\sc
iii} were considered.}
\end{figure*}
\begin{figure*}
\centering
{\resizebox{5.5in}{!}{\includegraphics{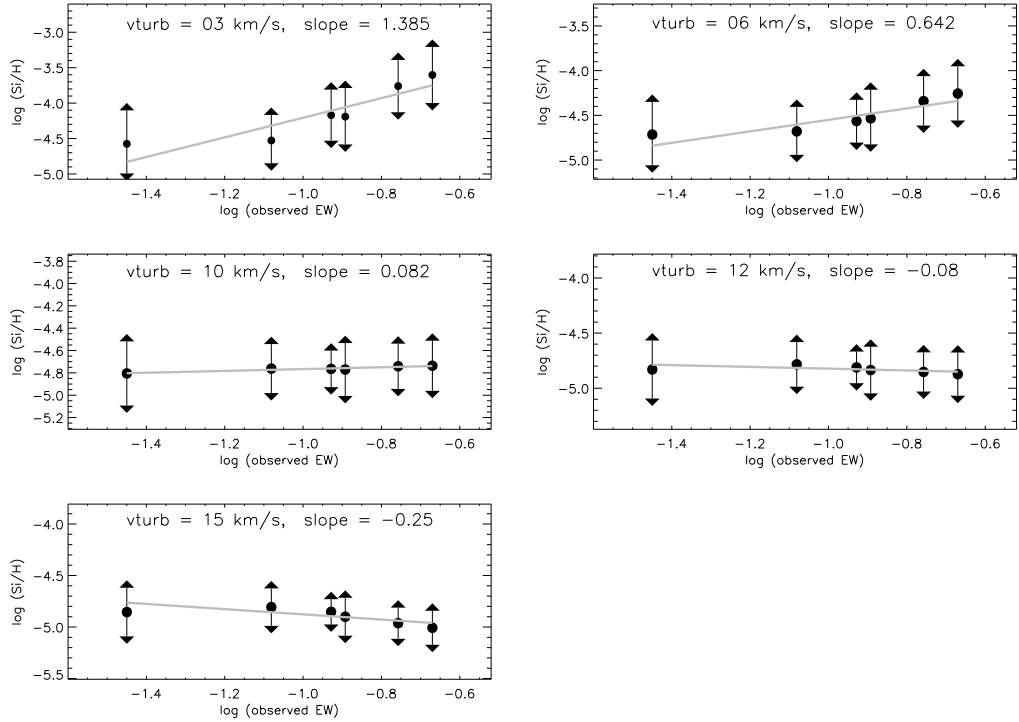}}}
\caption{\label{step2} Step 2 (for each microturbulence): we plot the Si
abundances and their uncertainties for all 6 lines for which we derived these
values in step 1, as a function of the observed EW. The least squares fit to
the abundances are shown in grey. The microturbulent velocity for which the
slope of this fit is zero (i.e. equal abundance from each line) gives an
estimate of the `interpolated' microturbulent velocity. The displayed example is
a synthetic simulation.}
\end{figure*}
\begin{figure}[ht!]
\centering
\begin{minipage}{3.4in}
\resizebox{3.4in}{!}{\includegraphics{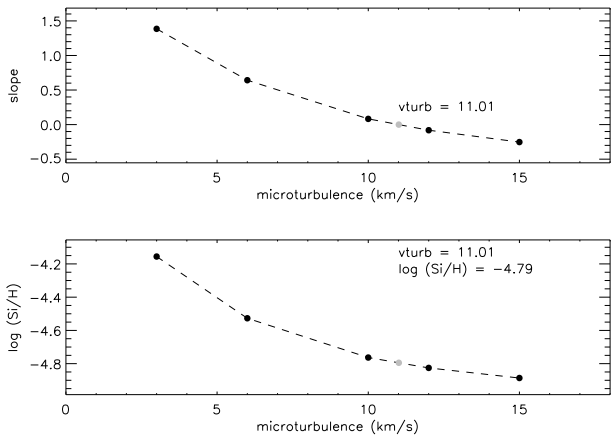}}
\end{minipage}
\caption{\label{step3} Steps 3 \& 4: the `interpolated' microturbulent velocity
and the accompanying `interpolated' Si abundance (indicated in grey) are derived
from the position where the slope of the best fit (step 2) is zero. The Si
abundances, given in the lower panel, are the \textit{mean} abundances derived
in step 2.  The displayed example is a synthetic simulation.}
\end{figure}

\begin{figure*}[ht!]
\centering
\resizebox{5in}{!}{\includegraphics{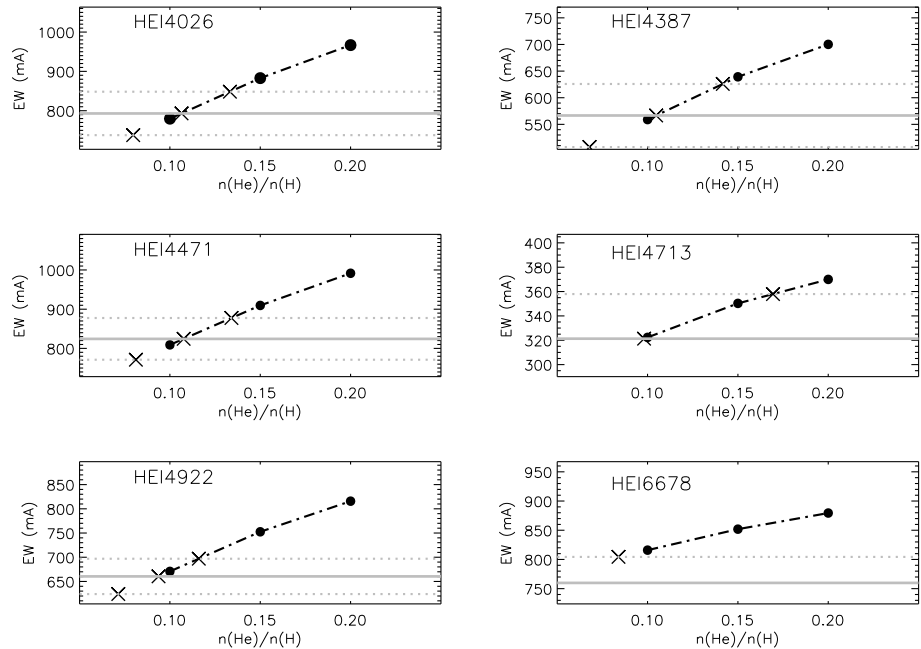}}
\caption{For each He line, the most suited He abundance is derived
through interpolation. The theoretically predicted values are shown as filled
circles, the interpolated values as crosses. The horizontal lines indicate the
observed equivalent width and the observed errors. The displayed example is
a synthetic simulation.}
\label{he_abun1}
\end{figure*}

\paragraph{Method 2: In case there is only one ionization stage of Si
available\label{teff2_cycle}} (mostly for late B-type stars, for which
we only have Si\,{\sc ii}, but also for few hotter objects around $\Teff =
23\,000~K \pm 3000~K$ with Si\,{\sc iii}) only,
we can no longer use the ionization balance to derive
information about the effective temperature, independent of the abundances
of He and Si, and independent of the microturbulence. Indeed, we have only two
\textit{known} factors (in the case of late B-type stars: EW of He\,{\sc
i} and EW of Si\,{\sc ii}) to derive four \textit{unknown} parameters
($\Teff$, $\xi$, $\logsi$, n(He)/n(H)). Fortunately, we can constrain the
effective temperature quite well.  For instance, in the cold B-type domain,
He\,{\sc i} is very sensitive to changes in effective temperature (see, e.g.,
Fig.\,2 in \citealt{Lefever2007a}), which means that we can use the joint
predictive power of He\,{\sc i} and Si\,{\sc ii} to derive a set of
\textit{plausible} values for $\Teff$. This is done by comparing the theoretical
EWs of all combinations of Si abundance, microturbulence and effective
temperature (within a range $\pm$ 4\,000~K around the currently investigated
$\Teff$) with the observed EWs, and selecting the `best' combinations from 
an appropriate log-likelihood function (for details, see
\citealt{Lefever2007PhD}). 
This method does not only work for the cool domain, but also for those hotter
objects where only Si\,{\sc iii} is present, in this case because the EW of
Si\,{\sc iii} varies much faster than the EW of He\,{\sc I}.

For each of the derived {\it plausible} $\Teff$ values, we still
have to find the corresponding Si (from the Si lines) and He abundances (from
the He lines), and the microturbulence (from the Si and/or He lines).
As the He~lines are liable to Stark broadening, we use only Si~lines to derive
the microturbulent broadening. This means that we essentially have to
derive only three unknown parameters from two known data: two from the Si~lines
($\logsi$ and $\xi$), and one more from the He~lines
(n(He)/n(H), assuming that the microturbulence in Si and He is
similar), which still leaves us with one free parameter for which we
will have to make an assumption in order to be able to fix the other two. We
have chosen to `fix' the Si abundance, as it has no direct influence on the
He~lines. However, we \textit{do} consider each of the three possibilities for
the Si~abundance in the grid, i.e., we consecutively consider the three cases in
which Si is depleted, solar and enhanced. This allows us to determine also the
microturbulent velocity, which is consequently used to fix the He~abundance from
the He~lines. 

To finally decide which of the various combinations matches the
observations best, a twofold check is applied. (i) all profile-sets are
inspected by eye (this, again, compromises a fully automatic procedure),
since parameter combinations leading to obvious mis-fits 
can be clearly excluded on this basis. (ii) Again, a log-likelihood procedure 
is used to find the best matching model among the list of final solutions,
combining the likelihoods of the EWs of the Si/He lines and of the profile
shapes of {\it all} lines, respectively (see \citealt{Lefever2007PhD}).

Even though, with this procedure, we are able to derive reasonably good
estimates for most of the stellar/wind parameters, we have no means to
restrict the Si abundance in a reliable way as it is possible within `method
1'. The outcome of maximizing the log-likelihood can give us only a general
idea of the abundance, but still leaves us with a larger uncertainty, which
has been accounted for in the adopted error budget (see
Table\,\ref{table_parameters}).

\subsection{Macroturbulence \vmacro \label{vmacro_cycle}}

The macroturbulence considered is characterized by a radial-tangential model
\citep{Gray1975}. In this model a time-independent velocity field is
assumed to be active and modeled by assuming that a fraction ($A_r$) of the
material is moving radially while the complementary
fraction moves tangentially ($A_t$). In our model for $\vmacro$, we use the
standard assumption that the radial and the tangential components are equal in
fraction. Assuming additionally that the distribution of the velocities in the
radial and the tangential directions are Gaussian, the effect of
macroturbulence can be introduced in the synthetic line profiles by means of a
convolution with a function, which consists of a Gaussian and an additional
term proportional to the error function (see equation (4) in
\citealt{Gray1975}). It changes the shape of the line profile, but leaves
the equivalent width untouched.

\begin{figure}[ht!]
\centering
\begin{minipage}{3.4in}
\resizebox{\hsize}{!}{\includegraphics{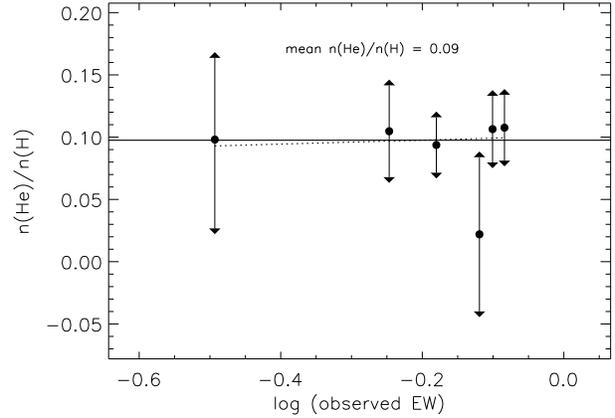}}
\end{minipage}
\caption{The finally provided He abundance is calculated as the mean over all lines.
The abundance derived from each line separately is represented by the filled 
circles, while the arrows show the derived errors. The best linear fit is
indicated as the dotted line, while the horizontal solid line shows the mean
value. The displayed example is a synthetic simulation.}
\label{he_abun2}
\end{figure}

\begin{figure}[ht!]
\centering
\begin{minipage}{3.4in}
\resizebox{3.0in}{!}{\includegraphics{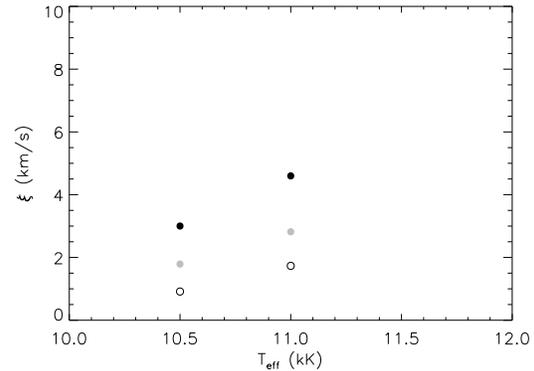}}
\end{minipage}
\caption{Once a set of possible effective temperatures is fixed from the joint
predictive power of He\,{\sc i} and Si\,{\sc ii} (see text), we can determine
for each of these temperatures (in this case: 10\,500 and 11\,000~K) the correct
microturbulence, under the assumption that we know the Si abundance. Resulting
combinations are shown for $\logsi$ = $-4.19$ (open circles), $-4.49$ (grey
filled circles) and $-4.79$ (black filled circles). The displayed example is a
synthetic
simulation.}
\label{sivt_onestage}
\end{figure}

\begin{figure}[t!]
\centering
\begin{minipage}{3.4in}
\resizebox{\hsize}{3.2in}{\includegraphics{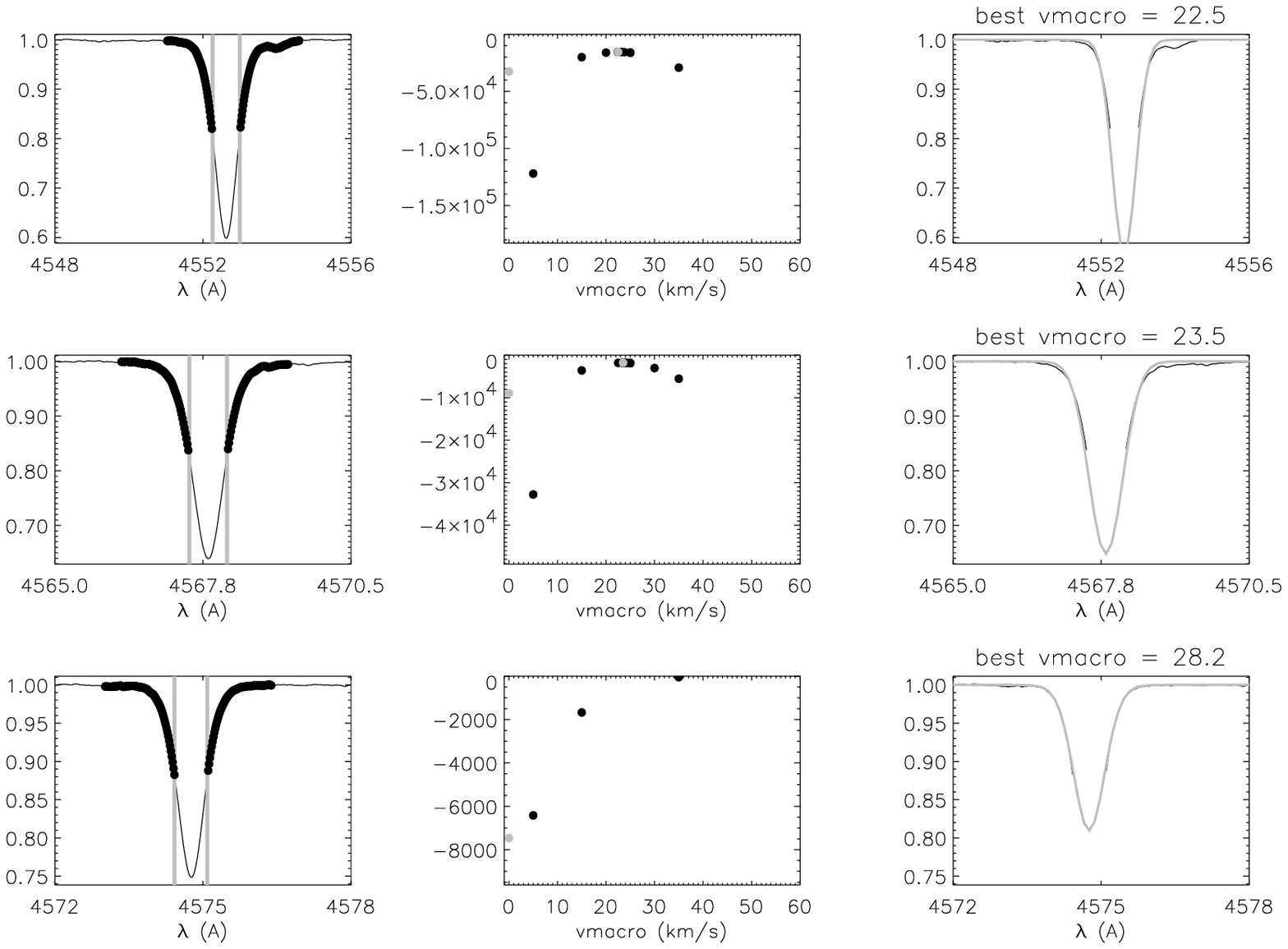}}
\end{minipage}
\caption{\label{vmacrodetermination}
Determination of the macroturbulent velocity of $\beta$ CMa for three different
Si lines. The panels should be read from left to right as follows.\newline
Left: The left and right edge of the line region  (i.e.\ outermost dots at
continuum level) are indicated by the user at the start of the procedure.
The blue and red edges (grey vertical lines) are determined as the point where
the flux is half of the minimal intensity. They determine how far the
wings extend towards the line center. The black dots represent the points that
are finally used to determine the macroturbulence.\newline
Middle: Obtained log-likelihood as a function of \vmacro\ (see
Eq.\,(\ref{Eq_chisq_vmacro})). Note the broad maximum in the
log-likelihood distribution, which gives rather large error bars.\newline
Right: The fit with the best \vmacro\, is shown as the grey profile.\newline
Note that the significant amount of macroturbulence might be explained by the
fact that this star exhibits non-radial oscillations \citep{Mazumdar2006}.}
\label{vmacro_illustration}
\end{figure}

Only weak lines should be used to determine \vmacro. For B-type
stars, Si
lines are the best suited, since they most obviously show the presence of
\vmacro\, in their wings. The user decides which Si lines should be used
throughout the full process. To determine the strength of \vmacro, we convolve
the Si profiles of the (at that instant) best fitting model with different values
of \vmacro\footnote{At this point the profile has already been convolved with the
appropriate rotational and instrumental profiles.}.
The consecutive values considered for the convolution are chosen using a
bisection method, with initial steps of 10~km\,s$^{-1}$.
The bisection continues as long as the stepsize is larger than or equal to
0.1 km\,s$^{-1}$, which will be the final precision of the 
\vmacro\,determination.
For each considered \vmacro-value, we compute its log-likelihood in order to
quantify the difference between the obtained synthetic profile and the observed
line profile, as follows:
\begin{equation}
\label{Eq_chisq_vmacro}
l \equiv \sum_{i=1}^{n}\left[- {\rm ln}(\sigma) -
{\rm ln}(\sqrt{2\,\pi}) - \frac{1}{2} \left(\frac{y_i -
\mu_i}{\sigma}\right)^2\right],
\end{equation}
with $n$ the considered number of wavelength points within the line profile,
$y_i$ the flux at wavelength point $i$ of the observed line profile, $\mu_i$ the
flux at point $i$ of the synthetic profile (convolved with the $\vmacro$ under
consideration), and $\sigma$ the noise, i.e.\ 1/SNR, of the considered line
profile.
A thorough discussion of why this particular likelihood function is suitable for
spectral line fitting can be found, e.g., in \citet{Decin2007}.

The same procedure is repeated for all considered line profiles, and the mean,
derived from the different lines, gives the final \vmacro. The error is set
by the maximal deviation of the derived $\vmacro$-values from the final
value. From Fig.\,\ref{vmacrodetermination}, it is clear that the shape of the
log-likelihood distribution near the maximum can be quite broad,
which indicates that the errors in $\vmacro$ with respect to the log-likelihood 
are significant.

A warning has to be made here. It was recently shown by \citet{Aerts2009} that
time-dependent line profile variations due to non-radial gravity-mode
oscillations, which are expected in B stars, offer a natural physical
explanation for the occurrence of macroturbulence. Such oscillations disturb
an appropriate estimate of $\vsini$, both from the Fourier method and from a
goodness-of-fit approach. The only way to assess the effect of oscillations on
the $\vsini$ determination is by taking an appropriate time series of line
profile variations (e.g., \citealt{Aerts2003}). Luckily, an inappropriate
$\vsini$ estimate, accompanying a line profile fit in terms of $\vmacro$, does
not have a serious effect on the determination of the other parameters of the
star from snapshot spectra, such that the approach implemented in AnalyseBstar 
is valid as long as the values of $\vsini$ and $\vmacro$ are not used for
physical interpretations. A large value for the macroturbulence is an
indication that time-dependent pulsational broadening may occur in the star
under investigation.

\subsection{Surface gravity \logg\,\label{logg_cycle}}

The surface gravity \logg\, is the result of fitting the Balmer lines. We mainly
use H$\gamma$ and H$\delta$, possibly complemented with
H$\beta$ in case of weak winds (i.e. if $\rm \logQ \leq -13.80$).
Since H$\epsilon$ is somewhat blended, and the merging is not always reliable
for this region (see above), this line will not be used as a 
primary gravity determinator, but will only serve as an additional check
(together with H$\beta$, in case of stronger winds). Since we want to define
the profile only from the extreme wings down to the strongest
curvature (see discussion of the different contributions below), we excluded
the central part of the line, accounting for several mechanisms, which
affect the core of the Balmer lines to some extent, e.g., rotation,
micro- and macroturbulence and thermal broadening. Once we removed the inner
part of the line profile, we compare the observed Balmer lines with the
synthetic line profiles, by considering all possibilities for $\logg$ which
are available at this given effective temperature gridpoint. We decide which
gravity is the best, by maximizing its log-likelihood (see
Eq.\,\ref{Eq_chisq_vmacro}), calculated over all Balmer lines. With this
procedure, we encountered the following problem: H$\gamma$ and H$\delta$
(our main gravity indicators) can sometimes be seriously affected by a large
amount of blends. Ignoring them can lead to a serious overestimation of the
gravity, since the log-likelihood function will take all these blends into
account (see Fig.\,\ref{hg_blends}). Therefore, we developed a procedure to
find the ``line continuum'', cutting away the blends. The procedure is
basically a sigma-clipping algorithm which keeps only those points that have
a flux higher than both neighboring points (the so-called `high points')
and where, at the same time, the difference is less than 1/SNR. After
removing all other flux points, only the ``line continuum'' is left (see
middle panel Fig.\,\ref{hg_blends}). We interpolate this continuum to obtain
the flux at all original wavelength points, and the inner part of the
line is added again (i.e.~the part between the last blend in the blue wing
and the first blend in the red wing, but obviously still without the central
core). 
Also all original flux points which deviate by less than 1/SNR from
the interpolated ``line continuum'' are included again. In this way we keep
only the flux points which really determine the shape of the wing to fit the
synthetic Balmer wings, and we are able to determine an accurate value for
\logg\,(see right panel Fig.\,\ref{hg_blends}). We realize that some points
that may be marked as local continuum in this way, may still be lower than
what the real local continuum level would be, because of the transition of
one blend into another. However, this will only be the case for very few
points, which will give no significant weight to the log-likelihood.

The finally accepted surface gravity will be this \logg\, which gives the
best fit to \textit{all} selected Balmer lines simultaneously. Half the
gridstep in \logg\,could thus be considered a good error estimate. However, to
account for the coupling with \Teff, we consider 0.1~dex as a more appropriate
error. This gravity should be corrected for centrifugal acceleration
($\rightarrow \loggc$) due to
stellar rotation by a factor $(\vsini)^2/\Rstar$, when calculating, e.g., the
mass of a star \citep[ and references therein]{Repolust2005}.

\begin{figure}[t!]
\centering
\begin{minipage}{3.4in}
\resizebox{3.4in}{!}{\includegraphics{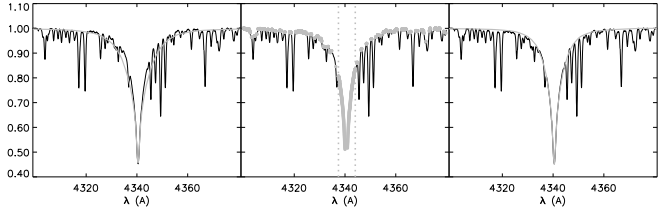}}
\end{minipage}
\caption{\label{hg_blends} Illustration of the effect of blends on the
determination of \logg, and the clipping algorithm to improve the fit, for
$\beta$ CMa.
Left: The many blends in the wings of the profile prevent an accurate least
squares fit and lead to too broad wings, implying too high a gravity.
Central: From the observed line profile (black), only the `high points' (grey
dots), which deviate less than 1/SNR from their neighbors, are kept,
complemented with the inner part of the line between the last blend in the blue
wing and the first blend in the red wing (indicated as grey vertical
lines), but still excluding the central core, which is affected by several
broadening mechanisms.
Right: After the removal of all blends through sigma clipping (see text), we
are able to obtain a good representation of the line wings (grey profile) and,
therefore, to derive a very accurate value for the gravity.}
\end{figure}

\subsection{Wind parameters}

A change in mass loss rate will mainly affect the shape of $\Ha$. For
cool objects and weak winds, $\Ha$ is nearly `photospheric' and will, in
essence, be an absorption profile, with more or less symmetric components,
since the `re-filling' by the wind is low. In case of stronger winds, $\Ha$
will take the shape of a typical P~Cygni profile.
For hot objects, where $\Ha$ is dominated by recombination processes,
and high mass-loss rates, the profile may even appear in full emission.

The wind parameter $\beta$ determines the velocity law, which directly
influences the density. The red wing of $\Ha$ is well suited to determine
$\beta$, since it is formed by emission processes alone, averaged over the
receding part of the (almost) complete wind.
For a fixed mass loss, a `slower' velocity law (i.e., a higher $\beta$ value)
will result in higher densities in the lower
atmosphere, close to the star. This enlarges the number of emitted photons
with velocities close to the line center, resulting in more emission. Around
the central wavelength, the absorption component of the line profile refills
and the emission component becomes stronger. Therefore the slope of the red wing
of the P Cygni profile becomes steeper. In this sense, the steepness of the red
wing is a measure for the value of $\beta$.\newline
We estimate the wind parameters $\logQ$ and $\beta$ by comparing
the observed \Ha\, profile with the different synthetic \Ha\, profiles
by making combinations of $\logQ$ and $\beta$.
We decided to make the determination of the wind parameters not too
sophisticated, by simply using the best values for $\logQ$ and $\beta$
available in the grid, without interpolation and further refinement.
Once we have selected for each $\logQ$ the best $\beta$, we decide on
the best combination by considering the best fit to the entire profile.

\begin{figure}[t!]
\centering
\begin{minipage}{3.4in}
\resizebox{3.4in}{4in}{\includegraphics{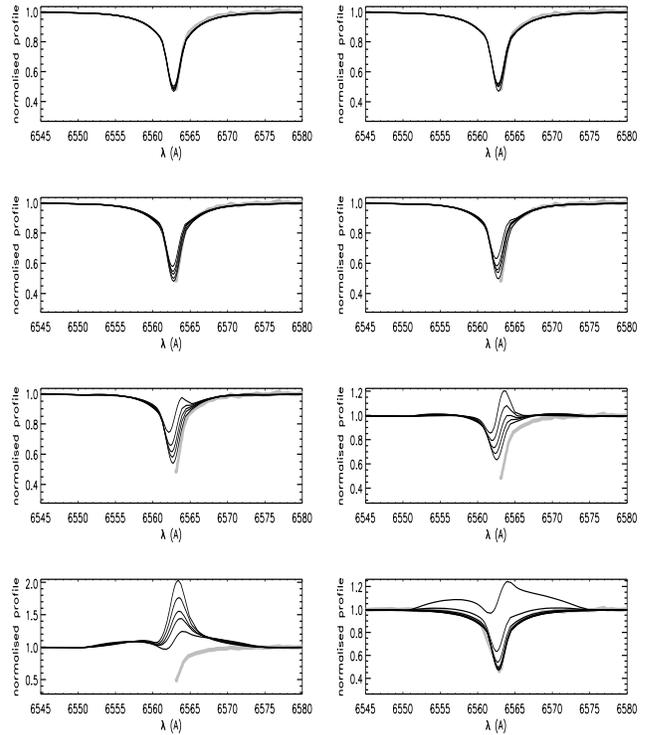}}
\end{minipage}
\caption{\label{ha_Qbeta} (Synthetic simulation) For each different wind
strength parameter $\logQ$ (7 values, from left to right and from
top to bottom: $\logQ = -14.30, -14.00, -13.80, -13.60, -13.40, -13.15,
-12.70$), we search for the best $\beta$ (in each panel, the 5
different values for $\beta$ are indicated: 0.9, 1.2, 1.5, 2.0, 3.0) by
comparing only the red wing (grey part of the profile). Then the
synthetic profiles of each best ($\logQ$, $\beta$) combination are
compared to the entire \Ha\,profile (bottom right).}
\end{figure}

\subsection{Some final remarks}

At the end of this iteration cycle, an acceptance test is performed.
When the method can run through the whole cycle without needing to update
any of the fit parameters, the model is accepted as a good model, and gets the
flag `2'. If a better model was found (i.e., when one or more parameters
changed), the initial model is rejected and gets flag `0', whereas the improved
model is added to the list of possibilities, and gets flag `1' (cf. `models to
check' in the flowchart diagram, Fig.\,\ref{page_flowchart}).
As long as the list with possibilities with flag `1' is not empty, the
parameters of the next possibility are taken as new starting values.

Line profile fits are always performed allowing for a small shift in
wavelength, or radial velocity
(5 wavelength points in either direction, corresponding to typically
5\,km\,s$^{-1}$), to prevent flux differences to add up in case of a small
radial velocity displacement.

\subsection{The underlying FASTWIND BSTAR06 grid \label{section_grid}}

Even though the majority of the CoRoT targets consisted of late B-type stars, we
nevertheless invested in a grid which covers the complete parameter space
of B-type stars. In this way, we hoped to establish a good starting point for
future (follow-up) detailed spectroscopic analysis of massive stars. The grid
has been constructed as representative and dense as possible for a wide
variety of stellar properties within a reasonable computation time. Hereafter we
give a short summary of the considered parameters. For a more detailed
description, we refer to \citet{Lefever2007a}.

\begin{itemize}
 \item 33 \textit{effective temperature} (\Teff): from 10\,000~K to
32\,000~K, in steps of 500~K below 20\,000~K and in steps of 1\,000~K above it.
 \item on average 28 \textit{surface gravities} (\logg) at each
effective temperature point: from $\logg$ = 4.5 down to 80\% of the Eddington
limit
 \item 1 `typical' value for the \textit{radius} (\Rstar) for each
($\Teff$, $\logg$)-gridpoint: approximative value from interpolation between
evolutionary tracks, keeping in mind that a rescaling to the `real' value is
required when analyzing specific objects. Indeed, the real radius can then, in
most cases, be determined from the visual magnitude, the distance of the star
and its reddening, and from the theoretical fluxes of the best fitting model.
 \item 3 values to fix the \textit{chemical composition}: n(He)/n(H) =
0.10, 0.15, and 0.20, and $\logsi$ = -4.19, -4.49, and -4.79 (i.e. enhanced,
solar, and depleted). The background elements (responsible, e.g., for radiation
pressure and line-blanketing) were assumed to have a solar composition.
 \item 7 values for the \textit{wind-strength} parameter $\logQ$
 \item 1 `typical' value for the \textit{terminal wind velocity}
($\vinf$): estimated from the relation between observed terminal velocity 
and photospheric escape velocity \citep[ and references
therein]{Kudritzki2000}, where the latter quantity has been calculated from
the actual grid parameters ($\logg, \Rstar$ and $\Teff$).
\end{itemize}

\end{document}